\documentclass{aa}  

\usepackage{natbib}
\usepackage{graphicx}
\usepackage{txfonts}
\usepackage[flushleft]{threeparttable}
\usepackage{multirow}
\usepackage{makecell}
\usepackage{supertabular}
\usepackage{amsmath}
\usepackage{graphicx}
\usepackage{textcomp}

\begin{document} 

\title{The EDIBLES survey}
\subtitle{VII. A survey of C$_2$ and C$_3$ in interstellar clouds}

   \author{
Haoyu Fan \inst{1,2} \and
Carlos M. R. Rocha \inst{3} \and
Martin Cordiner \inst{4, 5} \and
Harold Linnartz \inst{3} \and
Nick~L.~J.~Cox \inst{6} \and
Amin Farhang \inst{7,8} \and
Jonathan Smoker \inst{9,10} \and
Evelyne Roueff \inst{11} \and
Pascale Ehrenfreund \inst{12} \and
Farid Salama\inst{13} \and
Bernard H. Foing \inst{14,12} \and
Rosine Lallement \inst{15} \and
Heather MacIsaac \inst{1,2} \and 
Klay Kulik \inst{1,2} \and
Peter Sarre \inst{16} \and
Jacco Th. van Loon \inst{17} \and
Jan Cami \inst{1,2,18}
           }

   \institute{
{Department of Physics and Astronomy, The University of Western Ontario, London, ON N6A 3K7, Canada}
\and
{Institute for Earth and Space Exploration, The University of Western Ontario, London, ON N6A 3K7, Canada}
\and
{Laboratory for Astrophysics, Leiden Observatory, Leiden University, PO Box 9513, 2300 RA Leiden, The Netherlands}
\and
{Astrochemistry Laboratory, NASA Goddard Space Flight Center, Code 691, 8800 Greenbelt Road, Greenbelt, MD 20771, USA}
\and
{Department of Physics, The Catholic University of America,
Washington, DC 20064, USA}
\and
{Centre d’Etudes et de Recherche de Grasse, ACRI-ST, Av. Nicolas Copernic, Grasse 06130, France}
\and
{School of Astronomy, Institute for Research in Fundamental Sciences, 19395-5531 Tehran, Iran}
\and
{Department of Physics, University of Tehran, North Karegar Ave, 14395-547 Tehran, Iran}
\and
{European Southern Observatory, Alonso de Cordova 3107, Casilla 19001, Vitacura, Santiago, Chile}
\and
{UK Astronomy Technology Centre, Royal Observatory, Blackford Hill, Edinburgh EH9 3HJ, UK }
\and
{LERMA, Observatoire de Paris, PSL Research University, CNRS, Sorbonne Universite`s, 92190, Meudon, France}
\and
{Leiden Observatory, Leiden University, PO Box 9513, 2300 RA Leiden, The Netherlands}
\and
{NASA Ames Research Centre, Space Science \& Astrobiology Division, Moffett Field, California, USA}
\and
{ESTEC, ESA, Keplerlaan 1, 2201 AZ Noordwijk, The Netherlands}
\and
{GEPI, Observatoire de Paris, PSL Research University, CNRS, Universit\'e Paris-Diderot, Sorbonne Paris Cit\'e, Place Jules Janssen, 92195 Meudon, France}
\and
{School of Chemistry, The University of Nottingham, University Park, Nottingham NG7 2RD, UK}
\and
{Lennard-Jones Laboratories, Keele University, ST5 5BG, UK}
\and
{SETI Institute, 339 Bernardo Ave, Suite 200, Mountain View, CA, 94043, USA}
             }
   \date{}

 
  \abstract
{Small linear carbon chain radicals such as C$_2$ and C$_3$ act as both the building blocks and dissociation fragments of larger carbonaceous species. Their rotational excitation traces the temperature and density of local environments.
    However, these homo-nuclear di- and triatomic species are only accessible through their electronic and vibrational features because they lack a permanent dipole moment, and high signal-to-noise ratio data are necessary as the result of their generally low abundances in the interstellar medium (ISM).}
    {In order to improve our understanding of small carbonaceous species in the ISM, we carried out a sensitive survey of C$_2$ and C$_3$ using the ESO Diffuse Interstellar Bands Large Exploration Survey (EDIBLES) dataset. We also expanded our searches to C$_4$, C$_5$, and the $^{13}$C$^{12}$C isotopologue in the most molecule-rich sightlines.}
    {We fitted synthetic spectra generated following a physical excitation model to the C$_2$ (2-0) Phillips band to obtain the C$_2$ column density ($N$) as well as the kinetic temperature ($T_\textrm{kin}$) and number density ($n$) of the host cloud. The C$_3$ molecule was measured through its $\tilde{A} - \tilde{X}$ (000-000) electronic origin band system. We simulated the excitation of this band with a double-temperature Boltzmann distribution.}
    {We present the largest combined survey of C$_2$ and C$_3$ to date in which the individual transitions can be resolved. In total, we detected C$_2$ in 51 velocity components along 40 sightlines, and C$_3$ in 31 velocity components along 27 sightlines. Further analysis confirms the two molecules are detected in the same velocity components. We find a very good correlation between $N$(C$_2$) and $N$(C$_3$) with a Pearson correlation coefficient $r = 0.93$ and an average $N$(C$_2$)/$N$(C$_3$) ratio of 15.5$\pm$1.4. A comparison with the behaviour of the C$_2$ diffuse interstellar bands
(DIBs) shows that there are no clear differences among sightlines with and without detections of C$_2$ and C$_3$. This is in direct contrast to the better-studied non-C$_2$ DIBs, which have reduced strengths in molecule-rich environments, consistent with the idea that the C$_2$ DIBs are indeed a distinguishable DIB family. We also identify, for the first time, the $Q$(2), $Q$(3), and $Q$(4) transitions of the $^{13}$C$^{12}$C (2-0) Phillips band in the stacked average spectrum of molecule-rich sightlines, and estimate the isotopic ratio of carbon $^{12}$C/$^{13}$C to be 79$\pm$8, consistent with literature results. At this stage it is not yet possible to identify these transitions in individual sightlines. Our search for the C$_4$ and C$_5$ optical bands was unsuccessful; even in stacked spectra no unambiguous identification could be made.}
    {}

   \keywords{ISM: lines and bands - ISM: clouds – ISM: molecules
               }

   \titlerunning{The EDIBLES Survey. VII. A survey of C2 and C3 in interstellar clouds}
   \authorrunning{Fan et al.}
   
   \maketitle
%
\section{Introduction}\label{sec:intro}
Many carbon chain species have been detected via sub-millimetre and/or radio observations. However, due to the lack of a permanent dipole moment, bare carbon chains can only be accessed through their electronic and vibrational features. The simplest bare carbon molecule, interstellar C$_2$, was first detected towards the highly reddened star Cyg~OB2~12 through its \textit{A-X} (1-0) Phillips band in the near-IR \citep{1977ApJ...216L..49S}. This was followed by more extensive observations of other bands and towards more targets \citep[e.g.][]{1982ApJ...254..108H, 1984MNRAS.206..383V, 2007ApJS..168...58S, 2010A&A...518A..36W}. The population on the C$_2$ rotational excitation ladder is determined simultaneously by collisional and radiative processes \citep{1982ApJ...258..533V}, which makes C$_2$ a powerful diagnostic tool for deriving local environmental factors, such as the kinetic temperature and density within the interstellar cloud.

An emission band around 4051~\AA\ was first noted in the spectrum of
comet Tebbutt \citep{1881RSPS...33....1H}, and
\cite{1951ApJ...114..466D} subsequently assigned it to
C$_3$. Observations of this band in diffuse interstellar environments
have been difficult due to the much lower column density of C$_3$; the
first tentative detection was made in 1995 in the highly reddened
sightline towards HD~147889 \citep{1995ApJ...453..450H}. C$_3$
molecules were also observed with \textit{Herschel} in the $\nu_2$
bending mode towards Sagittarius B2 and star-forming cores
\citep{2000ApJ...534L.199C, 2010A&A...521L..13M}. To date, only a
handful of measurements of the C$_3$ 4051~\AA\ band are available in
the literature, and the data are mostly sightlines known to be
molecule-rich \citep{2001ApJ...553..267M, 2002A&A...384..629R,
  2002A&A...395..969G, Oka2003, 2003ApJ...595..235A,
  2014MNRAS.441.1134S}. In addition to the C$_3$ origin band,
\citet{2014MNRAS.441.1134S} also reported electronic transitions
involving vibrationally excited C$_3$ in its $\tilde{A}$ state. 
High-resolution spectra are required to accurately measure the C$_3$ column density. Once these measurements are available,  individual transitions can be resolved, allowing the rotational excitation to be characterised, and the contribution from the higher $J$-levels to be determined \citep[see e.g.][]{Oka2003}.

Small carbon molecules and chains are considered building blocks of
more complex carbon-bearing compounds \citep{2009A&A...495..513W}, and
column density ratios of different carbon chains help us constrain the
chemistry that drives their formation and destruction. The destruction
process may also involve the photo-dissociation of (larger) polycyclic
aromatic hydrocarbons \citep[PAHs;][]{HRODMARSSON2022116834}. C$_4$ and C$_5$ molecules may be abundant in protoplanetary nebulae and circumstellar shells around evolved stars \citep{1989Sci...244..562B, 2002ApJ...580L.157C, 2014MNRAS.444.3721H}, yet attempts to search for their spectral fingerprint in diffuse and translucent environments have been unsuccessful so far \citep{2002ApJ...566..332M, 2004ApJ...602..286M}. Additionally, the isotopic ratio of carbon may trace Galactic chemical evolution \citep{2005ApJ...634.1126M}. \cite{2019ApJ...881..143H} report a marginal detection of the $Q$(3) transition of the $^{13}$C$^{12}$C (0,0) Phillips band towards Cyg~OB2~12 \citep[see also][]{2001A&A...375..553G}, and \cite{2020A&A...633A.120G} reported the first detection of $^{13}$CCC and C$^{13}$CC through their ro-vibrational transitions at 1.9 THz. The measurement of an isotopologue also provides insight into local peculiarities since the isotope exchange rate that sets the isotopic fraction depends on environmental parameters such as temperature \citep[see e.g.][]{2021A&A...647A.142R, 2009A&A...503..323V}.

Since their first discovery in the 1920s \citep{1922LicOB..10..146H},
more than 500 diffuse interstellar bands (DIBs) have been identified
in the near-UV, optical, and near-IR regions
\citep{2008ApJ...680.1256H, 2009ApJ...705...32H,2011Natur.479..200G,
  2019ApJ...878..151F}. Despite this century-long history, the
carriers of DIBs remain unknown, except for the five near-IR DIBs that
were recently attributed to C$_{60}^+$ \citep{2015Natur.523..322C,
  2015ApJ...812L...8W, 2019ApJ...875L..28C,
  2022arXiv220312562E}. High-resolution observations reveal
substructures within several DIBs that suggest they have molecular origins \citep{1995MNRAS.277L..41S, 1998ApJ...495..941K, 2004ApJ...611L.113C}. Molecules of different types have been proposed as DIB carriers -- from complex structures, such as (charged) fullerenes and PAHs \citep[see e.g.][and their citations]{1994Natur.369..296F, 2011A&A...530A..26G, 2011ApJ...728..154S, 2014IAUS..297..364S}, to smaller molecules, such as carbon chains of different sizes (\citealt{2000ApJ...531..312M, 2013ApJ...773...42O, C4CS00049H}; see \cite{1995ARA&A..33...19H} for a review) -- despite the fact that none of their absorptions fully overlap with DIB features.

 C$_2$ and C$_3$ molecules are present in the dense molecular regions
 of diffuse and translucent cloud environments. Their detection
 provides physical information on interstellar medium (ISM)
 environmental conditions, such as the kinetic temperature inferred
 from their rotational excitation. The column density ratios between
 C$_2$ and the `first-generation molecules' H$_2$ and CH can trace the average ISM conditions of the sightline \citep{2003ApJ...584..339T, 2013MNRAS.428.1107W}, similar to the use of the molecular hydrogen fraction, $f_{\rm H_2}$, but presumably tracing deeper regions in the interstellar clouds \citep[e.g.][]{2017ApJ...850..194F}. This information in turn can be compared to the behaviour of DIBs and provides constraints on their carriers \citep[e.g.][]{1997A&A...326..822C, 1997A&A...327.1215S}. This may eventually lead to the identification of the C$_2$ DIBs, whose equivalent widths become larger when the sightline is more abundant in C$_2$ \citep{2003ApJ...584..339T}.

In this paper we present a survey of C$_2$ and C$_3$ signals in the EDIBLES data. This dataset is briefly introduced in Sect. \ref{sec:data}. We present our measuring methods and results for C$_2$ in Sect. \ref{sec:C2} and for C$_3$ in Sect. \ref{sec:C3}. These results are discussed in Sect. \ref{sec:comparison}, and we link these results to the behaviour of DIBs in Sect. \ref{sec:C2DIB}. Section \ref{sec:C4C5} describes our efforts to search for the $^{13}$C$^{12}$C, C$_4$, and C$_5$ transitions. We summarise our conclusions in Sect. \ref{sec:summary}.


\section{Observations and spectral data}\label{sec:data}
The ESO Diffuse Interstellar Bands Large Exploration Survey (EDIBLES)
is a sensitive spectroscopic survey of a large sample of DIB
sightlines towards early-type stars using the UV-visual echelle
spectrograph \citep[UVES;][]{UVES1,UVES2} mounted on the Very Large
Telescope (VLT). EDIBLES contains observations towards 123 sightlines
with high-spectral resolution (R$\sim$80,000 in the blue arm and
$\sim$100,000 in the red arm), high sensitivity (signal-to-noise ratio
S/N $\sim$400--1,000 in the optical region), and covering a wide
wavelength range (305--1042~nm). The program is described in
\citet{EDIBLESI_Data}, and we also refer the reader to earlier
publications that used EDIBLES data \citep[i.e.][]{EDIBLESII_C60,
  EDIBLESIII_C2, EDIBLESIV_OH, 2022A&A...662A..24M} for more details.

In this paper, we focus on the C$_2$ (2-0) Phillips system around 8757~\AA, and the C$_3 \: \tilde{A} - \tilde{X}$ origin band around 4051~\AA. All spectra used in this work were obtained with the 437+860 setting of UVES. For sightlines with multiple observations, we interpolated, normalised, and co-added all available data into a single and high-quality spectrum for the final analysis.

\section{C$_2$ models and measurements}\label{sec:C2}
\begin{figure*}
    \includegraphics[width=\textwidth]{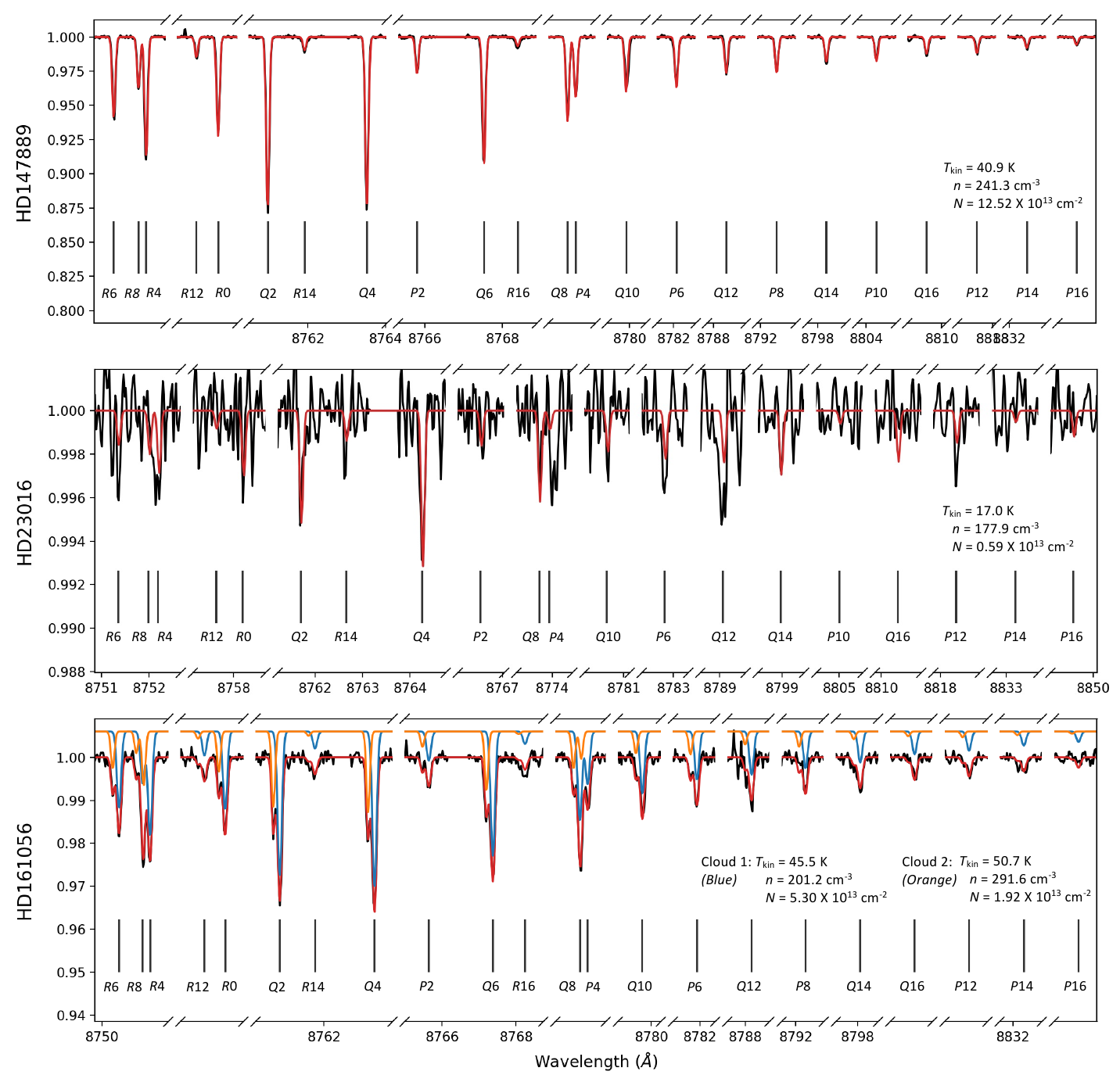}
    \caption{Spectral segments from three representative sightlines for the C$_2$ (A-X) Phillips (2-0) band. HD~147889 (top) has the highest C$_2$ column density in our data sample. HD~23016 (middle) shows a tentative detection, as the strongest $Q$(4) line is still below the 3$\sigma$ detection limit. We considered multiple rotational lines simultaneously when deciding about possible identifications. The C$_2$ molecules are detected in two velocity components in HD~161056 (bottom). The combined model (red) is plotted against the spectral data, while synthetic spectra for the two velocity components (blue and orange) are plotted separately with an offset. We focus on the spectral segments around the C$_2$ transitions, and the wavelength grid is hence discrete. Rotational lines being considered in each sightline may differ due to telluric interference and/or data quality. These rotational lines are labelled with vertical bars. The best-fit parameters for the kinetic temperature, $T_\textrm{kin}$, number density, $n$, and the column density of the velocity component, $N$, are listed in each panel.}
    \label{fig:C2_Demo}
\end{figure*}

Most studies target transitions from the C$_2$ $A ^1\Pi_{u} \leftarrow X ^1\Sigma _g^+$ Phillips system, especially the (1-0) band at 10144~\AA, the (2-0) band at 8757~\AA, and the (3-0) band at 7719~\AA\ \citep{1977ApJ...216L..49S, 1982ApJ...254..108H, 1984MNRAS.206..383V, 1986ApJ...307..332V, 1994ApJ...424..772F, 1999A&A...351..657G, 2012ApJ...749...48C}. Even though the EDIBLES data cover both the (2-0) and (3-0) bands, we only used the former, which is stronger and less affected by telluric lines. Figure \ref{fig:C2_Demo} shows spectral segments around the target transitions of the C$_2$ (2-0) band and the fitted model of three representative sightlines.

\subsection{Line list and modelling of the C$_2$ (2-0) Phillips band}\label{sec:C2Model}
We adopted the wavelengths and oscillator strengths ($f_{jj}$ values)
of the C$_2$ (2-0) Phillips band from \cite{2007ApJS..168...58S} for
\textit{J} up to 16 (Table \ref{tab:C2_linelist}). The $R$(2) and
$R$(10) transitions at 8753.949 and 8753.578~\AA\ were excluded due to
a nearby CCD artefact.

\begin{table}[ht]
        \centering
        \small
        \caption{Line list for the C$_2$ (2-0) Phillips bands. }
    \label{tab:C2_linelist}
        \begin{threeparttable}
      \begin{tabular}{ccc|ccc}
        \hline
        \multirow{2}{*}{Transition} & $\lambda_{jj}$\tnote{a} & $f_{jj}$ & \multirow{2}{*}{Transition} & $\lambda_{jj}$\tnote{a} & $f_{jj}$ \\
         & (\AA) & ($\times 10^{4}$) &  & (\AA) & ($\times 10^{4}$)\\
        \hline
        $R$(6) & 8750.848 & 4.31 & $Q$(8) & 8773.221 & 7.00 \\
        $R$(8) & 8751.486 & 4.12 & $P$(4) & 8773.430 & 2.34 \\
        $R$(4) & 8751.685 & 4.66 & $Q$(10) & 8780.141 & 7.00 \\
        $R$(10)\tnote{b} & 8753.578 & 4.00 & $P$(6) & 8782.308 & 2.69 \\
        $R$(2)\tnote{b} & 8753.949 & 5.60 & $Q$(12) & 8788.558 & 7.00 \\
        $R$(12) & 8757.127 & 3.92 & $P$(8) & 8792.648 & 2.88 \\
        $R$(0) & 8757.686 & 14.00 & $Q$(14) & 8798.459 & 7.00 \\
        $Q$(2) & 8761.194 & 7.00 & $P$(10) & 8804.499 & 3.00 \\
        $R$(14) & 8762.145 & 3.86 & $Q$(16) & 8809.841 & 7.00 \\
        $Q$(4) & 8763.751 & 7.00 & $P$(12) & 8817.826 & 3.08 \\
        $P$(2) & 8766.031 & 1.40 & $P$(14) & 8832.679 & 3.14 \\
        $Q$(6) & 8767.759 & 7.00 & $P$(16) & 8849.071 & 3.18 \\
        $R$(16) & 8768.627 & 3.82 \\
            \hline
          \end{tabular}
    \begin{tablenotes}
    \item Data are taken from \cite{2007ApJS..168...58S}.
      \item[a] Air wavelength.
      \item[b] Not included in the measurements due to adjacent CCD artefact.
    \end{tablenotes}
  \end{threeparttable}
\end{table}

\cite{1982ApJ...258..533V} provided a comprehensive model of the rotational excitation of C$_2$ by considering both radiative and collisional processes. The relative population of C$_2$ for different $J$-levels probes the kinetic temperature $T_\textrm{kin}$ as well as the number density $n$.  
\cite{2006ARA&A..44..367S} turned this model into an online C$_2$
Calculator \footnote{\url{http://dib.uiuc.edu/c2/}} that computes the
relative population for different $J$-levels at given the kinematic
temperature $T_\textrm{kin}$ and the H + H$_2$ number density $n$. We generated a finely sampled grid from this calculator, for $T_\textrm{kin}$ between 10 and 100~K with an interval of 1~K, and $n$ between 10 and 1000~cm$^{-3}$ with an interval of 10~cm$^{-3}$. The parameter grid is then interpolated using a 2D cubic spline.

We followed the routine described in \cite{EDIBLESIII_C2} to generate
synthetic C$_2$ spectra and assumed Gaussian optical depth profiles for all C$_2$ transitions \citep[see also][]{2007JChPh.126h4302K}. With the knowledge on the relative population on different $J$-levels, the total column density $N$, the Gaussian broadening parameter $b$, and the velocity offset $v_\textrm{off}$ (relative to the barycentric frame), we were then able to calculate the opacity spectrum for a velocity component. For each of the sightlines, opacity spectra for all velocity components where C$_2$ transitions can be discerned are co-added and then converted to flux units. This synthetic spectrum was finally convolved with a Gaussian kernel with a full width at half maximum (FWHM) of 2.8~km~$\cdot$~s$^{-1}$ that simulates the instrumental profile.

\subsection{C$_2$ measurements and results}\label{sec:C2Results}

Under typical ISM conditions, the $Q$(2), $Q$(4), and $Q$(6) are the strongest transitions of the C$_2$ (A-X) band. We first visually examined the presence of these absorption lines to screen the target sightlines for a more detailed model-fitting process. We paid close attention to the asymmetries and multiple peaks in the line profiles and compared them to the velocity components identified in the \ion{Na}{i} 3302~\AA\ doublet. Such efforts helped us determine whether the C$_2$ lines should be measured in multiple velocity components, each of which is characterised by its own set of parameters (i.e. the kinetic temperature $T_\textrm{kin}$, the number density $n$, the total column density $N$, the broadening factor $b$, and the velocity offset $v_\textrm{off}$ relative to the barycentric frame).

The rotational transitions of the  C$_2$ (2-0) Phillips band span a
$\sim$100~\AA\ wavelength range. Since they take up a very limited
fraction of the window, we chose to focus on spectral segments with
the target C$_2$ transitions without having to fit a global
continuum. These segments are 1.8~\AA\ or $\sim$60~km $\cdot$ s$^{-1}$
on each side of the target lines, and are combined when a segment
overlaps with another. We normalised these segments with third-order
polynomial continua. Telluric lines were identified from the
\texttt{HITRAN} database \citep{2021NatRP...3..302R} in this
process. In most cases, the telluric lines could be corrected by
fitting them with a Gaussian profile so their presence would not
affect the normalisation. If that was not possible, the spectral
segment was excluded from the fitting process. 

The synthetic spectrum was fitted to the normalised data using a
customised Python script based on the Sherpa package
\citep{sherpa}. The parameters in the model were optimised using a
Levenberg-Marquardt $\chi^2$ minimisation. The 1$\sigma$ errors were
estimated by independently tuning each parameter from its optimal
value until the sum of $\chi^2$ reached $\chi_{min}^2 + 1$.

The C$_2$ (2-0) Phillips band is detected at the 3$\sigma$ level in 37
sightlines, and tentatively in another 3 sightlines\footnote{We use
the term `tentative detection' when no individual transition reaches
the 3$\sigma$ detection limit set by the S/N. In such cases, we are
still able to extract some information by simultaneously fitting the
entire band, albeit with decreased accuracy}. Nine of these sightlines -- HD~41117, HD~61827, HD~112272, HD~147084, HD~149404, HD~161056, HD~183143, HD~186745, and HD~186841 -- contain two velocity components in which the target C$_2$ band is detected, while three components are detected towards HD~167971. Figure \ref{fig:C2_Demo} demonstrates spectral data and best-fit results of three representative sightlines and Appendix \ref{appx_C2} contains a gallery of all 40 sightlines. The fitted parameters and their 1$\sigma$ uncertainties are tabulated in Table \ref{table: C2 Detection}. In the table we also include literature values for targets that have been studied previously. For the sightlines in common with our study, both the kinetic temperatures and densities we derive are in good agreement with existing literature values, and typically with smaller uncertainties. 

\bigskip
\subsection{Temperatures and densities of C$_2$-bearing clouds}
\begin{figure}[t]
\resizebox{\hsize}{!}{\includegraphics{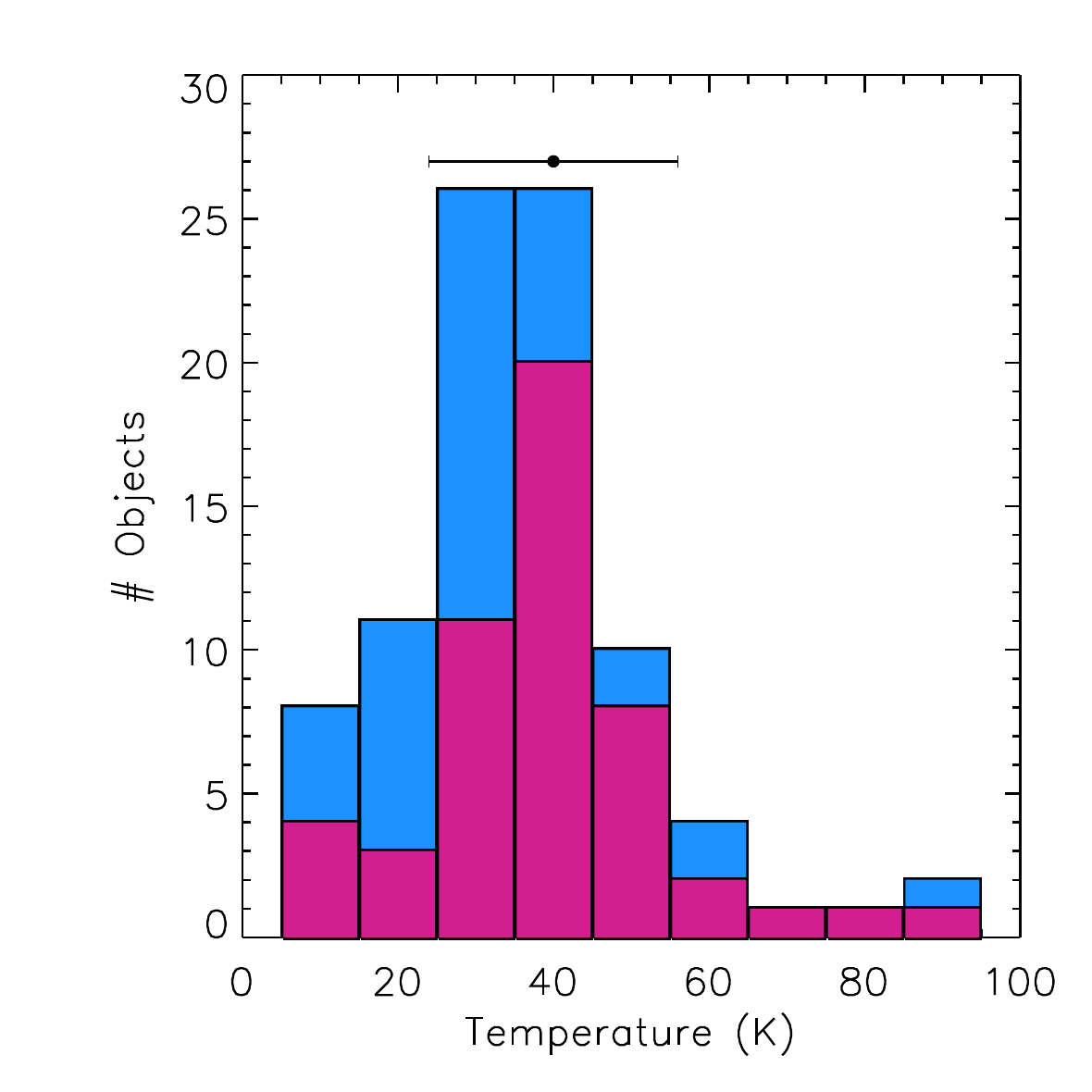}}
\resizebox{\hsize}{!}{\includegraphics{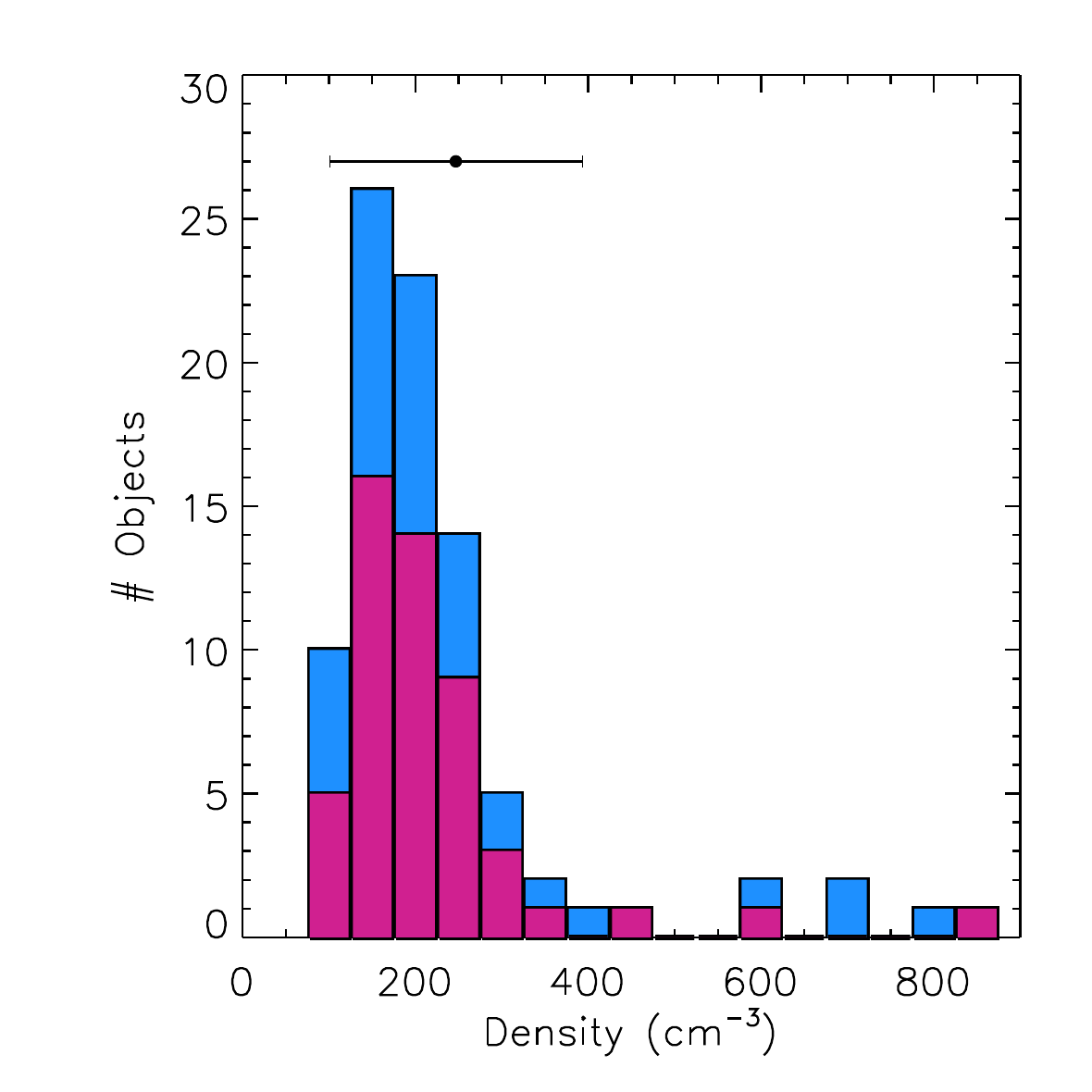}}
\caption{\label{Fig:histograms}Histograms of the kinetic temperatures (top) and densities (bottom) in C$_2$-bearing clouds. In both panels, the red histogram represents the values determined in this study; the blue histogram also includes literature values (see Table~\ref{table: C2 Detection}). The single data point with an error bar indicates the (unweighted) average of all values and the standard deviation. }
\end{figure}

We supplemented our derived values for $T_{\rm kin}$ and $n$ in 40 sightlines with literature values for an additional 36 targets (see the bottom part of Table~\ref{table: C2 Detection}) to study the distribution function of these parameters. 

Fig.~\ref{Fig:histograms} shows the distribution of the derived
kinetic temperatures and densities using results from our study (red)
as well as including literature values (blue). The temperature
distribution peaks in the 30-40~K range, with an unweighted average
kinetic temperature (using all values) of 40$\pm$16~K; we also note
that the median value is 40~K. Only four values are higher than
60~K. The density distribution is similarly strongly peaked in the
100-200~cm$^{-3}$ range. The unweighted average and standard deviation
is 247~$\pm$146~cm$^{-3}$, but the median value is 202.8~cm$^{-3}$ and
only six values are above 500~cm$^{-3}$. As can be ascertained from
Table~\ref{table: C2 Detection}, these values typically have large
uncertainties.

\section{C$_3$ models and measurements}\label{sec:C3}
\begin{figure*}
    \includegraphics[width=\textwidth]{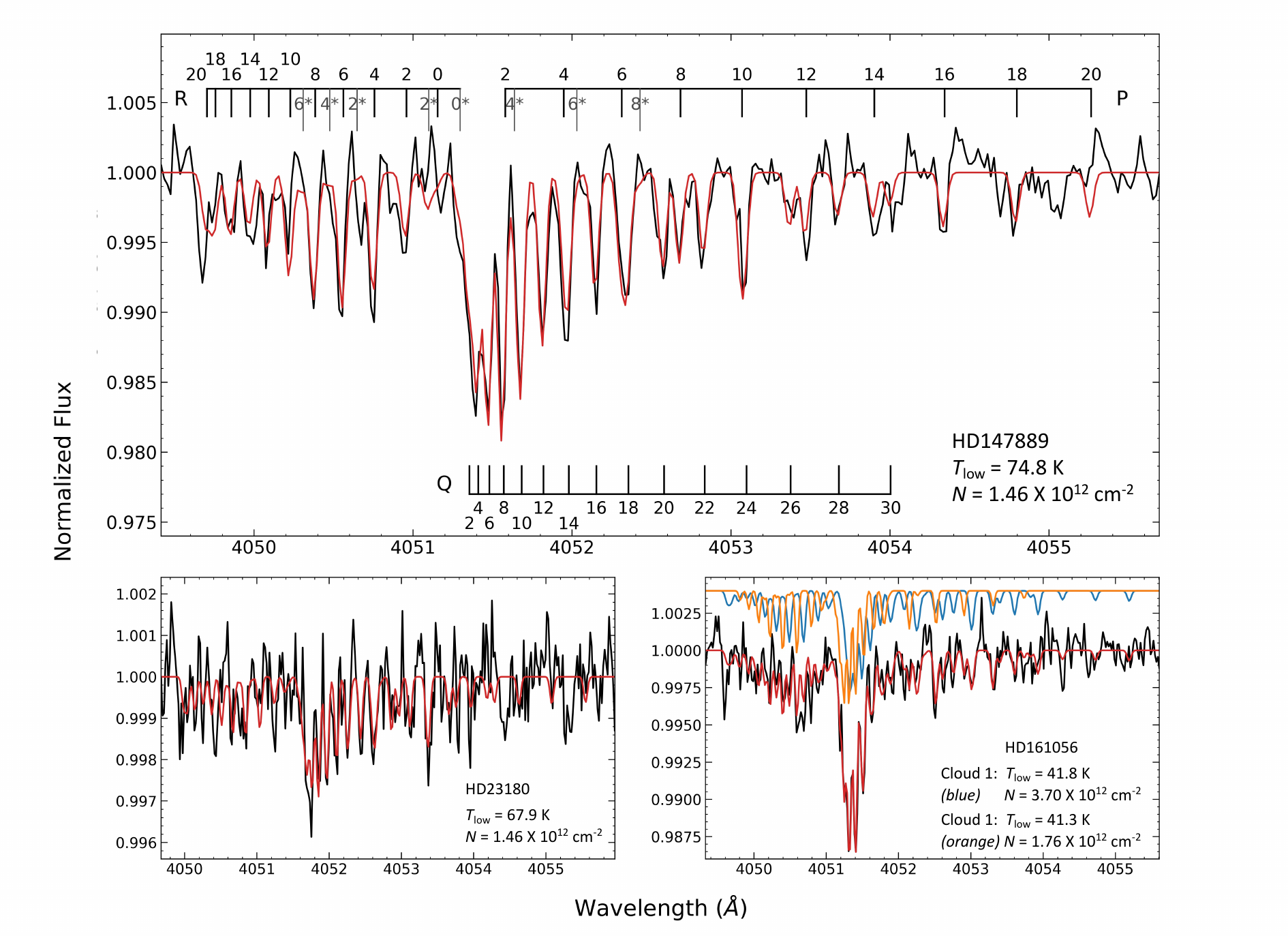}
    \caption{ C$_3 \ \tilde{A}^1\Pi_u - \tilde{X} ^1\Sigma_g^+$
      000-000 band in three representative sightlines. Top panel:
      HD~147889, in addition to harbouring the largest $N$(C$_2$),
      also harbours the largest $N$(C$_3$) in our data sample. The
      transitions are all labelled, including those to the original
      and the perturbed upper energy levels (the latter are
      highlighted with asterisks; refer also to
      Sect. \ref{sec:C3LineList} and Table
      \ref{tab:C3_linelist}). Bottom panels: The C$_3$ band is
      tentatively detected towards HD~23180, mainly through its
      piled-up Q branch in the middle. The sightline of HD~161056
      contains two velocity components of C$_3$. The synthetic spectra
      for the two components are plotted with a positive offset in
      different colours.}
    \label{fig:C3_Demo}
\end{figure*}

The strongest C$_3$ transitions in the optical region lie around 4051~\AA\ and are from the origin 000-000 band of the $\tilde{A}^1\Pi_u - \tilde{X} ^1\Sigma_g^+$ system. This band is characterised by a piled-up $Q$-branch in the centre, while the rotational lines from the $P$- and $R$-branches can be resolved in our data (Fig. \ref{fig:C3_Demo}). This allows us to determine the rotational excitation and thus the total column density of C$_3$ in a more accurate way by including the contributions from the higher \textit{J}-levels whose transitions are too weak to be detected. Some transitions involving vibrationally excited levels in the upper $\tilde{A}$ state  have recently been detected in diffuse interstellar environments \citep{2014MNRAS.441.1134S}. These transitions are also within the EDIBLES wavelength range, but in wavelength ranges where we lack sufficient S/N to detect these weak lines.  

\subsection{Line list of the C$_3 \: \tilde{A} - \tilde{X}$ band}\label{sec:C3LineList}
Wavelengths of the C$_3 \: \tilde{A} - \tilde{X}$ origin band transitions are reported in several works \citep[e.g.][]{C3wave1, C3wave2, C3wave3, 2023JMoSp.39111734M}. \cite{C3Perturb} report the presence of at least two perturbing energy levels that are close to the upper $\tilde{A}\ ^1\Pi_u$ state. These perturbing states would change the observation in two ways, where a) the effective $f_{jj}$ values of the original transitions are smaller than theoretically predicted, and b) some of these perturbed lines are strong enough to be detected in the spectrum. Such interference, especially when the lower \textit{J}-levels are involved, should be considered when simulating the C$_3$ spectrum \citep{2014MNRAS.441.1134S, HADDAD201441}.

\begin{table} [ht]
        \centering
        \small
        \caption{Line list used to simulate the C$_3 \: \tilde{A} - \tilde{X}$ 000-000 band. }
        \label{tab:C3_linelist}
        \begin{threeparttable}
      \begin{tabular}{ccc|ccc}
        \hline
        Rotational & $\lambda_{jj}$ & $f_{jj}$ & Rotational & $\lambda_{jj}$ & $f_{jj}$\\
        Line & (\AA) & ($\times 10^{3}$) & Line & (\AA) & ($\times 10^{3}$) \\
        \hline
        $R$(20)\tnote{a} & 4049.808 & 4.29 & $Q$(12) & 4051.918 & 7.97 \\
        $R$(18)\tnote{a} & 4049.861 & 4.32 & $P$(4) & 4052.054 & 1.57 \\
        $R$(16)\tnote{a} & 4049.962 & 4.36 & $Q$(14) & 4052.077 & 7.98 \\
        $R$(14) & 4050.075 & 3.49 & $Q$(6)\tnote{b} & 4052.122 & 0.28 \\
        $R$(12) & 4050.191 & 3.58 & $P$(4)\tnote{b} & 4052.180 & 0.86 \\
        $R$(10) & 4050.327 & 3.73 & $Q$(16) & 4052.257 & 7.98 \\
        $R$(6)\tnote{b} & 4050.401 & 0.58 & $P$(6) & 4052.412 & 2.56 \\
        $R$(8) & 4050.484 & 3.95 & $Q$(18) & 4052.459 & 7.99 \\
        $R$(4)\tnote{b} & 4050.567 & 0.49 & $P$(8)\tnote{b} & 4052.521 & 0.39 \\
        $R$(6) & 4050.661 & 4.21 & $Q$(20) & 4052.683 & 7.99 \\
        $Q$(2)\tnote{b} & 4050.746 & 0.45 & $P$(8) & 4052.782 & 2.82 \\
        $R$(4) & 4050.857 & 4.44 & $Q$(22)\tnote{a} & 4052.938 & 7.99 \\
        $R$(2) & 4051.055 & 3.78 & $P$(10) & 4053.169 & 2.87 \\
        $R$(2)\tnote{b} & 4051.190 & 2.08 & $Q$(24)\tnote{a} & 4053.201 & 7.99 \\
        $R$(0) & 4051.255 & 4.22 & $Q$(26) & 4053.479 & 7.99 \\
        $R$(0)\tnote{b} & 4051.396 & 10.58 & $P$(12) & 4053.577 & 2.87 \\
        $Q$(2) & 4051.448 & 6.87 & $Q$(28) & 4053.783 & 8.00 \\
        $Q$(4) & 4051.506 & 7.70 & $P$(14) & 4054.005 & 2.86 \\
        $Q$(6) & 4051.578 & 7.89 & $Q$(30)\tnote{a} & 4054.108 & 8.00 \\
        $Q$(8)\tnote{a} & 4051.670 & 7.95 & $P$(16)\tnote{a} & 4054.447 & 3.64 \\
        $Q$(10) & 4051.782 & 7.96 & $P$(18)\tnote{a} & 4054.902 & 3.68 \\
        $P$(2) & 4051.820 & 1.06 & $P$(20)\tnote{a} & 4055.369 & 3.71 \\
        \hline
          \end{tabular}
      \begin{tablenotes}
      \item The wavelengths and $f_{jj}$ values are taken from \cite{2014MNRAS.441.1134S} unless otherwise specified.
        \item[a] Wavelengths adopted from \cite{C3wave3} and $f_{jj}$ values based on theoretical calculations.
        \item[b] Perturbed transition.
      \end{tablenotes}
    \end{threeparttable}
\end{table}

Table \ref{tab:C3_linelist} lists the 44 transitions we considered to simulate the C$_3$ 4051~\AA\ band, including the original $Q$-branch lines for \textit{J} up to 30, the original $P$- and $R$-branch lines for \textit{J} up to 20, and eight strong perturbed lines reported in \cite{2014MNRAS.441.1134S}. The wavelengths of the target transitions are adopted from Table 3 of \cite{2014MNRAS.441.1134S}. For the rotational lines that are not included in that source, we used the values from \cite{C3wave3} after removing their correction of 0.04 $\textrm{cm}^{-1}$ on the wavenumbers. This treatment provides better agreement on the lower $J$-level transitions between  \cite{C3wave3} and \cite{2014MNRAS.441.1134S} (as well as \cite{C3wave1} and \cite{C3wave2}).

We adopted the more widely used oscillator strength of the C$_3$ 000-000 band as $f_{000} = 0.016$ \citep[e.g.][]{2001ApJ...553..267M, 2003ApJ...595..235A, Oka2003, 2014MNRAS.441.1134S}. However, a slightly smaller value of $f_{000} = 0.0146$ is used in some studies \citep[e.g.][]{2002A&A...384..629R, 2002A&A...395..969G} due to the uncertainties in the Frank-Condon factor \citep[see e.g.][]{1977CPL....50..344P, 1980MolPh..40...95J}. The C$_3$ column densities would be 10\% larger if this smaller $f_{000}$ value is used. Without the perturbed lines, the $f_{jj}$ values of the original transitions can be calculated as

\begin{equation}
\label{equ:C3fjj}
    f_{jj} = 
      \begin{cases}
        f_{000} \times \dfrac{J+2}{2(2J+1)} & \ \text{$R$-branch}\\[2ex]
        f_{000} \times \dfrac{1}{2} & \ \text{$Q$-branch}\\[2ex]
        f_{000} \times \dfrac{J-1}{2(2J+1)} & \ \text{$P$-branch.}
      \end{cases}       
\end{equation}

Due to the presence of perturbed lines, the effective $f_{jj}$ values are different from the theoretical calculations.
We used the experimentally determined $f_{jj}$ values from \cite{2014MNRAS.441.1134S}, except for transitions not listed there. In that case, we used the theoretical results of Eq. \ref{equ:C3fjj}. These transitions are mostly from the higher \textit{J}-levels with limited population, and the interference from the perturbed lines becomes very subtle.

\subsection{Excitation and modelling of C$_3$}\label{sec:C3_excitation}
Under typical diffuse interstellar conditions, C$_3$ molecules have a
relatively short lifetime, and collisional processes may not
significantly change the population of the higher \textit{J}-levels
\citep{2002A&A...384..629R, 2014MNRAS.441.1134S}. Observations suggest
that the rotational excitation of C$_3$ is characterised by two
temperatures with the transition point around $J$ = 14. The excitation
temperature of lower \textit{J}-levels, $T_\textrm{low}$, is governed
more by collisions and is typically a few tens of kelvin, whilst the
temperature for the higher \textit{J}-levels, $T_\textrm{high}$, is
governed more by radiative processes and is usually a few hundred
kelvin \citep[see e.g.][]{2003ApJ...595..235A, 2001ApJ...553..267M,
  2013MNRAS.428.1107W}. The much higher excitation for the higher
$J$-levels leads to a shallow decrease in their population, and in
total the $J$\ >\ 14 levels may contribute $\sim$15 \% of the
total column density \citep{2003ApJ...595..235A}.

To simulate the rotational excitation of the C$_3$ molecules and obtain the relative population of different $J$-levels, we composed a two-temperature Boltzmann distribution model with the following steps:
\begin{enumerate}
    \item Take the $J$ = 14 population as reference, by defining $N$($J$=14) = 1.0.
    \item Calculate the relative population of the lower $J$-levels \{$N_\textrm{low}$\} from the Boltzmann distribution of temperature $T_\textrm{low}$: \{$N_\textrm{low}$\}=\{$N_\textrm{low}$($J$=0), $N_\textrm{low}$($J$=2), $N_\textrm{low}$($J$=4), ..., $N_\textrm{low}$($J$=12), $N$($J$=14)\}.
    \item Similarly, calculate the relative population of higher $J$-levels (for $J$ up to 100), \{$N_\textrm{high}$\}=\{$N$($J$=14), $N_\textrm{high}$($J$=16), ..., $N_\textrm{high}$($J$=100)\}, from the Boltzmann distribution of temperature $T_\textrm{high}$.
    \item Combine \{$N_\textrm{low}$\} and \{$N_\textrm{high}$\} into \{$N_\textrm{J}$\}, where \{$N$\}=\{$N_\textrm{low}$($J$=0), $N_\textrm{low}$($J$=2), ..., $N_\textrm{low}$($J$=12), $N$($J$=14), $N_\textrm{high}$($J$=16), ..., $N_\textrm{high}$($J$=100)\}.
    \item Scale \{$N_\textrm{J}$\} so it adds up to unity.
\end{enumerate}

\noindent Despite the approximation, this model provides results that agree well with the literature (see the next subsection). The rotational energies of the different $J$-levels required in the Boltzmann distribution are calculated using the molecular constants in \cite{C3wave3} as follows:

\begin{equation}
    \begin{array}{c}
    E_J \ = \ B \times J(J+1) \ - \ D \times [J(J+1)]^2 \ + \ H \times [J(J+1)]^3 \\[1.0ex]
    B \ = \ 0.4305883 \ \textrm{cm}^{-1} \\[0.5ex]
    D \ = \ 1.437 \ \times 10^{-6}\ \textrm{cm}^{-1} \\[0.5ex]
    H \ =\ 1.129 \ \times 10^{-10}\ \textrm{cm}^{-1}.
    \end{array}
\end{equation}

The column density of C$_3$ is typically one order of magnitude lower than C$_2$ and for the much weaker C$_3$ absorption strengths optically thin conditions can be assumed (Fig. \ref{fig:C3_Demo}). \cite{1979CPL....60..502B} report the radiative lifetime of C$_3$ as approximately 200~ns, 
and natural broadening can hence be ignored. We thus used simple Gaussian profiles in flux units to simulate the rotational lines of the C$_3$ 4051~\AA\ band, and rotational lines from the same velocity component would share the same FWHM and velocity offset ($v_\textrm{off}$, relative to barycentric frame) parameters. We also set a lower boundary of 3.0 km $\cdot$ s$^{-1}$ for the FWHM, so the Gaussian profiles have at least the instrumental width, in addition to physical processes that broaden the C$_3$ transitions.

\subsection{Measurements and results}\label{C3Result}

\begin{table*}
        \centering
        \caption{Best-fit parameters of the C$_3$ models in 27 sightlines.}
        \label{table: C3 Detection}
        \begin{threeparttable}
        \begin{tabular}{c|ccccc|ccc} 
                \hline
                 & \multicolumn{5}{c|}{This Work\tnote{b}} & \multicolumn{3}{c}{Literature\tnote{c}} \\
                Sightline\tnote{a} & $T_\textrm{low}$ & $T_\textrm{high}$ & $N$ & FWHM & $v_\textrm{off}$ & $T$ & $N$ & Source \\
                & (K) & (K) & ($10^{12}\ \textrm{cm}^{-2}$) & (km $\cdot$ s$^{-1}$) & (km $\cdot$ s$^{-1})$ & (K) & ($10^{12}\ \textrm{cm}^{-2}$) & \\
                \hline
                  &   &   &   &   &   &   &   & \\
                HD~23180\tnote{d} & 74.8$\substack{+12.0 \\ -8.8}$ & 275.5$\substack{+60.2 \\ -49.3}$ & 1.46$\substack{+0.12 \\ -0.12}$ & 4.47$\substack{+0.49 \\ -0.42}$ & 13.2$\substack{+1.2 \\ -1.2}$ & 127.3$\pm$3.9 & 1.27$\pm$0.13 & A03 \\
                
                HD~24398\tnote{d} & 145.3$\substack{+105.5 \\ -41.5}$ & 137.4$\substack{+45.0 \\ -34.3}$ & 1.37$\substack{+0.15 \\ -0.15}$ & 4.22$\substack{+0.64 \\ -0.52}$ & 13.2$\substack{+1.3 \\ -1.3}$ & 132.0$\pm$10.3 & 1.30$\pm$0.35 & A03 \\
                 &   &   &   &   &   & & 1.00 & O03 \\
                 &   &   &   &   &   & 50.0 & 1.37 & R02 \\
                 &   &   &   &   &   &  & 1.00 & M01 \\ [1ex]
                 
                HD~27778 & 49.2$\substack{+3.5 \\ -3.1}$ & 253.1$\substack{+55.8 \\ -49.0}$ & 2.79$\substack{+0.16 \\ -0.16}$ & 5.56$\substack{+0.61 \\ -0.49}$ & 14.7$\substack{+1.3 \\ -1.3}$ & & 1.20$\pm$0.30 & O03 \\
                
                HD~45314 & 59.5$\substack{+6.5 \\ -5.4}$ & 143.7$\substack{+39.4 \\ -33.7}$ & 1.80$\substack{+0.12 \\ -0.12}$ & 4.66$\substack{+0.45 \\ -0.38}$ & 17.0$\substack{+1.2 \\ -1.2}$ &   &   &  \\
                
                HD~61827 (Comp.1) & 36.3$\substack{+2.7 \\ -2.5}$ & 179.4$\substack{+71.5 \\ -60.7}$ & 2.47$\substack{+0.12 \\ -0.12}$ & 3.91$\substack{+0.28 \\ -0.24}$ & 41.1$\substack{+1.1 \\ -1.1}$ &   &   &  \\
                HD~61827 (Comp.2) & 54.6$\substack{+6.4 \\ -5.2}$ & 196.7$\substack{+56.8 \\ -46.5}$ & 1.83$\substack{+0.14 \\ -0.14}$ & 3.11$\substack{+0.31 \\ -0.26}$ & 34.5$\substack{+1.1 \\ -1.1}$ &   &   &  \\
                
                HD~63804 & 54.9$\substack{+5.9 \\ -4.8}$ & 283.2$\substack{+68.4 \\ -57.4}$ & 4.56$\substack{+0.34 \\ -0.34}$ & 5.60$\substack{+0.73 \\ -0.57}$ & 38.6$\substack{+1.3 \\ -1.4}$ &   &   &  \\
                
                HD~73882\tnote{d} & 30.0$\substack{+1.2 \\ -1.1}$ & 286.7$\substack{+77.3 \\ -70.1}$ & 3.34$\substack{+0.10 \\ -0.10}$ & 3.58$\substack{+0.12 \\ -0.12}$ & 20.4$\substack{+1.0 \\ -1.0}$ &   &   &  \\
                
                HD~80558\tnote{d} & 40.3$\substack{+4.8 \\ -3.7}$ & 1000.0 & 1.68$\substack{+0.24 \\ -0.24}$ & 6.48$\substack{+1.46 \\ -0.92}$ & 19.6$\substack{+2.2 \\ -2.1}$ &   &   &  \\
                
                HD~144217\tnote{d} & 15.1$\substack{+6.2 \\ -4.1}$ & 100.0 & 0.36$\substack{+0.09 \\ -0.08}$ & 3.46$\substack{+1.93 \\ -0.85}$ & -10.5$\substack{+1.7 \\ -1.6}$ & & <0.38 & G02 \\
                
                HD~147683 & 87.4$\substack{+19.8 \\ -13.6}$ & 115.0$\substack{+30.4 \\ -25.2}$ & 1.64$\substack{+0.12 \\ -0.12}$ & 3.98$\substack{+0.40 \\ -0.33}$ & -2.2$\substack{+1.2 \\ -1.2}$ &   &   &  \\
                
                HD~147888 & 58.1$\substack{+6.4 \\ -5.3}$ & 167.0$\substack{+46.1 \\ -39.7}$ & 1.35$\substack{+0.09 \\ -0.09}$ & 3.89$\substack{+0.33 \\ -0.28}$ & -8.9$\substack{+1.1 \\ -1.1}$ &   &   &  \\
                
                HD~147889 & 67.9$\substack{+3.1 \\ -2.8}$ & 215.3$\substack{+17.1 \\ -16.0}$ & 7.74$\substack{+0.21 \\ -0.21}$ & 4.14$\substack{+0.14 \\ -0.14}$ & -8.4$\substack{+1.0 \\ -1.0}$ & \multicolumn{2}{c}{tentative detection} & H95 \\
                
                HD~147933 & 63.4$\substack{+7.4 \\ -6.0}$ & 165.6$\substack{+43.6 \\ -37.0}$ & 1.74$\substack{+0.12 \\ -0.12}$ & 3.67$\substack{+0.33 \\ -0.28}$ & -9.2$\substack{+1.1 \\ -1.1}$ &   &   &  \\
                
                HD~148184 & 140.7$\substack{+35.4 \\ -22.7}$ & 222.6$\substack{+29.3 \\ -25.6}$ & 2.60$\substack{+0.14 \\ -0.14}$ & 3.74$\substack{+0.28 \\ -0.24}$ & -12.7$\substack{+1.1 \\ -1.1}$ & & 3.10$\pm$0.50 & G02 \\
                
                HD~149404 (Comp.1) & 47.4$\substack{+4.9 \\ -4.1}$ & 243.2$\substack{+79.7 \\ -67.1}$ & 1.16$\substack{+0.09 \\ -0.09}$ & 5.25$\substack{+0.80 \\ -0.61}$ & -1.2$\substack{+1.3 \\ -1.4}$ &   &   &  \\
                HD~149404 (Comp.2) & 66.5$\substack{+10.0 \\ -7.5}$ & 274.2$\substack{+75.0 \\ -60.6}$ & 1.10$\substack{+0.10 \\ -0.10}$ & 4.22$\substack{+0.47 \\ -0.40}$ & -20.0$\substack{+1.3 \\ -1.3}$ &   &   &  \\  [1ex]

                HD~149757 & 72.1$\substack{+8.8 \\ -7.0}$ & 162.1$\substack{+32.0 \\ -27.7}$ & 1.60$\substack{+0.10 \\ -0.10}$ & 4.62$\substack{+0.42 \\ -0.35}$ & -15.0$\substack{+1.2 \\ -1.2}$ &   & 1.60$\pm$0.00 & O03 \\
                 &   &   &   &   &   &   & 2.01$\pm$0.20 & G02 \\
                 &   &   &   &   &   &   & 1.83$\pm$0.00 & R02 \\
                 &   &   &   &   &   &   & 1.60$\pm$0.00 & M01 \\ [1ex]
                 
                HD~152248\tnote{d} & 57.5$\substack{+6.6 \\ -5.1}$ & 552.0$\substack{+115.0 \\ -94.7}$ & 1.61$\substack{+0.14 \\ -0.14}$ & 4.12$\substack{+0.64 \\ -0.49}$ & 2.3$\substack{+1.2 \\ -1.2}$ &   &   &  \\
                
                HD~154043\tnote{d} & 56.8$\substack{+8.0 \\ -6.2}$ & 244.1$\substack{+71.3 \\ -56.8}$ & 1.70$\substack{+0.15 \\ -0.15}$ & 5.09$\substack{+0.78 \\ -0.59}$ & -12.2$\substack{+1.4 \\ -1.4}$ &   &   &  \\
                
                HD~154368 & 42.1$\substack{+1.2 \\ -1.1}$ & 440.4$\substack{+40.4 \\ -38.0}$ & 4.84$\substack{+0.14 \\ -0.14}$ & 3.91$\substack{+0.14 \\ -0.12}$ & -4.7$\substack{+1.0 \\ -1.0}$ &   &   &  \\
                
                HD~161056 (Comp.1) & 44.5$\substack{+1.5 \\ -1.3}$ & 913.4$\substack{+78.4 \\ -71.2}$ & 4.08$\substack{+0.16 \\ -0.18}$ & 4.92$\substack{+0.33 \\ -0.28}$ & -13.2$\substack{+1.1 \\ -1.1}$  &   &   &  \\
                HD~161056 (Comp.2) & 42.6$\substack{+3.0 \\ -2.6}$ & 100.0 & 1.94$\substack{+0.08 \\ -0.09}$ & 3.79$\substack{+0.21 \\ -0.19}$ & -19.9$\substack{+1.1 \\ -1.1}$ &   &   &  \\  [1ex]

                HD~169454 &23.9$\substack{+0.4 \\ -0.4}$ & 1000.0 & 6.45$\substack{+0.13 \\ -0.13}$ & 3.93$\substack{+0.09 \\ -0.09}$ & -9.5$\substack{+1.0 \\ -1.0}$ & 22.4$\pm$1.0 & 6.61$\pm$0.19 & S14 \\
                 &   &   &   &   &   & 23.4$\pm$1.4 & 2.24$\pm$0.66 & A03 \\
                 &   &   &   &   &   & & 4.30$\pm$0.50 & O03 \\ [1ex]
                 
                HD~179406 & 57.2$\substack{+3.7 \\ -3.3}$ & 105.3$\substack{+20.6 \\ -18.7}$ & 2.58$\substack{+0.09 \\ -0.09}$ & 4.24$\substack{+0.21 \\ -0.19}$ & -13.2$\substack{+1.1 \\ -1.1}$ &   & 1.30$\pm$0.70 & O03 \\
                 &   &   &   &   &   &   & 2.90 & R02 \\
                 &   &   &   &   &   &   & 2.00 & M01 \\ [1ex]
                 
                HD~185859\tnote{d} & 65.1$\substack{+11.3 \\ -8.3}$ & 220.0$\substack{+71.3 \\ -56.1}$ & 1.24$\substack{+0.12 \\ -0.12}$ & 4.59$\substack{+0.71 \\ -0.54}$ & -8.1$\substack{+1.3 \\ -1.3}$ &   &   &  \\
                
                HD~186745 & 50.2$\substack{+3.6 \\ -3.2}$ & 315.0$\substack{+63.5 \\ -55.6}$ & 3.64$\substack{+0.22 \\ -0.22}$ & 6.45$\substack{+0.94 \\ -0.71}$ & -10.8$\substack{+1.4 \\ -1.4}$ &   &   &  \\
                
                HD~186841 (Comp.1) & 72.4$\substack{+9.5 \\ -7.3}$ & 271.7$\substack{+52.3 \\ -44.1}$ & 3.12$\substack{+0.22 \\ -0.22}$ & 5.16$\substack{+0.57 \\ -0.47}$ & -11.8$\substack{+1.3 \\ -1.3}$ &   &   &  \\
                HD~186841 (Comp.2) & 41.4$\substack{+5.6 \\ -4.4}$ & 528.2$\substack{+232.7 \\ -176.5}$ & 1.50$\substack{+0.20 \\ -0.20}$ & 4.07$\substack{+0.85 \\ -0.59}$ & -5.3$\substack{+1.4 \\ -1.4}$ &   &   &  \\
                
                HD~203532 & 47.3$\substack{+2.1 \\ -1.9}$ & 231.1$\substack{+33.1 \\ -30.4}$ & 2.78$\substack{+0.09 \\ -0.09}$ & 4.10$\substack{+0.19 \\ -0.16}$ & 14.2$\substack{+1.1 \\ -1.1}$ &   &   &  \\
                
                HD~210121 & 61.8$\substack{+2.9 \\ -2.7}$ & 140.7$\substack{+15.5 \\ -14.5}$ & 4.98$\substack{+0.14 \\ -0.14}$ & 3.86$\substack{+0.12 \\ -0.12}$ & -14.7$\substack{+1.0 \\ -1.0}$ &   & 1.90$\pm$0.50 & O03 \\
                 &   &   &   &   &   &   & 4.56 & R02 \\
                 &   &   &   &   &   &   & \\
                \hline
        \end{tabular}
  \begin{tablenotes}
    \item[a] Best-fit results are listed in separate entries for sightline harbouring multiple velocity components.
    \item[b] The Gaussian broadening factor FWHM has a lower boundary of 3.00 km $\cdot$ s$^{-1}$ to reflect instrumental profile; the fitting boundaries of the excitation temperature for the higher $J$-levels $T_\textrm{high}$ are between 100 and 1,000~K; the velocity offset $v_\textrm{off}$ is relative to the barycentric frame.
    \item[c] Reference codes are: H95, \cite{1995ApJ...453..450H}; M01, \cite{2001ApJ...553..267M}; G02, \cite{2002A&A...395..969G}; R02, \cite{2002A&A...384..629R}; A03, \citet{2003ApJ...595..235A}; O03, \cite{Oka2003}; S14, \cite{2014MNRAS.441.1134S}; Column densities from G02 and R02 have been converted to $f_{000} = 0.016$. The excitation temperature for the lower $J$-levels is listed if the reference uses double-temperature model.
    \item[d] Tentative detection, no individual transition exceeds the 3$\sigma$ detection limit.
  \end{tablenotes}
  \end{threeparttable}
\end{table*}

As for C$_2$, we first visually examined the target C$_3$ transitions,
especially the more strongly piled-up Q branch contour. The C$_3 \:
\tilde{A} - \tilde{X}$ 000-000 band spans a narrow wavelength window
of $\sim$6~\AA, and we used a cubic spline continuum to normalise the
entire wavelength region. The compact distribution of the transitions
hampers the identification of possible extra velocity components. We
thus referred to the C$_2$ measurements and the \ion{Na}{i}
3302~\AA\ doublet of the same sightline to determine the number of
components needed for the model. Each of these components is
characterised by its own set of $T_\textrm{low}$, $T_\textrm{high}$,
column density $N$, Gaussian width factor FWHM, and $v_\textrm{off}$,
which is relative to the barycentric frame. We followed the same
routine as C$_2$ for parameter optimisation and error estimation (see
Sect. \ref{sec:C2Results}).

The target C$_3$ band is detected at the 3$\sigma$ level in 19
sightlines, and tentatively in another 8 sightlines\footnote{We use
the term `tentative detection' when the deepest point of the synthetic
spectrum is shallower than 3$\sigma$ from the continuum level.}. These
include 4 sightlines where C$_3$ is measured in two velocity
components (HD~61827, HD~149404, HD~161056, and HD~186841). In Table
\ref{table: C3 Detection} we summarise the C$_3$ measurements and
compare to available literature values, and we find good agreement in
most of the cases, especially with results from high-resolution
data. The C$_2$ and C$_3$ detection are clearly linked; all sightlines
with a C$_3$ detection (Table \ref{table: C3 Detection}) also exhibit
C$_2$ absorption. The best-fit models are visualised for three
representative targets in Fig. \ref{fig:C3_Demo} and the full gallery
is available in Appendix \ref{app_C3}.

The temperature parameters may be less well determined when $N$(C$_3$) is small, especially for $T_\textrm{high}$. Assuming average excitation conditions where $T_\textrm{low}$ = 50~K and $T_\textrm{high}$ = 250~K, the higher $J$ levels harbour $\sim$25$\%$ of the total population. This leaves less than $5 \%$ of the total equivalent width for each of the $J$-levels considered in our model, which is further divided into the $P$, $Q$, and $R$ transitions. These transitions are then very weak and provide limited constraints on minimising the $\chi^2$ during the fitting process.

\section{C$_2$ versus C$_3$}\label{sec:comparison}
\begin{figure} [ht]
    \includegraphics[width=\columnwidth]{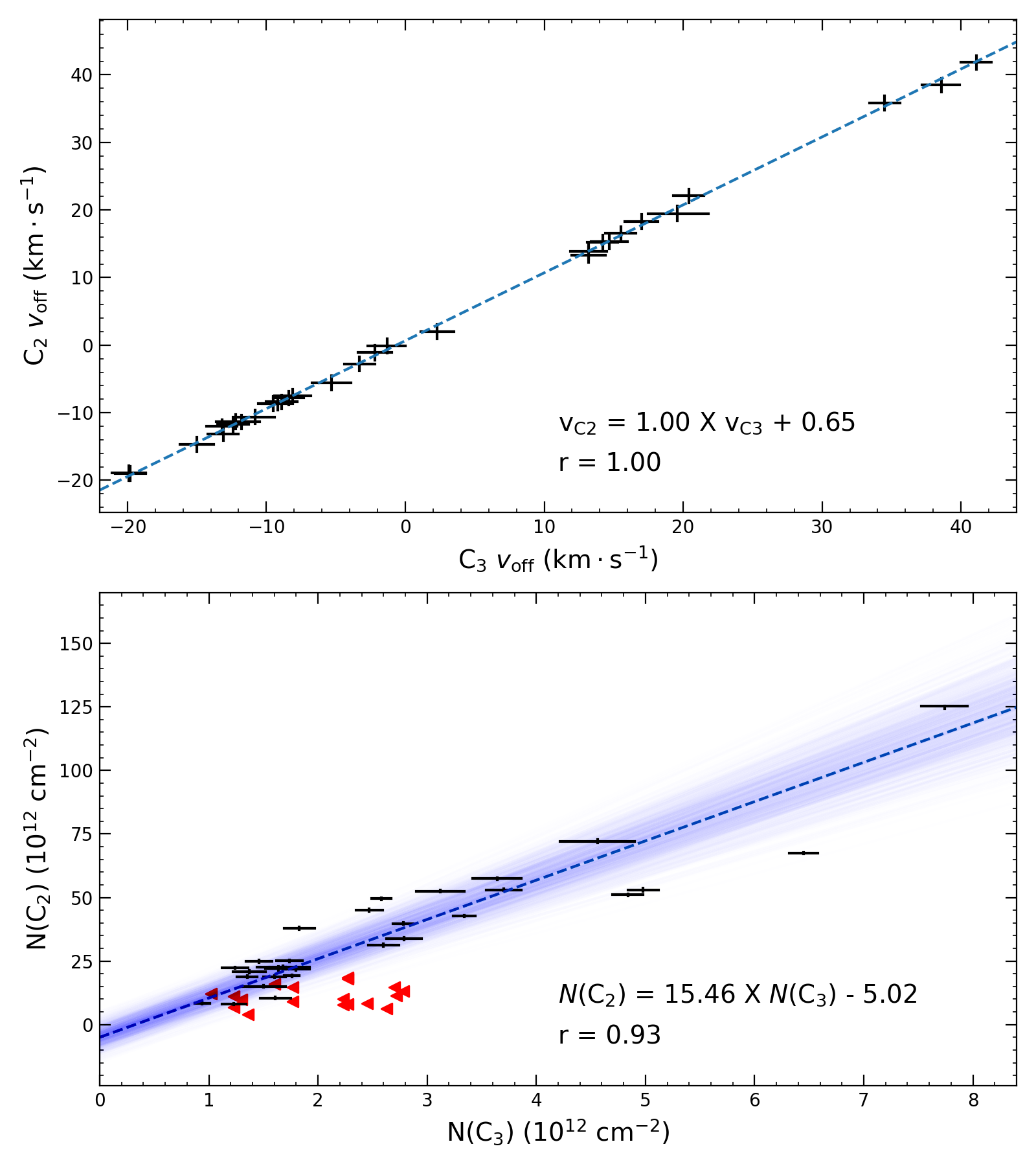}
    \caption{Comparison between the C$_2$ and C$_3$ measurements. Each point represents a velocity component for which both molecules are detected. Upper panel: The velocity offsets of C$_2$ and C$_3$ (relative to the barycentric frame) are well aligned, suggesting the two molecules are detected in the same velocity component. Lower panel: $N$(C$_2$) and $N$(C$_3$) have a good correlation, with r = 0.93. The dashed blue line represents the best linear fit from orthogonal distance regression. The 1$\sigma$ confidence interval of the fit is shown as the shadow. The red triangles represent the 3$\sigma$ upper limits of $N$(C$_3$) in velocity components without C$_3$ detection. These upper limits are not considered in the linear regression.}
    \label{fig:C2vsC3}
\end{figure}

In this section we compare the velocity offsets and column densities of C$_2$ and C$_3$ obtained in our survey (Fig. \ref{fig:C2vsC3}). Since there may be multiple velocity components in some sightlines, each data point in the plot represents an individual ISM cloud. We find the velocity offsets of C$_2$ and C$_3$ agree with each other (within uncertainties), which suggests the two molecules are detected in the same velocity components, and validates further comparison of their column densities.

The column densities of C$_2$ and C$_3$ show a tight correlation among
the 30 velocity components from 26 sightlines with $r$ = 0.93
(Fig. \ref{fig:C2vsC3}, lower panel). We find the best-fit line to
have a relatively small intercept. The slope of the best-fit line,
indicating the $N$(C$_2$)/$N$(C$_3$) ratio, is 15.5$\pm$1.4. While it
is widely agreed that $N$(C$_3$) should be one magnitude lower than
$N$(C$_2$), it used to be difficult to determine the
$N$(C$_2$)/$N$(C$_3$) ratio with good accuracy due to the required
high quality data. In a pioneering survey by \cite{Oka2003}, the
authors also found a good correlation between $N$(C$_2$) and
$N$(C$_3$), and estimated a $N$(C$_2$)/$N$(C$_3$) ratio of
$\sim$40. However, since the individual rotational lines could not be
resolved in their medium resolution data (R$\sim$38,000),
\cite{Oka2003} focused on the piled-up Q branch and assumed it
contributed half of the total intensity of the band. This
approximation may underestimate contributions from the higher
$J$-levels and hence $N$(C$_3$) \citep[][]{2003ApJ...595..235A}.

The $N$(C$_2$)/$N$(C$_3$) ratio of 15.5 in this work is slightly larger than the value of $\sim$10 reported by \citet{2014MNRAS.441.1134S}, and is in fact close to the prediction for a typical diffuse interstellar cloud with $A_V \sim$2.0~mag in the model presented by \cite{2014MNRAS.441.1134S}. This is approximately the extinction of our targets \citep{EDIBLESI_Data}. However, the $N$(C$_2$)/$N$(C$_3$) ratio can be quite different in circumstellar environments, presumably related to the much higher density and/or enhanced radiation field. For example, literature reports column densities of both C$_2$ and C$_3$ to be $\sim$10$^{15} \textrm{cm}^{-2}$ towards the well-studied carbon star IRC+10216 \citep[see e.g.][]{1988Sci...241.1319H, 1997A&A...323..469B, 2000ApJ...534L.199C, 2014MNRAS.444.3721H}, and $N$(C$_2$)/$N$(C$_3$) $\sim$1. 

In diffuse and translucent clouds the C$_2$ molecules may be formed bottom-up along the reaction C$^+$ + CH $\rightarrow$ C$_2^+$ + H. The product then turns into C$_2$ molecules via multiple channels that consist of series of abstractions of H atoms and dissociative recombinations \citep[e.g.][]{1989ApJ...338..140F, 1994ApJ...424..772F, 2007ApJS..168...58S}. \cite{Oka2003} suggested that C$_3$ molecules are formed via dissociative recombination of C$_3$H$^+$ \citep{Oka2003}, while \cite{2001PCCP....3.2038B} advocate for non-negligible contribution from neutral-neutral reactions between C and C$_2$H$_2$. It is also possible for the two molecules to be formed from detachment of larger molecules in a top-down manner. In either case, there must be a strong chemical link between C$_2$ and C$_3$ since their column densities are tightly correlated.


\section{C$_2$, DIBs, and C$_2$ DIBs}\label{sec:C2DIB}
\cite{2003ApJ...584..339T} introduced the C$_2$ DIBs as `a class of
weak, narrow bands whose normalised equivalent widths $W$(X)/$W$(6196)
are well correlated, specifically with $N$(C$_2$)/$E_{B-V}$ via power
laws'. The C$_2$
DIBs demonstrate different behaviour compared to the `regular'
non-C$_2$ DIBs, in particular in dense environments \citep[see
  e.g.][]{2017ApJ...850..194F}. Contrary to their name, the C$_2$
DIBs may correlate better with $E_{B-V}$ than with $N$(C$_2$)
\citep{EDIBLESIII_C2}. The strong C$_2$ DIBs $\lambda\lambda$4963 and
4984 are detected in sightlines exposed to intense radiation fields
and with low $f_\textrm{H2}$ values \citep[see
  e.g.][]{2017ApJ...850..194F, 2019ApJ...878..151F}, suggesting that
their carriers are not necessarily directly chemically linked to C$_2$
or other molecules.

In previous sections we carefully examined all 123 EDIBLES targets and detected the C$_2$ (2-0) Phillips band and the C$_3 \: \tilde{A} - \tilde{X}$ origin band in a large number of sightlines. The detection of C$_2$ and C$_3$ molecules usually suggests dense and molecular environments along the sightlines (see also Appendix \ref{app_C2_dense}). We refer to the 41 sightlines that show C$_2$ features as `C$_2$ sightlines' and the rest as `non-C$_2$ sightlines'. This section discusses how C$_2$ and non-C$_2$ DIBs behave in these sightlines. Since DIB measurements are not the primary focus of this work, we adopted the DIB equivalent widths ($W$(DIBs)) from \cite{2017ApJ...850..194F}. Readers are referred to the original paper for details on how the measurements were obtained.

\begin{figure*} [bt]
    \centering
    \includegraphics[width=0.8\textwidth]{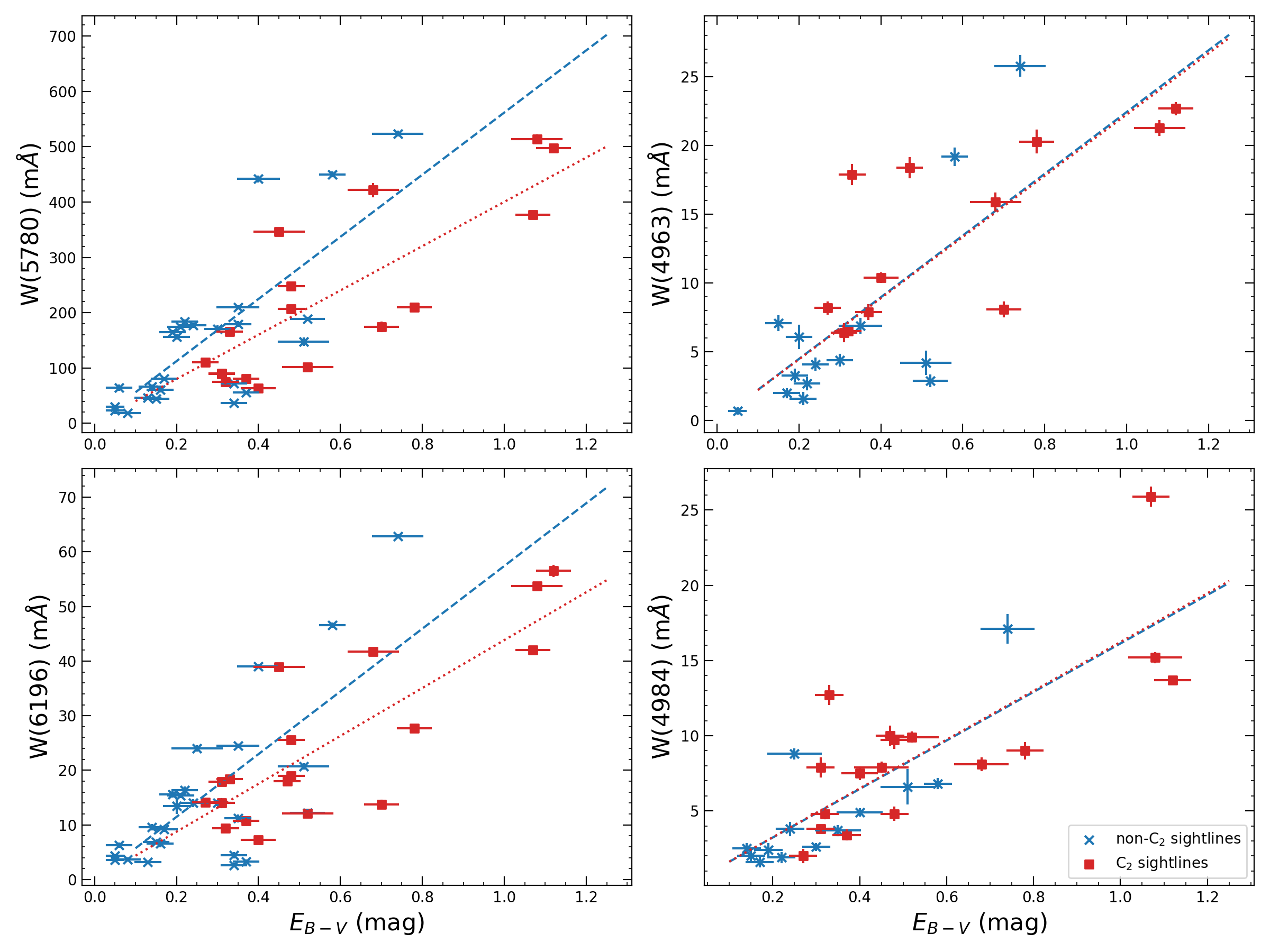}
    \caption{ Behaviour of non-C$_2$ DIBs $\lambda\lambda$5780 and
      6196 (left panels), and C$_2$ DIBs $\lambda\lambda$4963 and 4984
      (right panels) in C$_2$ and non-C$_2$
      sightlines. The non-C$_2$ DIBs show the `skin effect' and have
      reduced strengths in the C$_2$ sightlines. However, the
      behaviour of the C$_2$ DIBs is rather similar in the two types
      of sightlines. They may trace deeper regions of a cloud than the
      non-C$_2$ DIBs, but not as deep as the C$_2$ molecules. The DIB
      equivalent width data are gathered from
      \cite{2017ApJ...850..194F}.}
    \label{fig:C2DIB_behaviour}
\end{figure*}

Figure \ref{fig:C2DIB_behaviour} compares the general behaviour of four DIBs, in terms of $W$(DIBs) - $E_\textrm{B-V}$ correlations, between the C$_2$ and non-C$_2$ sightlines. The comparisons are made for two non-C$_2$ DIBs $\lambda\lambda$5780 and 6196, and two of the well-recognised C$_2$ DIBs $\lambda\lambda$4963 and 4984. Both non-C$_2$ DIBs have reduced strengths in the C$_2$ sightlines, which was described as the `skin effect' in early DIB observations \citep[e.g.][]{1966ApJ...144..921W, 1975ApJ...196..489S, 1984ApJ...283...98M, 1995ARA&A..33...19H}. This observational fact suggests that the inner parts of the clouds have a limited contribution to the column density of the general DIB material \citep{1974ApJ...194..313S}. As a more quantitative analysis, \cite{2017ApJ...850..194F} reported `lambda-shaped' behaviour of non-C$_2$ DIBs where their normalised strengths ($W$(DIB)/$E_\textrm{B-V}$) decrease as a function of $f_\textrm{H2}$ for $f_\textrm{H2}$ > $\sim$\ 0.3. This is also approximately the range of the $f_\textrm{H2}$ values for the C$_2$ sightlines we identified (Appendix \ref{app_C2_dense}). 

In direct contrast, the C$_2$ DIBs $\lambda\lambda$4963 and 4984
showed undifferentiated behaviour among C$_2$ and non-C$_2$
sightlines. This is consistent with the observations that $W$(C$_2$
DIBs)/$E_\textrm{B-V}$ remains relatively constant (with large
scatter) in sightlines with large $f_\textrm{H2}$ values
\citep{2017ApJ...850..194F}. The C$_2$ DIBs are originally defined by
an enhanced ratio of $W$(C$_2$ DIBs)/$W$(6196) in the C$_2$ sightlines \citep{2003ApJ...584..339T}. This increment is more likely the result of decreased $W$(6196) rather than increased $W$(C$_2$ DIBs), and provides evidence that the C$_2$ DIBs may trace denser environments than the non-C$_2$ DIBs. However, since the C$_2$ DIBs are not enhanced in the C$_2$ sightlines, their carriers may not reside in the deep layers of the ISM clouds where C$_2$ and other second-generation molecules are found \citep[e.g. CN and CO; see][]{2022MNRAS.510.3546F}.

\section{One step further: Search for more complicated carbon chains}\label{sec:C4C5}

\subsection{Rare C$_2$ isotopologue ($^{13}$C$^{12}$C)}
The positions of the transitions from the $^{13}$C$^{12}$C (0-0) and (1-0) Phillips $A\,^{1}\Pi$--$X\,^{1}\Sigma^{+}$ bands were first provided by \cite{AMI83:257} using Fourier spectroscopy. \cite{HAM019:143} recently reported the first tentative detection of $^{13}$C$^{12}$C in the ISM using these data. This marginal detection is for the $Q$(3) transition of the (0-0) band and is made towards the heavily reddened sightline of Cyg~OB2~12. To the best of our knowledge, no dedicated laboratory studies are yet available on the $^{13}$C$^{12}$C (2-0) bands. We calculated the required transition wavelengths and $f_{jj}$ values using the well-known isotopic dependence of the Dunham vibration-rotation coefficients \citep{WAT80:411}. 

For this purpose, we followed the same approach as described in \cite{RAM014:5} and \cite{ROU012:A24} and utilised the molecular parameters of the main $^{12}$C$_{2}$ isotopologue \citep{WAN015:064317} and the isotopic relationships \citep{RAM014:5,ROU012:A24,WAT80:411}: 

\begin{center}
   \begin{align}
   \begin{split}
   \label{equ:13C12C}
       \;\;\;\;\;\;\;\;\;\;\;\;\;\;\;\;\;\;\;\;\;\;\;\;\;\;\;\;\;\;\;\;\;\;\;\;\;\omega_{e}'\!&=\!\rho\omega_{e} \\
        \omega_{e}x_{e}'\!&=\!\rho^{2}\omega_{e}x_{e} \\
        \omega_{e}y_{e}'\!&=\!\rho^{3}\omega_{e}y_{e} \\ 
        \omega_{e}z_{e}'\!&=\!\rho^{4}\omega_{e}z_{e} \\
        \omega_{e}a_{e}'\!&=\!\rho^{5}\omega_{e}a_{e} \\ 
        B_{e}'\!&=\!\rho^{2}B_{e} \\
        \alpha_{e}'\!&=\!\rho^{3}\alpha_{e} \\ 
        \gamma_{e}'\!&=\!\rho^{4}\gamma_{e} \\ 
        \delta_{e}'\!&=\!\rho^{5}\delta_{e} \\ 
        \epsilon_{e}'\!&=\!\rho^{6}\epsilon_{e} \\
        D_{e}'\!&=\!\rho^{4}D_{e} \\
        \beta_{e}'\!&=\!\rho^{5}\beta_{e} \\
        \zeta_{e}'\!&=\!\rho^{6}\zeta_{e} \\
        q'\!&=\!\rho^{4}q, \\
        \alpha^{q}{'}\!&=\!\rho^{5}\alpha^{q} \\
        q_{D}'\!&=\!\rho^{8}q_{D}. \\
    \end{split}
\end{align} 
\end{center}

\noindent Here the primed and unprimed quantities are equilibrium spectroscopic constants for $^{13}$C$^{12}$C and $^{12}$C$_{2}$, respectively (as defined in \citealt{WAN015:064317}) and $\rho\!=\!\sqrt{\mu_{\mathrm{main}}/\mu_{\mathrm{rare}}}\!=\!0.98052,$ with the $\mu$ being the reduced mass. The resulting primed (equilibrium) constants of $^{13}$C$^{12}$C (Eq. \ref{equ:13C12C}) were then used as input into standard spectroscopic formulas \citep[see e.g. pages 3--4][]{WAN015:064317} to introduce the proper vibrational dependence into the relevant mass-scaled parameters. Their final values are gathered in Table \ref{tab:specconst}. The list with calculated line positions was subsequently generated using the final spectroscopic constants in Table \ref{tab:specconst} and \texttt{PGOPHER} \citep[see Table \ref{tab:13C2_linelist};][]{WES017:221}.

\begin{table}
\centering
\caption{\footnotesize Mass-scaled 
spectroscopic constants of $^{13}$C$^{12}$C in the (2-0) Phillips system.}
\label{tab:specconst}
\begin{threeparttable}
\begin{tabular}{ccccc}
\hline 
\multirow[c]{2}{*}{Constants\tnote{a}}  & \multicolumn{2}{c}{$X\,^{1}\Sigma^{+}$($v=0$)} & & \multicolumn{1}{c}{$A\,^{1}\Pi$($v=2$)}                          \\
\cline{2-3} \cline{5-5}                   
 & {This work\tnote{b}} & {Literature\tnote{c}} & & {This work\tnote{b}} \\[0.5ex] 
\hline 
      Band origin             &           &           & & 11354.65876 \\
      $B_{v}$                 &  1.74133  &  1.741417   & &     1.51402 \\
      $D_{v}\!\times\!10^{6}$ &  6.46215  &  6.530    & &     6.06544 \\
      $q_{v}\!\times\!10^{4}$ &           &           & &    -1.77837 \\
      $q_{D}\!\times\!10^{9}$ &           &           & &     4.31457 \\
\hline
\end{tabular}
\begin{tablenotes}[flushleft]
  \item All values are in cm$^{-1}$. 
  \item[a] {$B_{v}$ and $D_{v}$ are rotational and centrifugal distortion parameters, while $q_{v}$ and $q_{D}$ are $\Lambda$-type doubling constants. $v$ define vibrational quantum numbers; see~\citet{WAN015:064317} for details.}
  \item[b]{Calculated from $^{12}$C$_{2}$~\citep{WAN015:064317} and isotopic relations (see the main text).}
  \item[c]{Laboratory spectroscopic data; see \citet{AMI83:257}.}
\end{tablenotes}
\end{threeparttable}
\end{table}

We employed the methodology described in \cite{RAM014:5} to determine the $f_{jj}$ values for the $^{13}$C$^{12}$C (2-0) Phillips bands. Briefly, the potential energy curves for each electronic state are represented by Rydberg-Klein-Rees (RKR) classical turning points \citep{RKR014}. These are calculated using the \texttt{RKR1} program \citep{RKR014} and the mass-scaled spectroscopic constants of the upper ($A\,^{1}\Pi$) and lower ($X\,^{1}\Sigma^{+}$) $^{13}$C$^{12}$C states (Eq. \ref{equ:13C12C}). With the corresponding ab initio electronic transition dipole moment function reported by \cite{YUR018:3397}, Einstein A-coefficients ($A_{J' \rightarrow J''}$) were then determined using the \texttt{LEVEL} code \citep{LER017:167}, and finally converted into the $f_{jj}$ values following the usual expression \citep{RAM014:5}:

\begin{equation}
f_{J' \leftarrow J''}=1.49919368\frac{1}{\tilde{\nu}^2}\frac{(2J'+1)}{(2J''+1)}A_{J' \rightarrow J''}.  
\end{equation} 

\noindent Here, $\tilde{\nu}$ is the transition wavenumber (in $\mathrm{cm^{-1}}$) associated with the lower ($J''$) and upper ($J'$) ro-vibronic $J$ states. 

Unlike the regular C$_{2}$ molecule, $^{13}$C$^{12}$C has additional transitions from odd $J$-levels due to its hetero-nuclear nature. Our simulation suggests that the $Q$(1) - $Q$(4) transitions are the strongest under typical excitation conditions with $T_\textrm{exc} \sim$ \ 60~K. Their calculated wavelengths and $f_{jj}$ values are summarised in Table \ref{tab:13C2_linelist}. We also included the calculated (following the same approach described above) $f_{jj}$ values of the regular $^{12}$C$_{2}$ molecule for comparison. As expected, they differ by less than $1.2\%$. We also note that our calculated $f_{jj}$ values for the $^{12}$C$_{2}$ molecule reproduce to within $0.4\%$ those reported by \cite{2007ApJS..168...58S}, hence further supporting the reliability of our approach.

\begin{table} [ht]
        \centering
        \small
        \caption{Calculated line list of the $^{13}$C$^{12}$C (2-0) Phillips band.}
        \label{tab:13C2_linelist}
        \begin{threeparttable}
      \begin{tabular}{c|c|ccc}
        \hline
        \multirow{2}{*}{Transition} & $\lambda_{jj}$ \tnote{a,b} & \multicolumn{3}{c}{$f_{jj}\ (\times 10^{4})$ } \\
         & (\AA) & $^{13}$C$^{12}$C\tnote{b} & $^{12}$C$_{2}$\tnote{b} & Literature\tnote{c} \\
        \hline
        $Q$(1) & 8804.892 & 7.11 & - & - \\ 
        $Q$(2) & 8805.597 & 7.11 & 7.03 & 7.00 \\
        $Q$(3) & 8806.655 & 7.11 & - & - \\
        $Q$(4) & 8808.065 & 7.11 & 7.03 & 7.00 \\
        \hline
          \end{tabular}
      \begin{tablenotes}
          \item[a] Air wavelength.
          \item[b] Calculated from molecular constants of $^{12}$C$_{2}$ \citep{WAN015:064317} and isotopic relations. See the main text for more details.
          \item[c] $f_{jj}$ values of $^{12}$C$_{2}$ from \cite{2007ApJS..168...58S}.
      \end{tablenotes}
    \end{threeparttable}
\end{table}

\begin{figure} [ht]
    \includegraphics[width=\columnwidth]{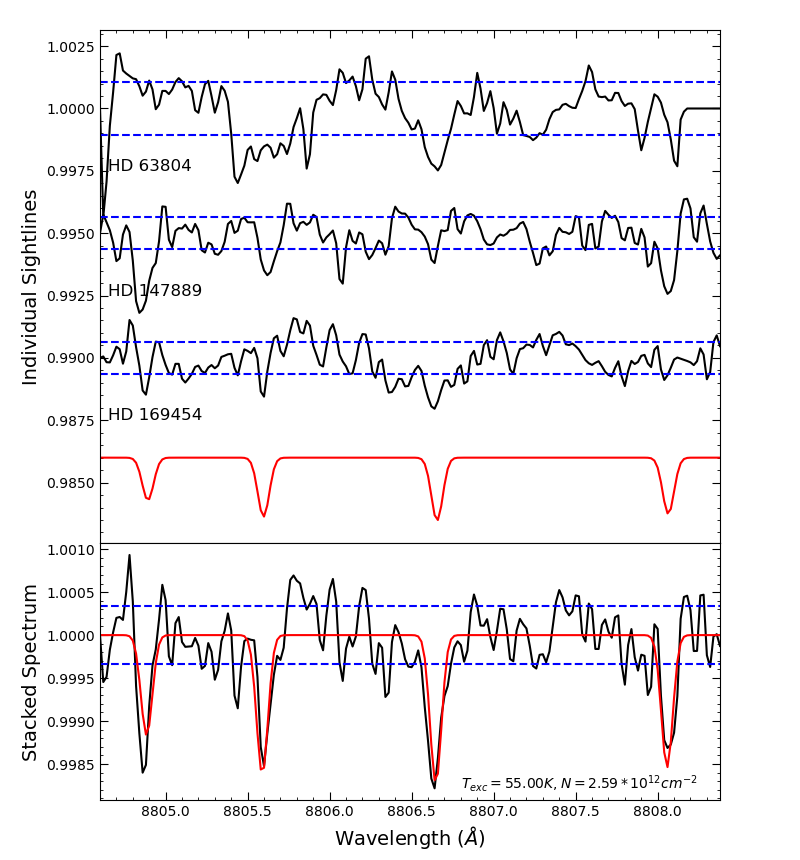}
    \caption{Search for $^{13}$C$^{12}$C (2-0) Phillips transitions. Top panel: Spectral segments of HD~63804, HD~147889, and HD~169454 around the target wavelength region, and simulated data (red) for $^{13}$C$^{12}$C with $T_\textrm{exc}$ = 55~K. From left to right, the four transitions are $Q$(1), $Q$(2), $Q$(3), and $Q$(4). Unfortunately, we cannot identify these features in any individual sightline. Bottom panel: Stacked (average) data of the three sightlines. We have a 3$\sigma$ detection of the $Q$(2) and $Q$(3) transitions, and a marginal detection of the $Q$(4) transition. The estimated column density of $^{13}$C$^{12}$C is $2.30\pm0.31 \times 10^{12} \textrm{cm}^{-2}$ in the stacked spectrum, corresponding to a $^{12}$C/$^{13}$C ratio of 79$\pm$8. The dashed blue lines around the spectra represent 1$\sigma$ uncertainties.}
    \label{fig:13C2}
\end{figure}

We focused on the sightlines of HD~63804, HD~147889, and HD~169454 because of their large C$_2$ column densities and good S/N in the target region. These factors make them plausible candidates to search for the $^{13}$C$^{12}$C transitions, yet we cannot make robust detection on the target $^{13}$C$^{12}$C transitions in any of these sightlines (Fig. \ref{fig:13C2}, upper panel). To further increase the sensitivity of our data, we continued by performing a S/N-weighted spectral stacking for the three sightlines. The measured S/N of the stacked spectrum reaches $\sim$\ 3,000, and the $Q$(2), $Q$(3), and $Q$(4) transitions are detected in the stacked spectrum (Fig. \ref{fig:13C2}, lower panel). This detection validates our theoretical prediction on the wavelengths, which is based solely on isotopic relations \citep{ROU012:A24, RAM014:5}. 

The stacking process, however, averages different ISM conditions of
the three sightlines, and we cannot obtain detailed excitation
information for $^{13}$C$^{12}$C in the stacked data. Because of this,
the best approach to even calculating the column density of
$^{13}$C$^{12}$C requires analysis of a wide range of probable
excitation conditions especially regarding the excitation
temperature. We compiled a series of models with $T_\textrm{exc}$
between 40 and 80~K (with an interval of 5~K), and obtained the
relative strengths of the $Q$(1) to $Q$(4) transitions from
\texttt{PGOPHER}. We assumed a Gaussian profile with instrumental
widths for each of the transitions. The best fit with the minimal
$\chi^2$ occurs when $T_\textrm{exc}$ = 55~K with
$N$($^{13}$C$^{12}$C) = $2.30 \times 10^{12}\ \textrm{cm}^{-2}$. The
standard deviation of $N$($^{13}$C$^{12}$C) among models of different
$T_\textrm{exc}$ is $0.31 \times 10^{12}\ \textrm{cm}^{-2}$ ($\sim 13
\%$), representing the uncertainty due to the unknown excitation
conditions.

The average $N$(C$_2$) is $8.92 \times 10^{13}\ \textrm{cm}^{-2}$ for the three sightlines\footnote{During the stacking process, the spectral data were weighted by their S/Ns. The same weight was applied to calculate the average $N$(C$_2$) in the three sightlines. We also tried to measure $N$(C$_2$) in the stacked spectrum around the $^{12}$C$_{2}$ transitions, but the fitting was unsuccessful due to the mixture of ISM conditions.}. The $N$(C$_2$)/$N$($^{13}$C$^{12}$C) ratio is then $\sim39\pm4$, or a $^{12}$C/$^{13}$C ratio of $\sim79\pm8$. This is very close to the same ratio of 72$\pm$26 observed towards the circumstellar shell of the post-asymptotic-giant-branch star HD~56126 \citep{1998ApJ...508..387B}. The $^{12}$C/$^{13}$C ratio can also be obtained from other carbon-bearing molecules. For examples, \cite{2005ApJ...634.1126M} reported $\sim$68 by considering CO, H$_2$CO, and CN, and \cite{2008A&A...477..865S} find 76$\pm$2 using CH$^+$ measurements of the local ISM. Again the findings of this work are well aligned with the above literature results.

The $^{12}$C/$^{13}$C ratio traces the large-scale Galactic chemical evolution as well as local peculiarities of the ISM conditions \citep[e.g.][]{2005ApJ...634.1126M, 2011ApJ...731...38F, 2012ApJ...747...55L}. \cite{2021A&A...647A.142R} discussed the temperature dependence of the isotopic fractionation of carbon. Their calculation suggests that a smaller $^{12}$C/$^{13}$C ratio can be expected in sightlines characterised by lower kinetic temperature, since the forward isotope-exchange reaction $^{13}$C + $^{12}$C$_2 \ \rightarrow\ ^{13}$C$^{12}$C + $^{12}$C is exothermic by 26.4~K and the reverse endothermic process becomes largely inefficient at low temperatures. We note in passing that the $^{13}$C$^{+}$ + $^{12}$C$_{2}$ reaction, despite not being considered by \cite{2021A&A...647A.142R} in the context of dense molecular clouds, is also expected to play a key contribution to the overall carbon isotopic fractionation, particularly under the diffuse ISM environments \citep{2020MNRAS.498.4663L, 2020A&A...640A..51C}. Compared to $T_\textrm{kin}$ = 20~K, the $^{12}$C/$^{13}$C ratio can be a factor of 2 greater when $T_\textrm{kin}$ = 100~K. The identification in stacked data cannot assign the $^{12}$C/$^{13}$C ratio to specific ISM conditions, but such dependences may be examined in future studies.

\subsection{$C_4$ and $C_5$}

\begin{table*}
        \centering
        \small
        \caption{Upper limits of $N$(C$_4$) and $N$(C$_5$).}
        \label{tab:C4C5UL}
        \begin{threeparttable}
      \begin{tabular}{cc|cccc}
        \hline
        \multicolumn{2}{c|}{Species} & \multicolumn{4}{c}{Sightlines} \\
        \hline
        \multirow[c]{7}{*}{C$_4$} &  & HD~147889 & HD~169454 & HD~210121 & Stacked \\
         & $\lambda_{jj}$ (\AA) & \multicolumn{4}{c}{3789} \\
         & FWHM\tnote{a,b}\ \ \ \    (\AA) & \multicolumn{4}{c}{0.24} \\
         & $f_{jj}$\tnote{b} & \multicolumn{4}{c}{0.0006} \\
         & $W_\textrm{max}$\tnote{c} \ (m\AA) & 2.1 & 1.7 & 1.6 & 1.0\\
         & $N_\textrm{max}$ \ (cm$^{-2}$) & $2.7\times 10^{13}$ & $1.8\times 10^{13}$ & $2.1\times 10^{13}$ & $1.2\times 10^{13}$\\
         & $N$(C$_3$)/$N_\textrm{max}$ & 0.28 & 0.35 & 0.23 & 0.50 \\
        \hline
        \multirow[c]{7}{*}{C$_5$} &  & HD~154368 & HD~169454 & HD~203532 & Stacked \\
         & $\lambda_{jj}$ (\AA) & \multicolumn{4}{c}{5109} \\
         & FWHM\tnote{a,b} \ \ \ \ (\AA) & \multicolumn{4}{c}{0.13} \\
         & $f_{jj}$\tnote{b} & \multicolumn{4}{c}{0.001} \\
         & $W_\textrm{max}$\tnote{c} \  (m\AA) & 0.74 & 0.84 & 0.44 & 0.39\\
         & $N_\textrm{max}$ \ (cm$^{-2}$) & $5.9\times 10^{12}$ & $7.6\times 10^{12}$ & $3.6\times 10^{12}$ & $3.1\times 10^{12}$\\
         & $N$(C$_3$)/$N_\textrm{max}$ & 0.81 & 1.18 & 0.78 & 0.66\\
         \hline
          \end{tabular}
      \begin{tablenotes}
          \item[a] For the most prominent feature.
          \item[b] From \cite{2002ApJ...566..332M}.
          \item[c] Assuming the most prominent feature has Gaussian profile and contributes to half of the total absorption.
      \end{tablenotes}
    \end{threeparttable}
\end{table*}

\begin{figure*}
    \centering
    \includegraphics[width=15cm]{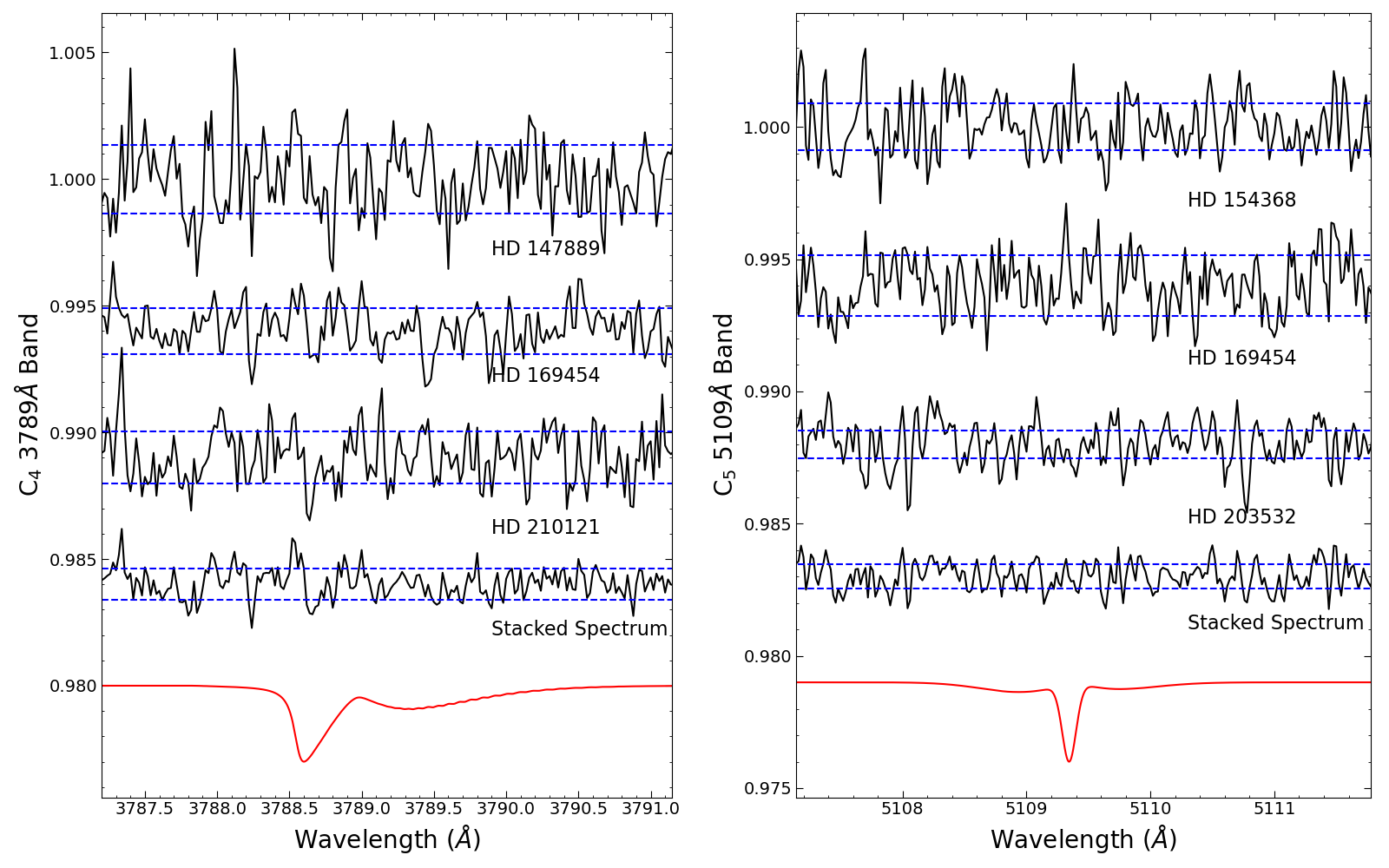}
    \caption{Spectral data around the C$_4$ $^3\Sigma_u^- - ^3\Sigma_g^-$ 3789~\AA\ band (left) and the C$_5$ 5109~\AA\ band, possibly from the $^1\Pi_u - X ^1\Sigma_g^+$ system (right). Their lab spectra are shown at the bottom as reference (red). The sightlines are selected by their measured $N$(C$_3$) and the S/N of the wavelength region, which is indicated by the dashed blue lines. We cannot find the target transition in either individual sightlines nor their stacked spectrum. Upper limits on $N$(C$_4$) and $N$(C$_5$) can be estimated from their most prominent feature, and are around $10^{13}$ cm$^{-2}$ and $10^{12}$ cm$^{-2}$, respectively.}
    \label{fig:C4C5}
\end{figure*}

The C$_4$ and C$_5$ molecules are the simplest bare carbon chains after C$_2$ and C$_3$. In the optical region, the strongest C$_4$ band occurs at 3789~\AA\ and is between the $^3\Sigma_u^-$ and $^3\Sigma_g^-$ states. The laboratory spectrum of this band is characterised by the piled-up $R$-branch to the blue and more expanding $P$-branch contour to the red (\citealt{2000JChPh.112.9777L}; Fig. \ref{fig:C4C5}, left panel). The strongest C$_5$ optical band is at 5109~\AA\ and \cite{1996JChPh.104.4954F} assigned this band to the $^1\Pi_u - X ^1\Sigma_g^+$ system. However, \cite{2001CPL...337..368H} noted a discrepancy of $\sim$0.5~eV in the transition energy and suggested the 5109~\AA\ band to be associated with formally forbidden $^1\Sigma^-_u$ and $^1\Delta_u$ states. Nevertheless the most prominent feature of this band is the piled-up Q-branch in the middle while the contours of the $P$- and $Q$-branches are much shallower (Fig. \ref{fig:C4C5}, right panel). 

Similar to our search for $^{13}$C$^{12}$C, we searched for these optical bands of C$_4$ and C$_5$ in candidate sightlines characterised by large $N$(C$_3$) and high S/N in the wavelength region (namely, HD~147889, HD~169454, and HD~210121 for the C$_4$ 3789~\AA\ band and HD~154368, HD~169454, and HD~203532 for the C$_5$ 5109~\AA\ band). However, neither of the target transitions can be identified in individual sightlines or their stacked spectra (Fig. \ref{fig:C4C5}). We instead estimated the upper limits of $N$(C$_4$) and $N$(C$_5$) by considering the most prominent feature in their band profile (i.e. the $R$-branch contour for C$_4$ 3789~\AA\ band and the $Q$-branch contour for C$_5$ 5109~\AA\ band) while assuming it contributes 50\% of the total absorption. The results are summarised in Table \ref{tab:C4C5UL}.

The C$_5$ molecules have been detected in earlier work via their IR
transitions in the carbon-rich circumstellar shell around IRC
+10216. \cite{1989Sci...244..562B} suggested $N$(C$_5$) might be
$\sim$\ 1/10 of $N$(C$_3$), yet a more recent observation provided a
much smaller $N$(C$_5$)/$N$(C$_3$) ratio of 1/68
\citep{2014MNRAS.444.3721H}. Theoretical models also predict
$N$(C$_4$) and $N$(C$_5$) to be one to two orders of magnitude smaller
than $N$(C$_3$) \citep[e.g.][]{1998ApJ...509..932T,
  2014MNRAS.441.1134S}, and the much smaller $f_{jj}$ values of the
C$_4$ and C$_5$ optical bands make their detection even more
difficult. These $f_{jj}$ values are also inconsistent in the
literature. For example, \cite{2000CPL...324..195M} reported $f_{jj}
\sim$0.0001 for the C$_4$ 3789~\AA\ band, which is merely one-sixth of
the value used in Table \ref{tab:C4C5UL} \citep[see
  also][]{TheAbsorptionSpectrumofC4, 1988JChPh..88.1066P}. Under the
best case scenario where $N$(C$_4$) and $N$(C$_5$) is $\sim$ 1/10 of
$N$(C$_3$) and their transitions have the largest $f_{jj}$ values
reported in the literature, the S/N of the most molecule-rich
sightlines should be boosted by a factor of 2--10 to increase the
chance for a successful C$_4$ or C$_5$ detection.

\section{Summary}\label{sec:summary}
We conducted an extensive survey of C$_2$ and C$_3$ signatures in the EDIBLES dataset, using the C$_2$ $A ^1\Pi_{u} \leftarrow X ^1\Sigma _g^+$ (2-0) Phillips band and the C$_3 \ \tilde{A}\ ^1\Pi_u - \tilde{X} ^1\Sigma_g^+$ 000-000 electronic origin band. Our data have high S/N and can resolve individual ro-vibronic transitions of the above bands, which allowed us to characterise the rotational excitation and obtain more accurate column densities. The results and conclusions we have reached are as follows:

\begin{enumerate}
    \item We detect the C$_2$ (2-0) Phillips band in 51 velocity components along 40 sightlines; 48 velocity components have detections at the 3$\sigma$ level or better, and three sightlines have tentative detections. The data were interpreted  following the excitation model of \cite{1982ApJ...258..533V}, and we were able to obtain information about the kinetic temperature and density in addition to the column density.
    \item We detect the C$_3$ $\tilde{A} - \tilde{X}$ 000-000 band in 31 velocity components along 27 sightlines; 23 velocity components in 19 sightlines have detections at the 3$\sigma$ level or better, and eight sightlines have tentative detections of C$_3$. Due to the lack of a detailed excitation model, we assumed a double-temperature Boltzmann distribution and report the excitation temperatures and column density.
    \item The C$_2$ and C$_3$ transitions have consistent velocity offsets within their uncertainties, suggesting that the two species are detected in the same velocity components. We find a good correlation between $N$(C$_2$) and $N$(C$_3$), namely $r = 0.93$. The average $N$(C$_2$)/$N$(C$_3$) ratio in our survey is 15.5$\pm$1.4, which is lower than a previously reported value of $\sim$\ 40 (where individual C$_3$ transitions were not resolved) and slightly higher than the value of $\sim$10 reported by \citet{2014MNRAS.441.1134S}.
    \item We confirm that the C$_2$/C$_3$ detections occur in $\zeta$-type sightlines characterised by large $f_{H2}$ values and low $W$(5780)/$W$(5797) ratios. We compared the behaviour of DIBs in C$_2$ and non-C$_2$ sightlines. Unlike regular DIBs, the C$_2$ DIBs exhibit a similar behaviour in the two types of sightlines. They may trace denser regions in the ISM clouds than the regular DIBs, yet not as deep as where C$_2$ molecules are found.
    \item We searched for the optical bands of C$_4$, C$_5$, and the singly substituted isotopologue of C$_2$ ($^{13}$C$^{12}$C) in sightlines with high C$_2$/C$_3$ column densities. We identified the $^{13}$C$^{12}$C (2-0) Phillips band in a stacked spectrum via its $Q$(2), $Q$(3), and (tentatively) $Q$(4) transitions. The estimated $^{12}$C/$^{13}$C ratio is 79$\pm$8. For C$_4$ and C$_5$, only upper limits to the column density can be derived, as in earlier unsuccessful surveys.
\end{enumerate}

The high spectral resolution and high S/N offered by the EDIBLES
dataset thus result in accurate determinations of the physical
parameters that describe the denser parts of diffuse clouds, and
provides us with insight into the chemistry in these regions. 

\section*{Acknowledgements}

HF and JC acknowledge support from an NSERC Discovery Grant and a Western SERB Accelerator Award. CMRR gratefully acknowledges the financial support from the European Union’s Horizon 2020 research and innovation program under the Marie
Sklodowska-Curie grant agreement no. 894321.

\bibliographystyle{aa}


\begin{appendix}
\onecolumn

\section{Fitted C$_2$ models}\label{appx_C2}
In this appendix we present plots of the best-fit model of the C$_2$ (2-0) Phillips band compared to the normalised flux for the 40 sightlines listed in Table \ref{table: C2 Detection}. Readers are referred to Sect. \ref{sec:C2} for details about the model.  

\begin{figure*} [hb]
\resizebox{\hsize}{!}{\includegraphics{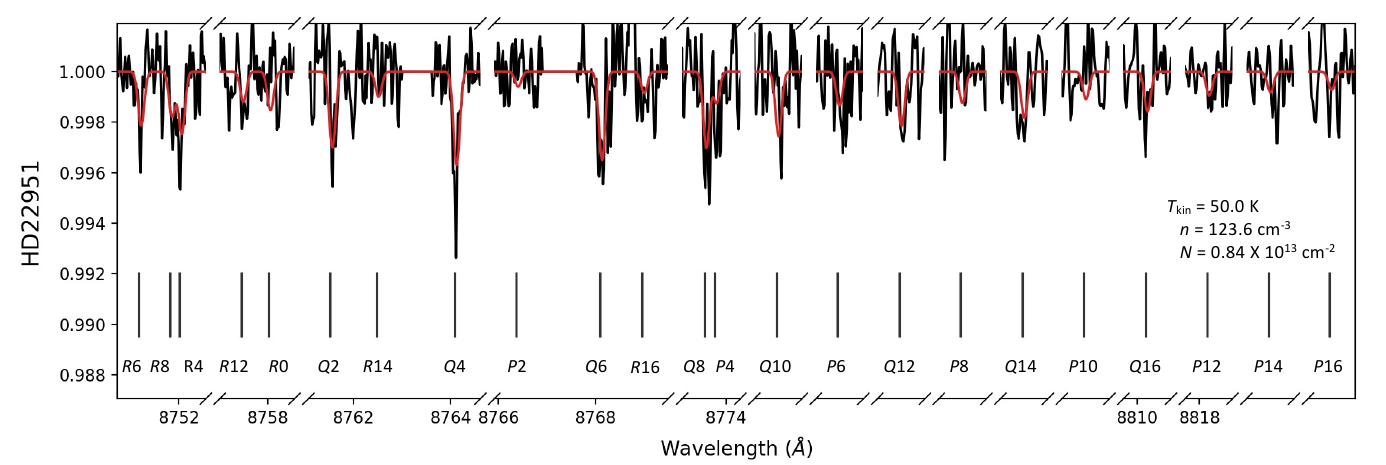}}
\resizebox{\hsize}{!}{\includegraphics{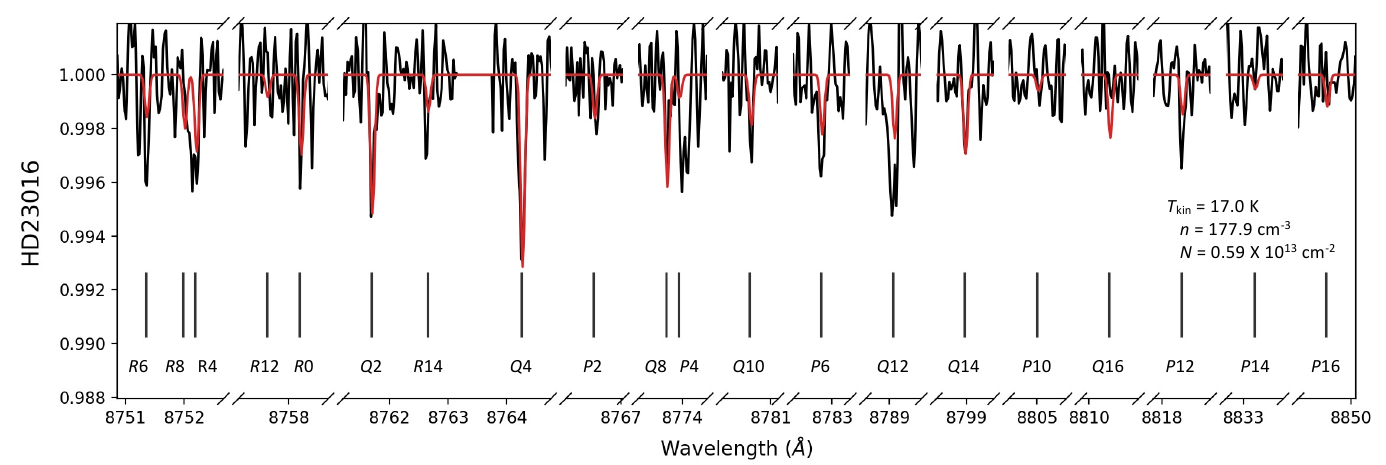}}
\resizebox{\hsize}{!}{\includegraphics{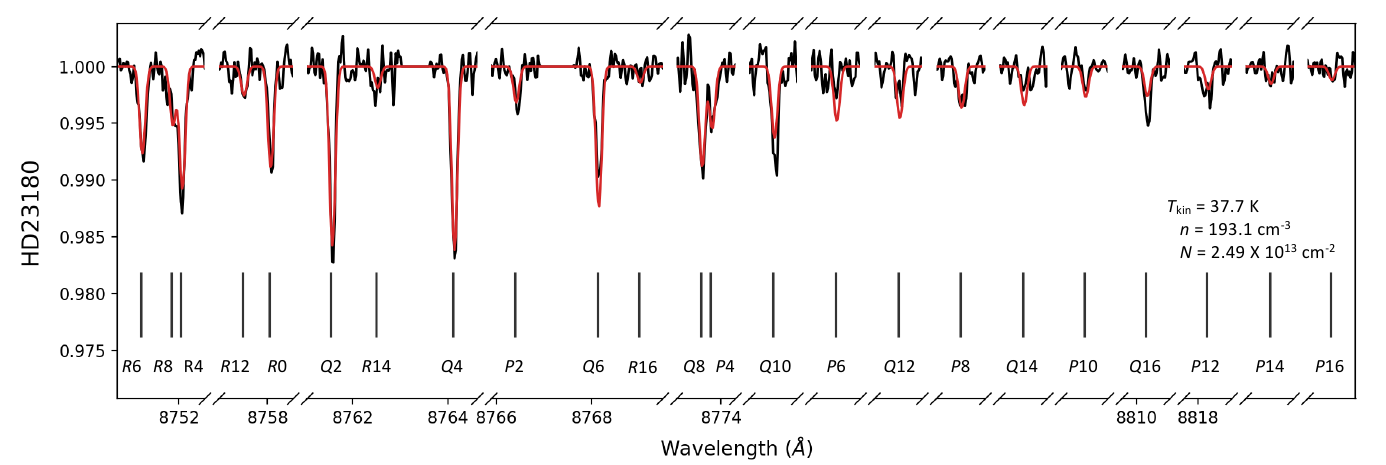}}
    \caption{\label{Fig:fits_c2_app}Normalised flux (black) and fitted model of the C$_2$ (2-0) Phillips band (red) for HD~22951 (top), HD~23016 (middle), and HD~23180 (bottom). The best-fit model parameters $T_\textrm{kin}$, $n$, and $N$ are given in the legend. As in Fig.~\ref{fig:C2_Demo}, the plots focus on segments around the C$_2$ transitions being considered for the fitting, and the wavelength grid is not continuous.}
\end{figure*}

\begin{figure*} [ht]
    \includegraphics[width=.96\textwidth]{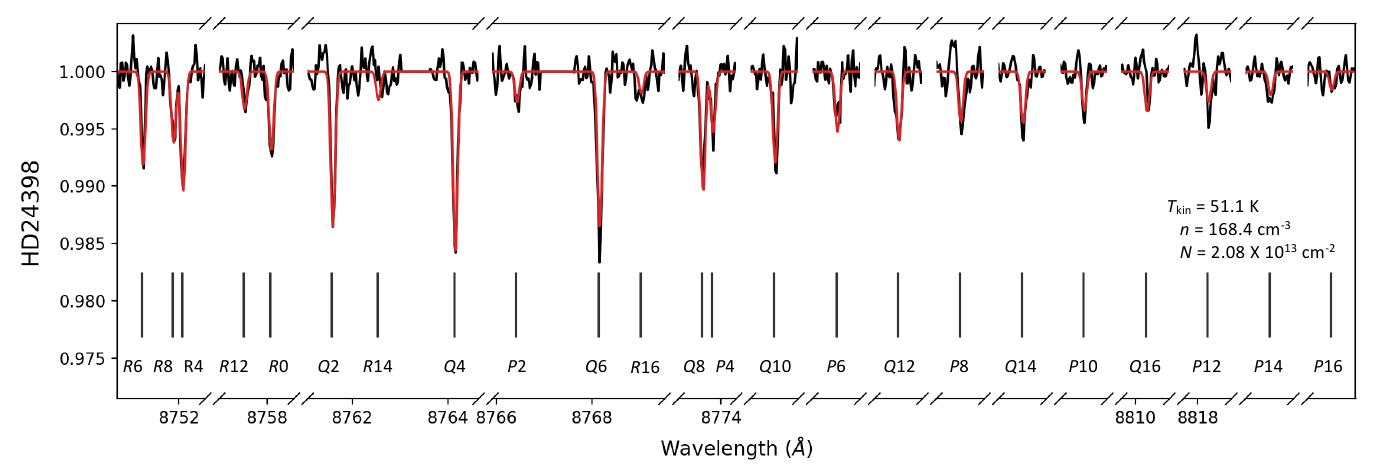}
    \includegraphics[width=.96\textwidth]{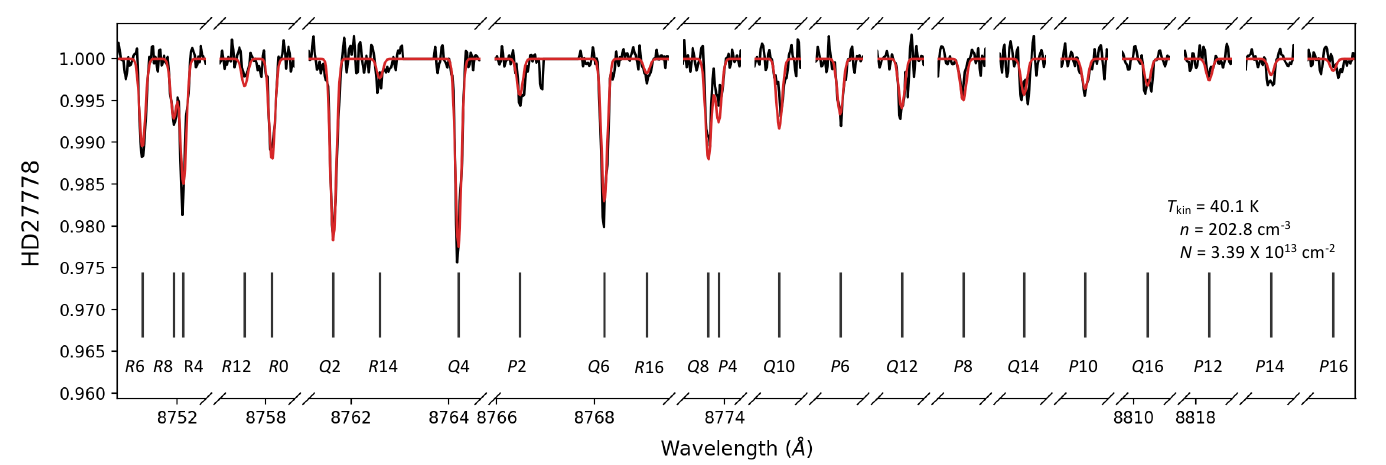}
    \includegraphics[width=.96\textwidth]{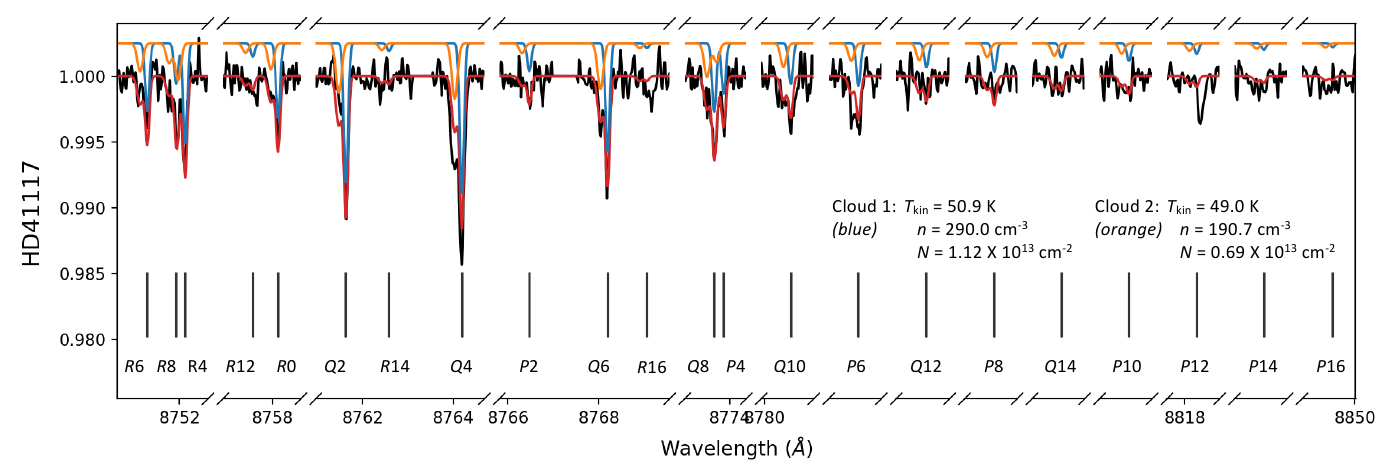}
    \includegraphics[width=.96\textwidth]{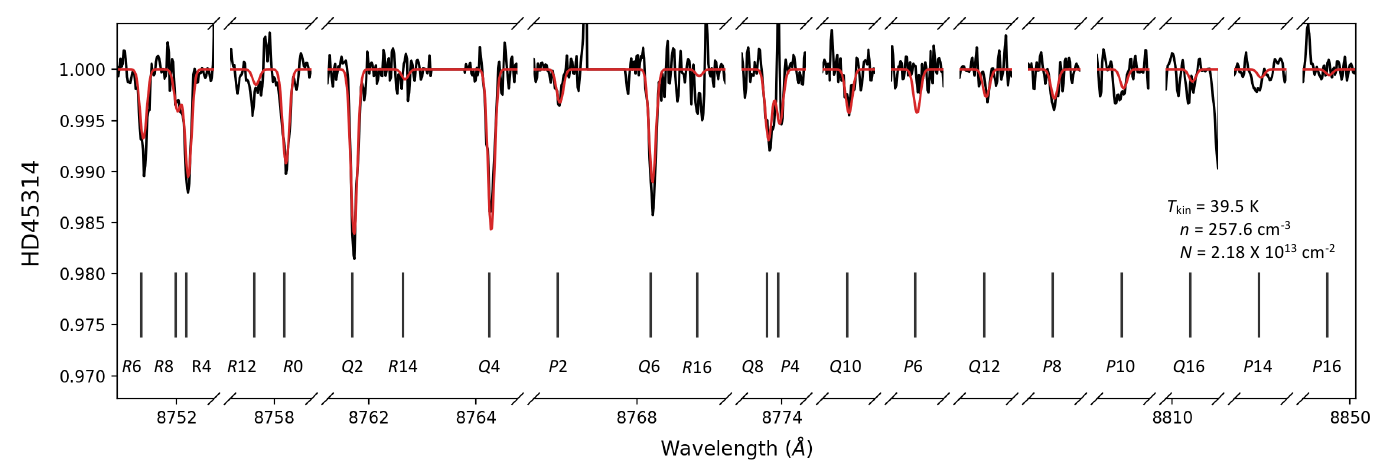}
\caption{Same as Fig.~\ref{Fig:fits_c2_app} but for HD~24398, HD~27778, HD~41117, and HD~45314. For HD~41117, synthetic spectra for individual velocity components are shown in orange and blue.}
\end{figure*}

\begin{figure*} [ht]
    \includegraphics[width=.96\textwidth]{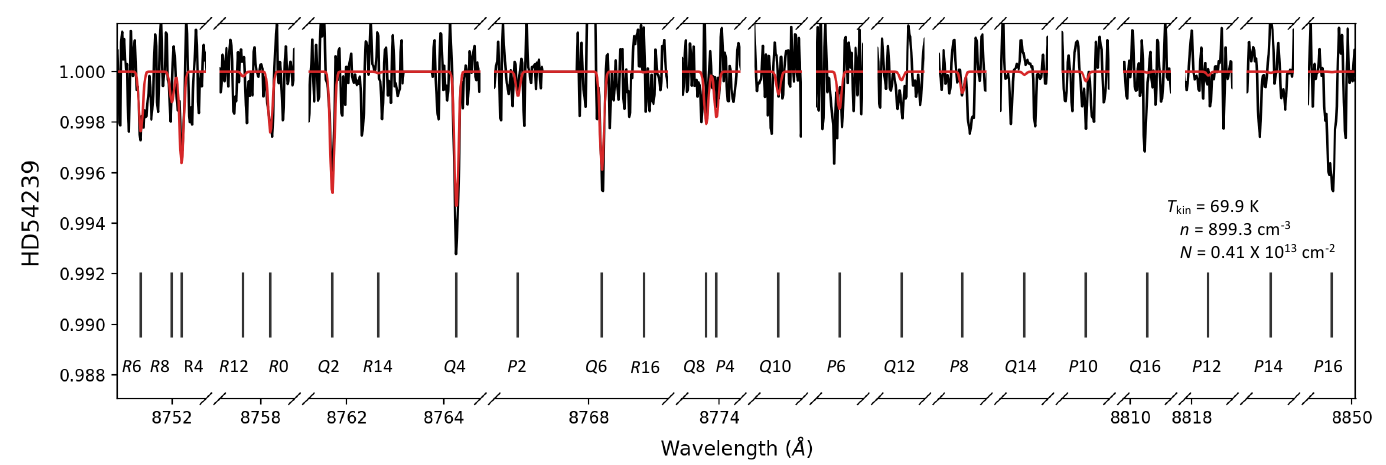}
    \includegraphics[width=.96\textwidth]{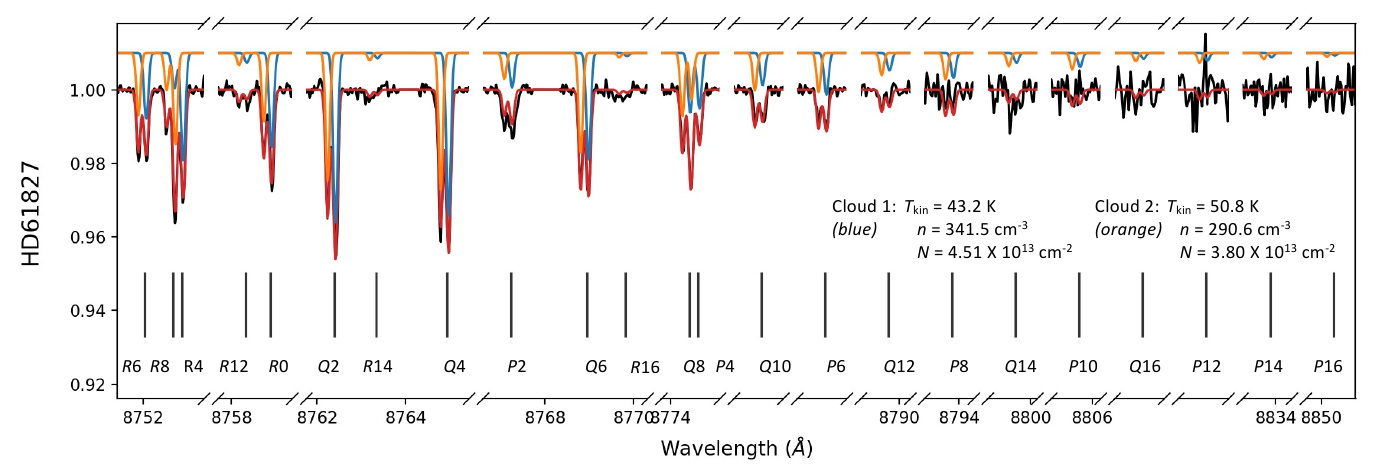}
    \includegraphics[width=.96\textwidth]{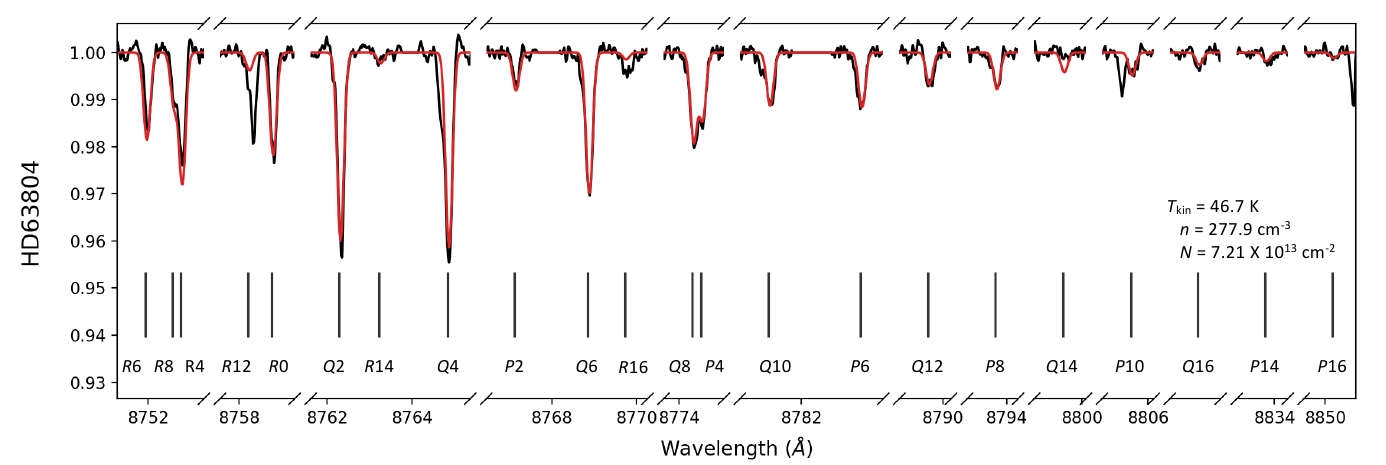}
    \includegraphics[width=.96\textwidth]{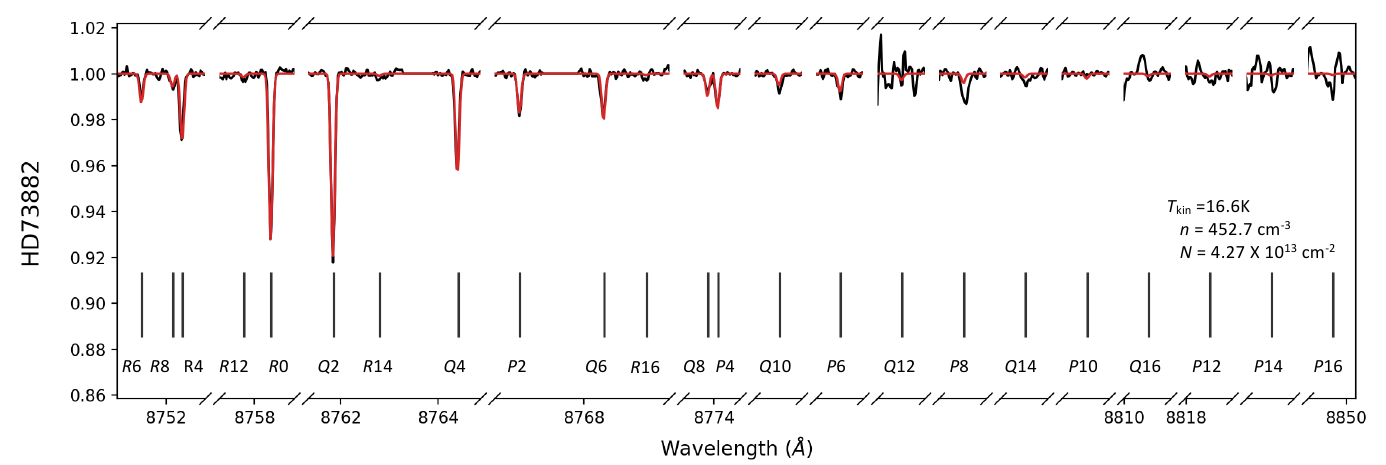}
\caption{Same as Fig.~\ref{Fig:fits_c2_app} but for HD~54239, HD~61827, HD~63804, and HD~73882. For HD~61827, synthetic spectra for individual velocity components are shown in orange and blue.}

\end{figure*}

\begin{figure*} [ht]
    \includegraphics[width=0.96\textwidth]{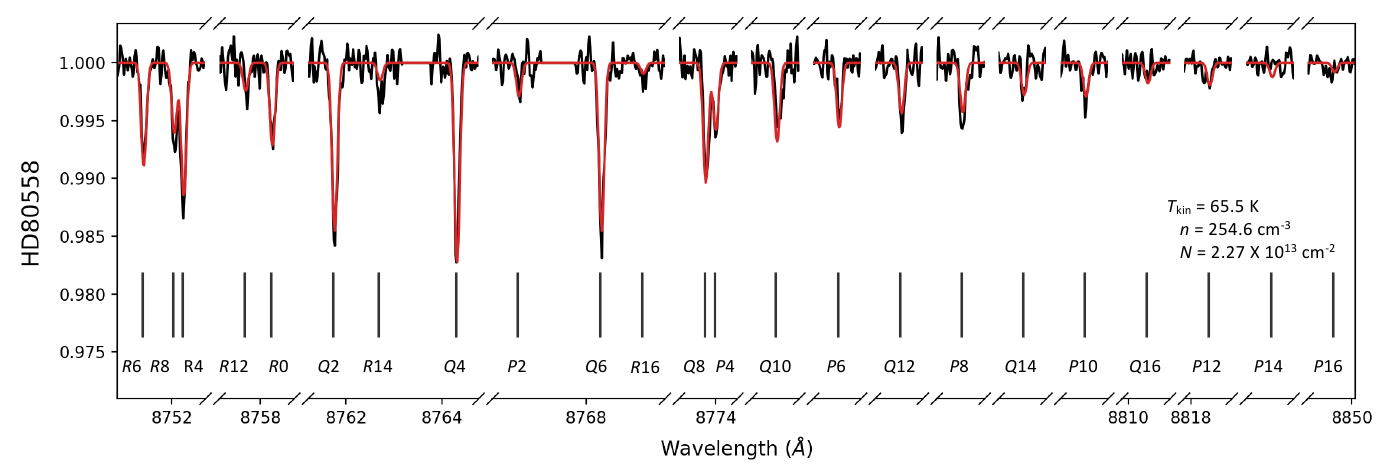}
    \includegraphics[width=0.96\textwidth]{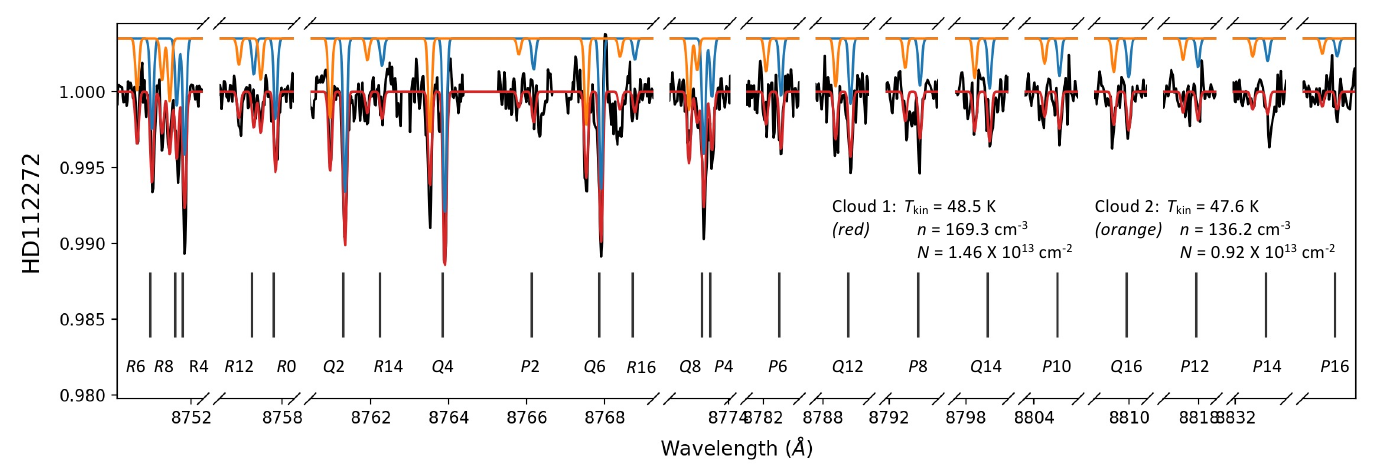}
    \includegraphics[width=0.96\textwidth]{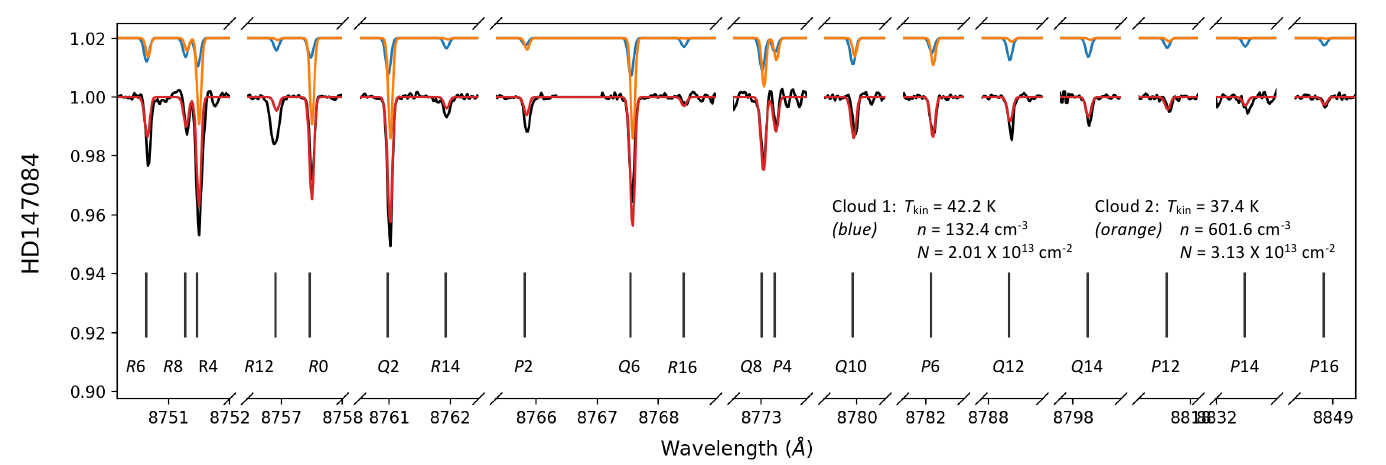}
    \includegraphics[width=0.96\textwidth]{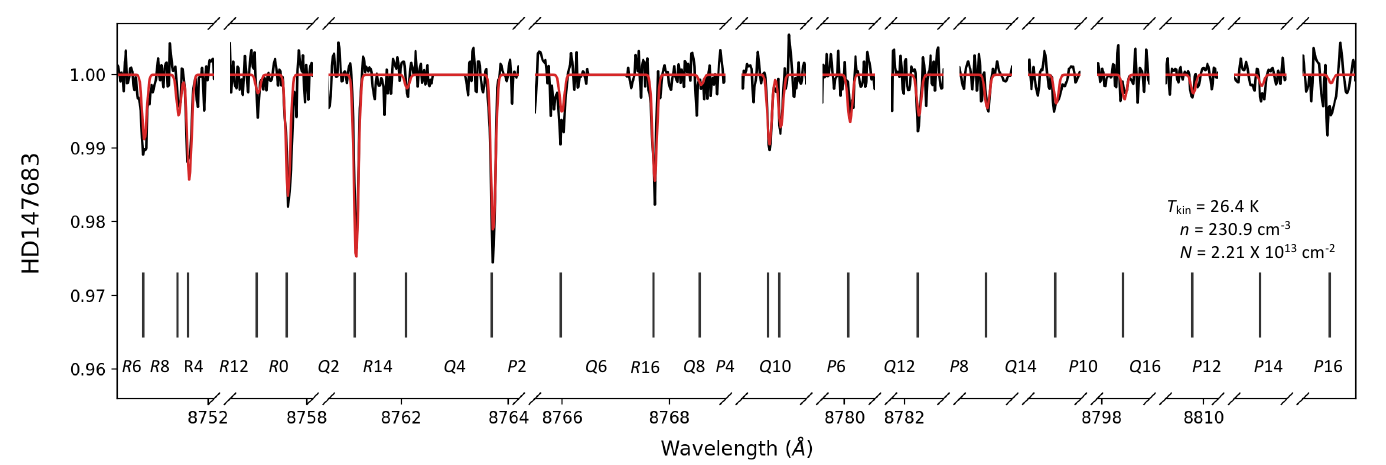}
\caption{Same as Fig.~\ref{Fig:fits_c2_app} but for HD~80558, HD~112272, HD~147084, and HD~147683. For HD~112272 and HD~147084, synthetic spectra for individual velocity components are shown in orange and blue.}    
\end{figure*}

\begin{figure*} [ht]
    \includegraphics[width=0.96\textwidth]{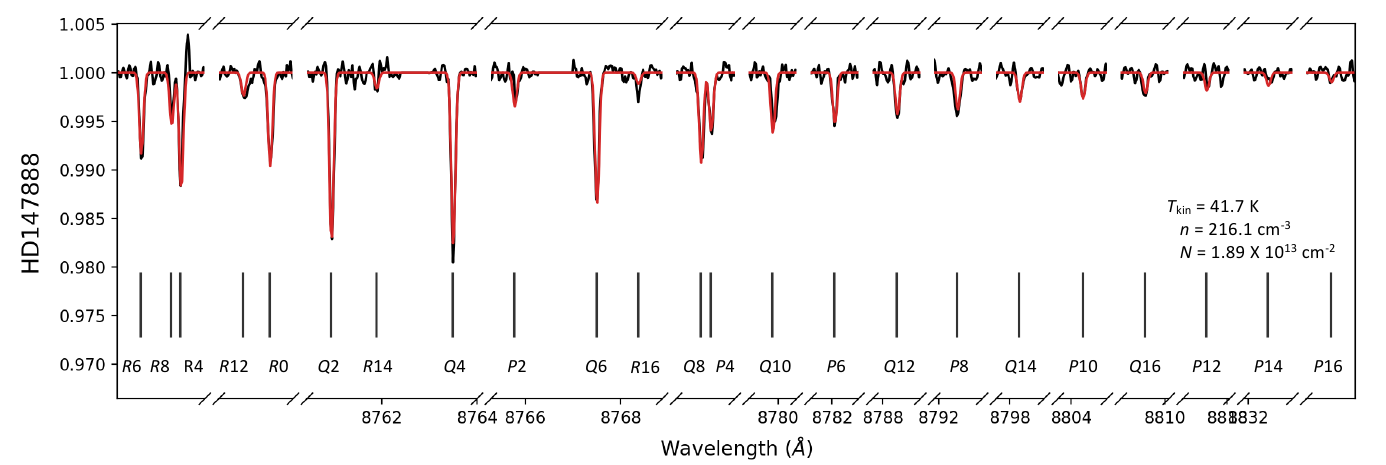}
    \includegraphics[width=0.96\textwidth]{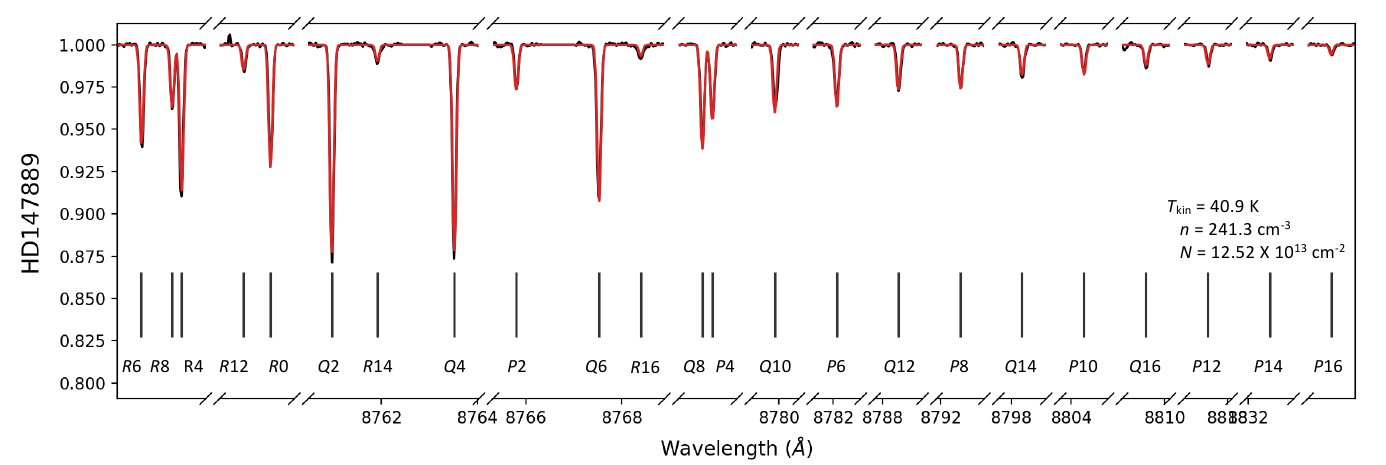}
    \includegraphics[width=0.96\textwidth]{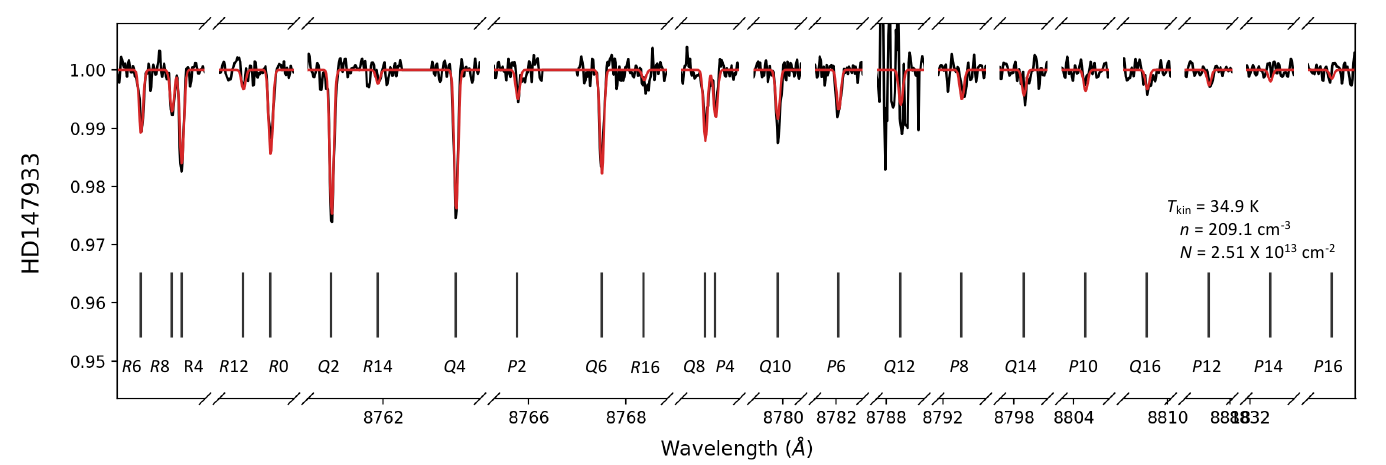}
    \includegraphics[width=0.96\textwidth]{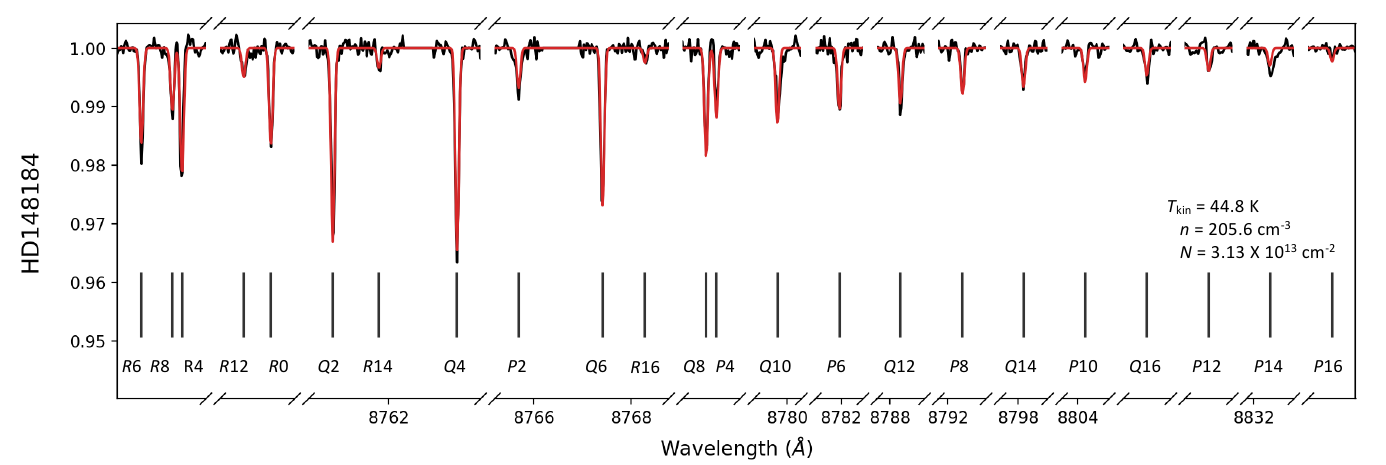}
\caption{Same as Fig.~\ref{Fig:fits_c2_app} but for HD~147888, HD~147889, HD~147933, and HD~148184. }    

\end{figure*}

\begin{figure*} [ht]
    \includegraphics[width=0.96\textwidth]{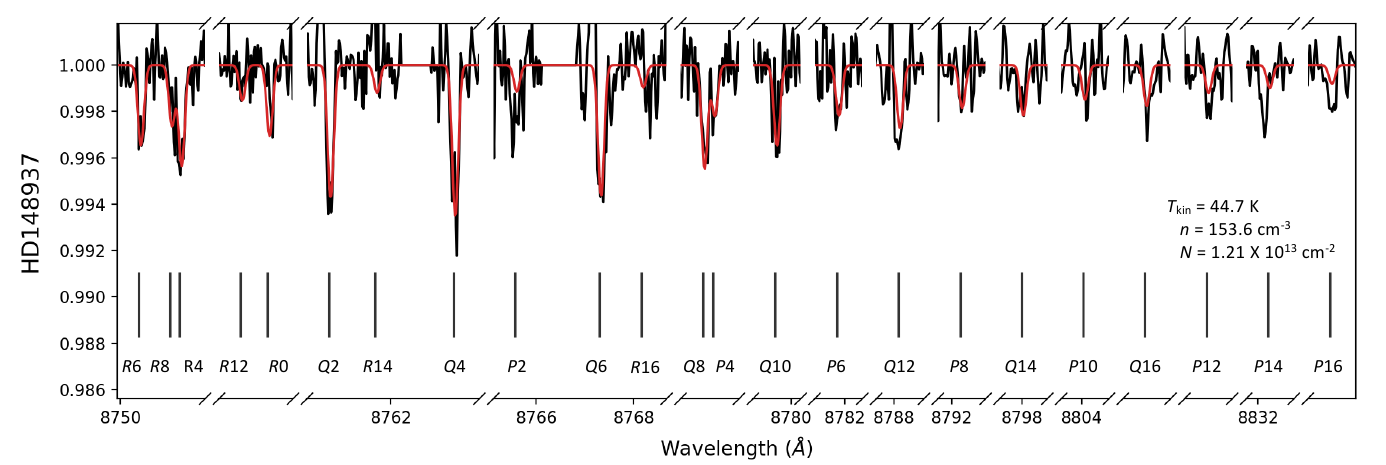}
    \includegraphics[width=0.96\textwidth]{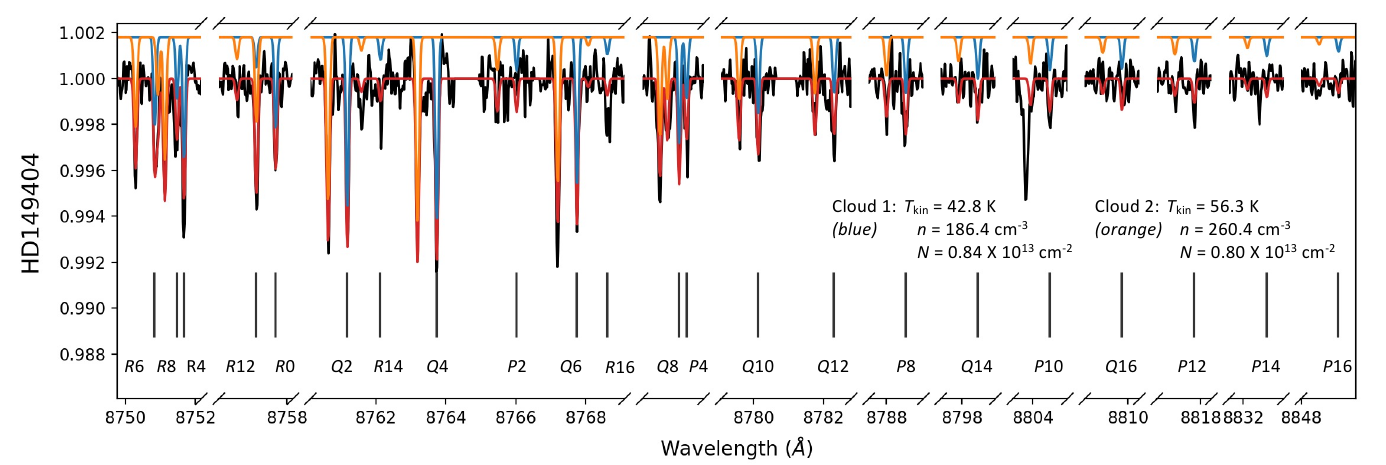}
    \includegraphics[width=0.96\textwidth]{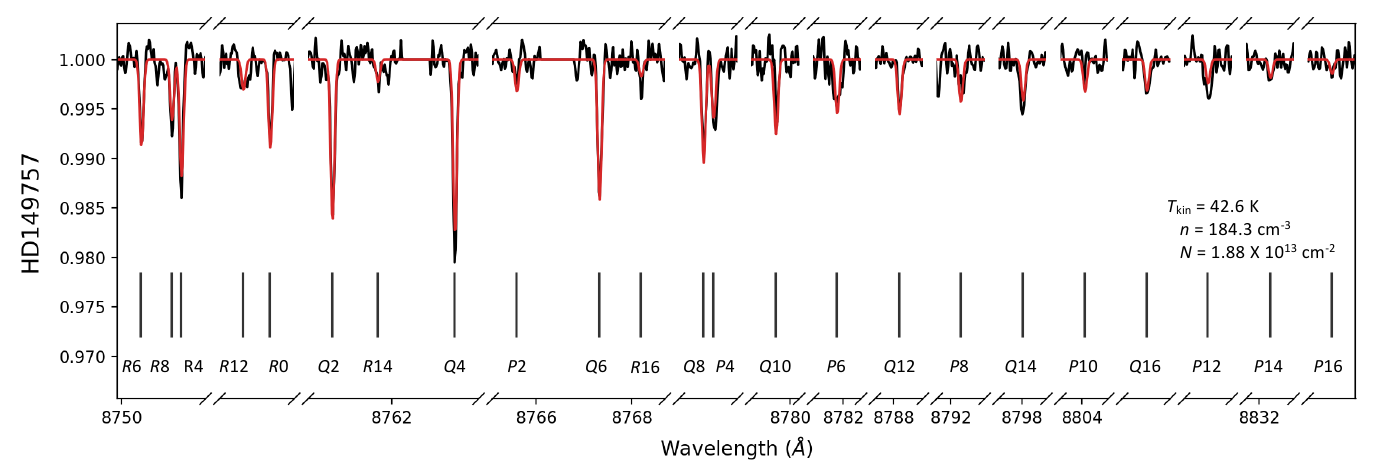}
    \includegraphics[width=0.96\textwidth]{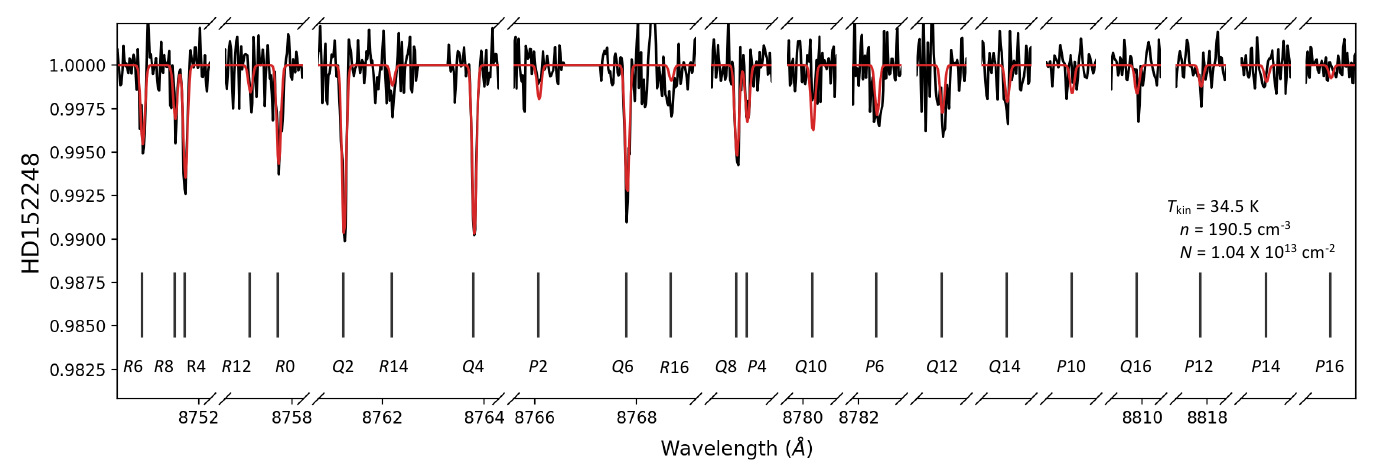}
\caption{Same as Fig.~\ref{Fig:fits_c2_app} but for HD~148937, HD~149404, HD~149757, and HD~152248. For HD~149404, synthetic spectra for individual velocity components are shown in orange and blue.}    

\end{figure*}

\begin{figure*} [ht]
    \includegraphics[width=0.96\textwidth]{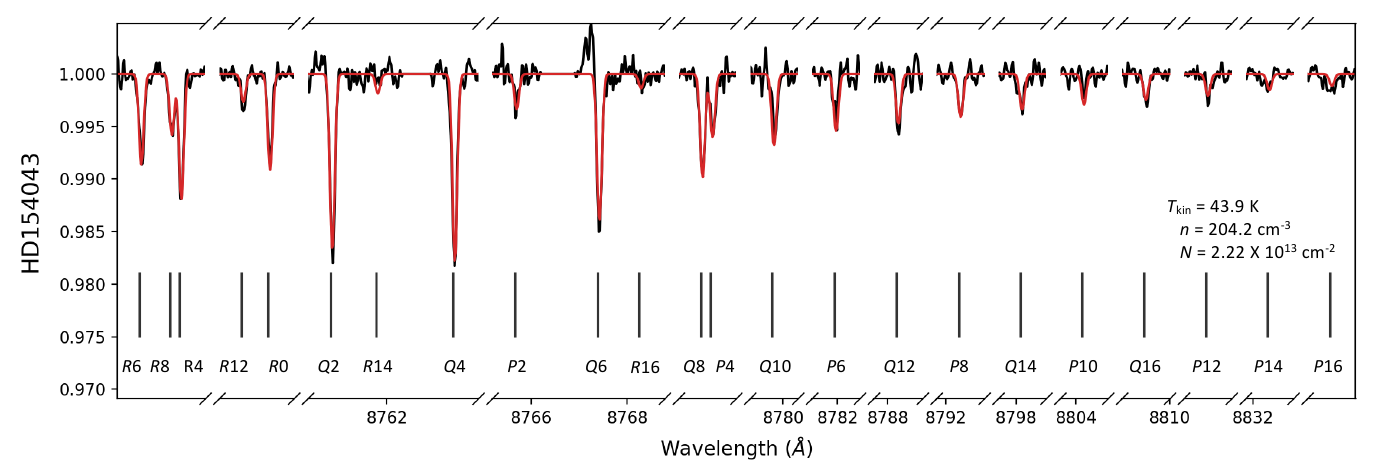}
    \includegraphics[width=0.96\textwidth]{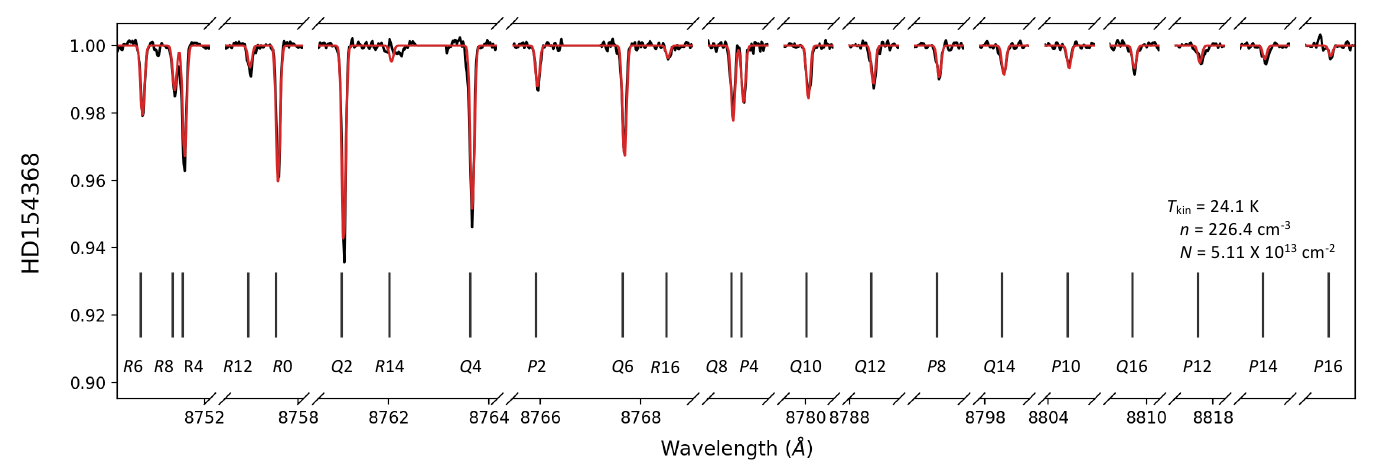}
    \includegraphics[width=0.96\textwidth]{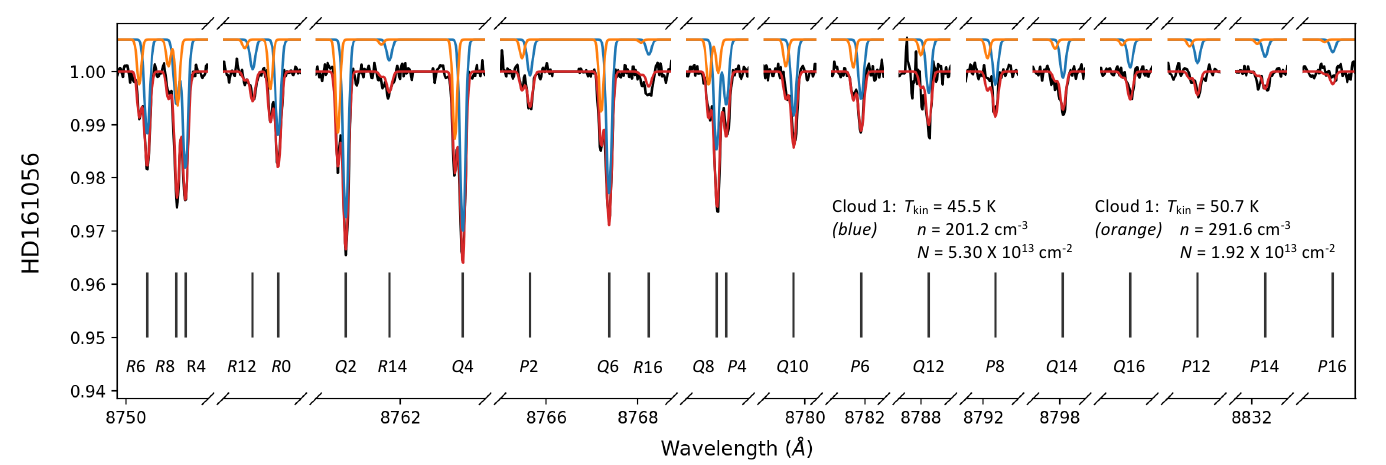}
    \includegraphics[width=0.96\textwidth]{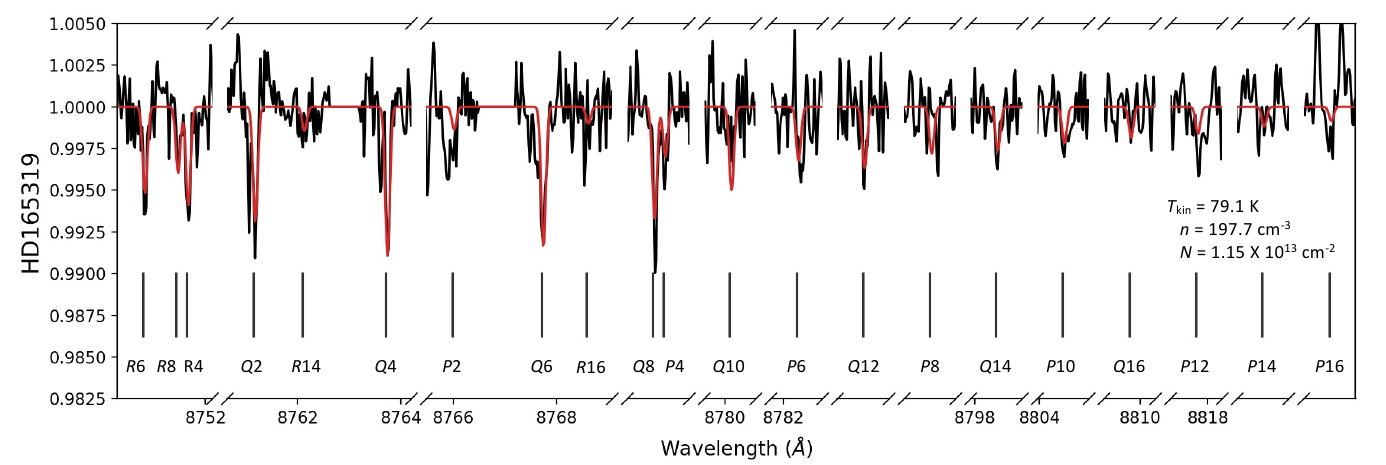}
\caption{Same as Fig.~\ref{Fig:fits_c2_app} but for HD~154043, HD~154368, HD~161056, and HD~165319. For HD~161056, synthetic spectra for individual velocity components are shown in orange and blue.}    

\end{figure*}

\begin{figure*} [ht]
    \includegraphics[width=0.96\textwidth]{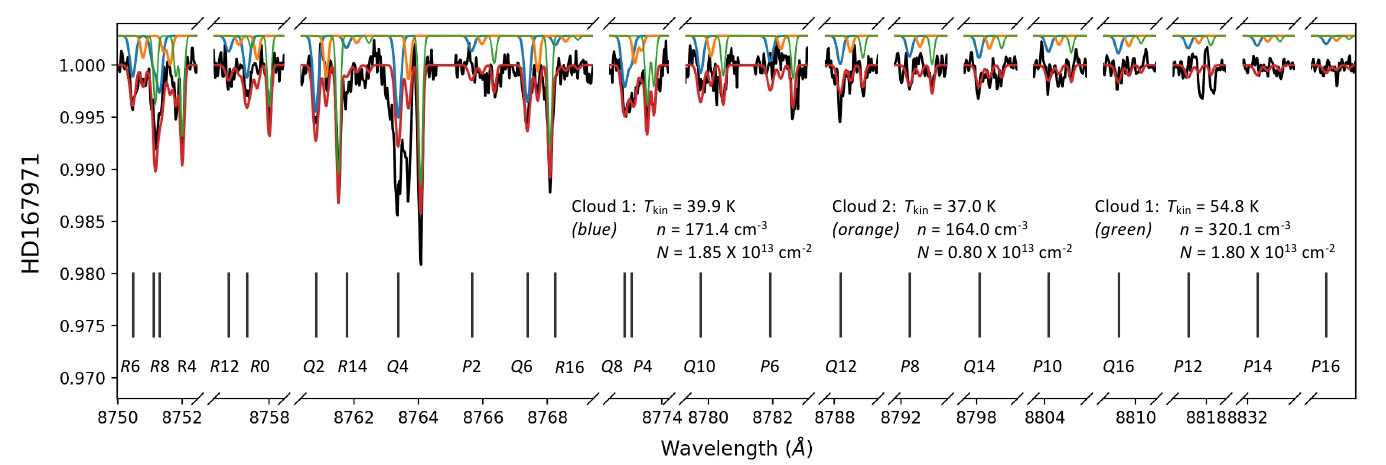}
    \includegraphics[width=0.96\textwidth]{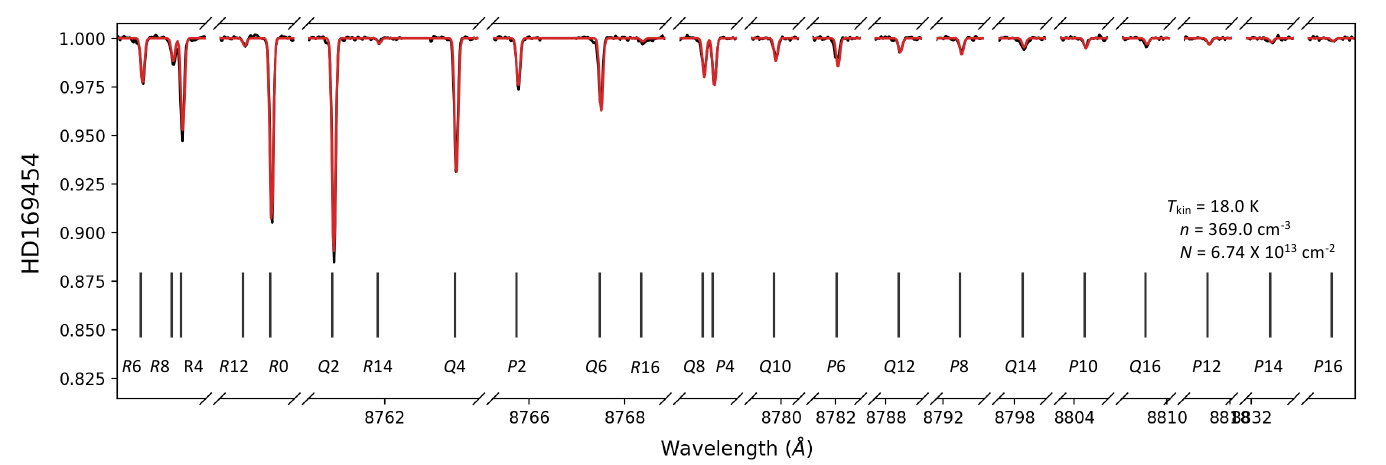}
    \includegraphics[width=0.96\textwidth]{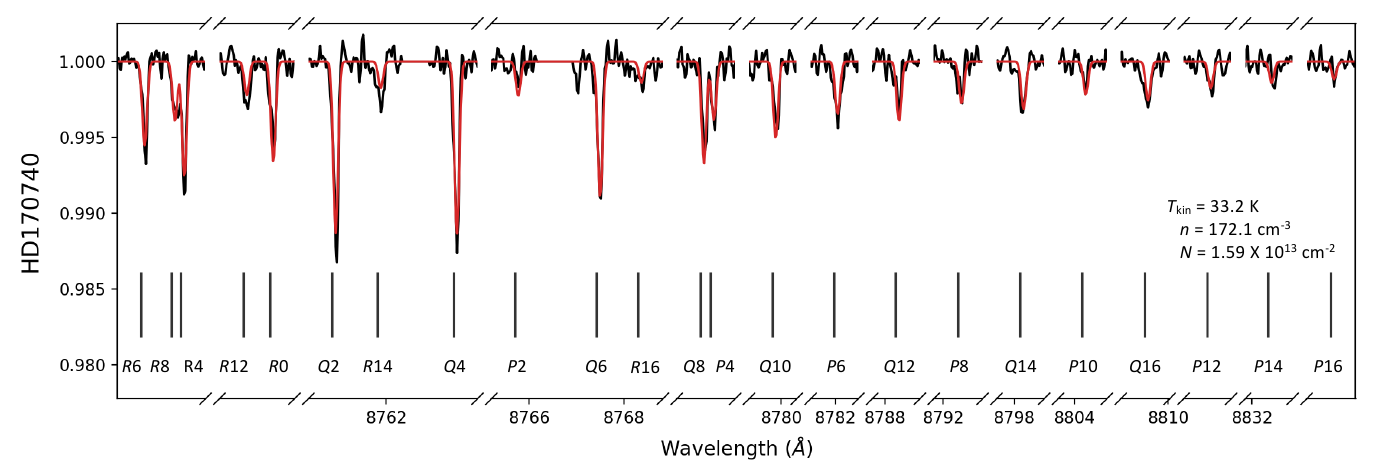}
    \includegraphics[width=0.96\textwidth]{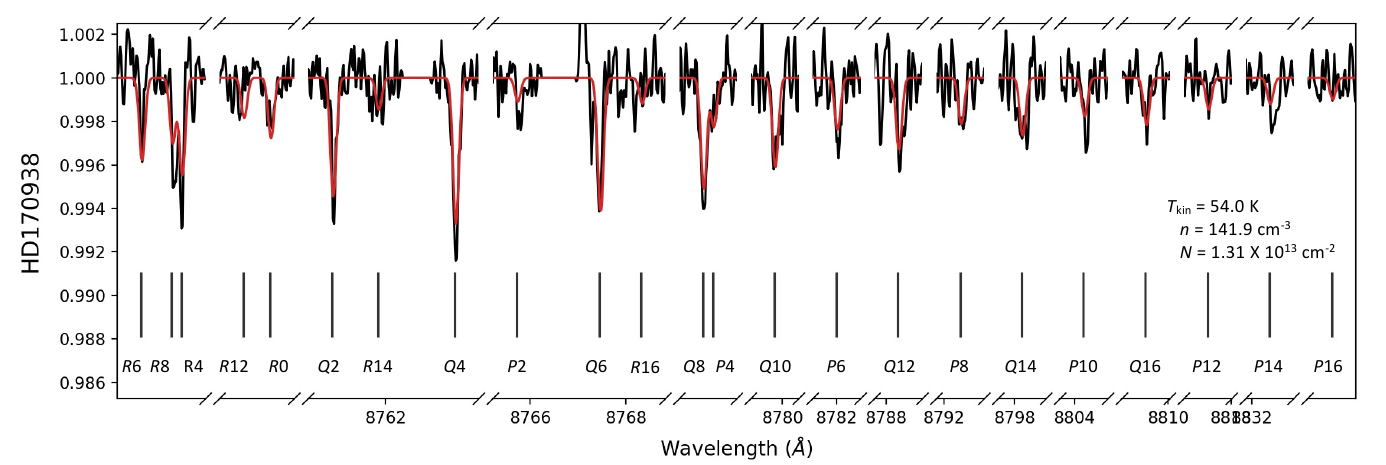}
\caption{Same as Fig.~\ref{Fig:fits_c2_app} but for HD~167971, HD~169454, HD~170740, and HD~170938. For HD~167971, synthetic spectra for individual velocity components are shown in orange, blue, and green.}    

\end{figure*}

\begin{figure*} [ht]
    \includegraphics[width=0.96\textwidth]{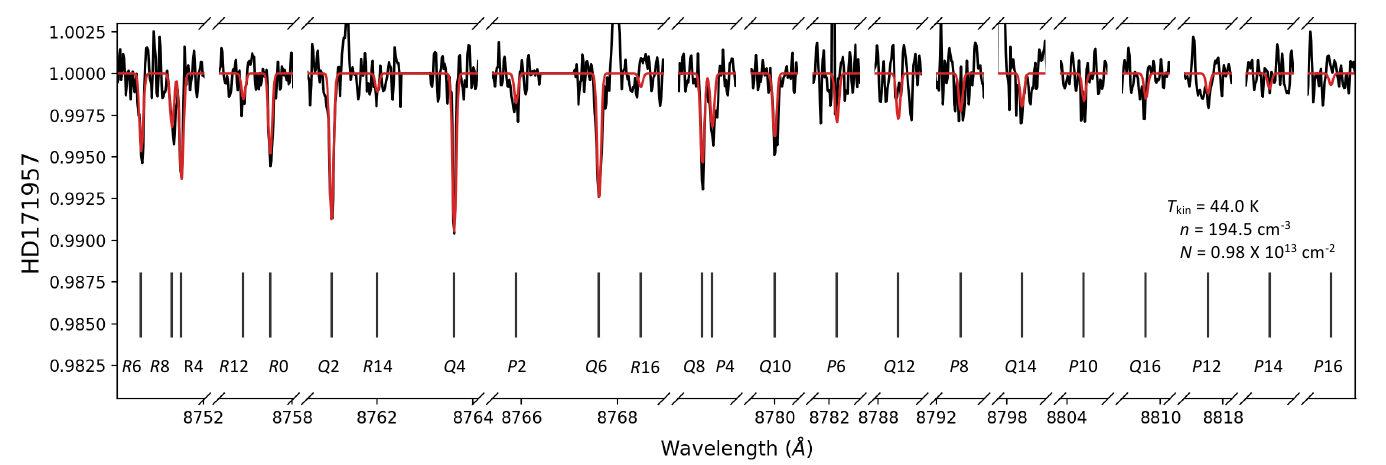}
    \includegraphics[width=0.96\textwidth]{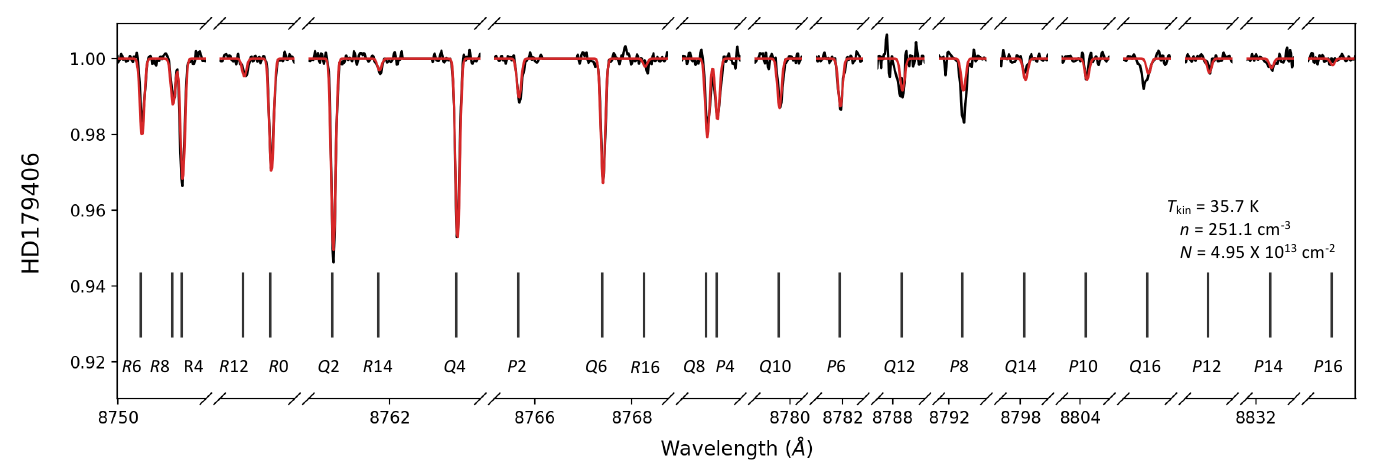}
    \includegraphics[width=0.96\textwidth]{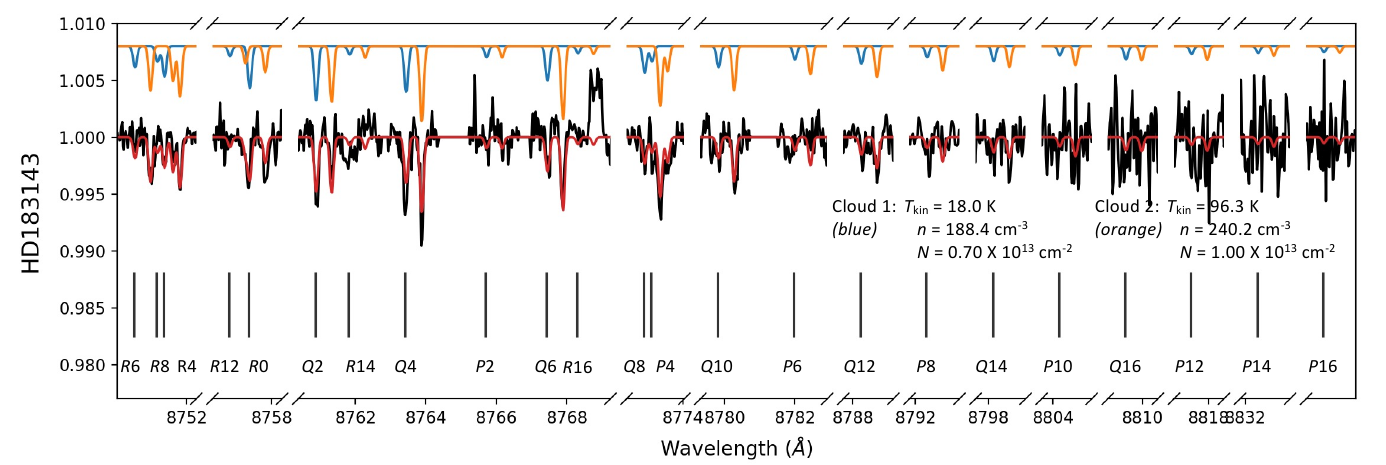}
    \includegraphics[width=0.96\textwidth]{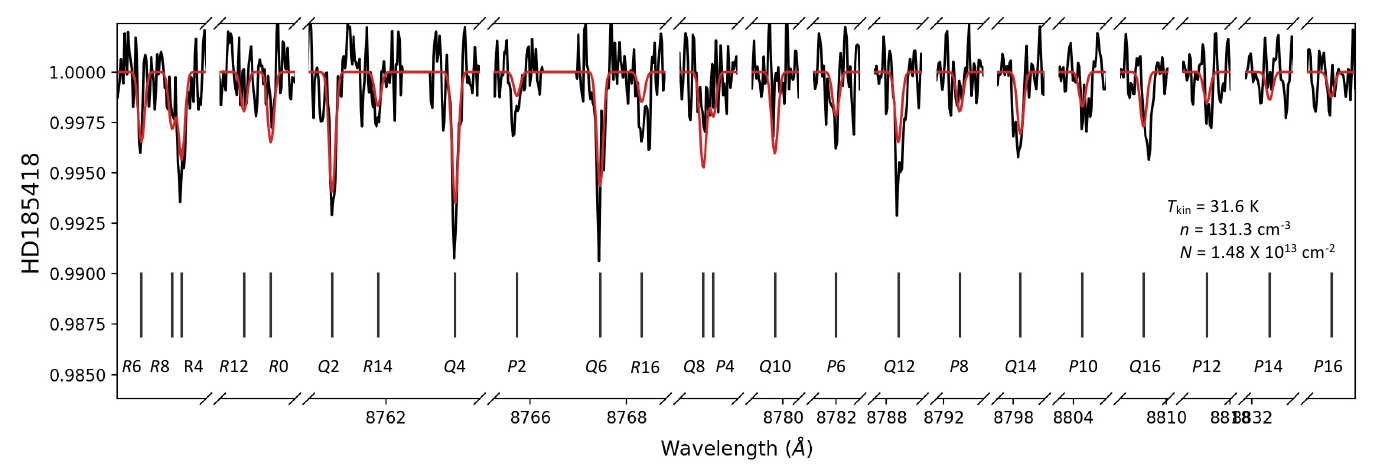}
\caption{Same as Fig.~\ref{Fig:fits_c2_app} but for HD~171957, HD~179406, HD~183143, and HD~185418. For HD~183143, synthetic spectra for individual velocity components are shown in orange and blue.}    
\end{figure*}

\begin{figure*} [ht]
    \includegraphics[width=0.96\textwidth]{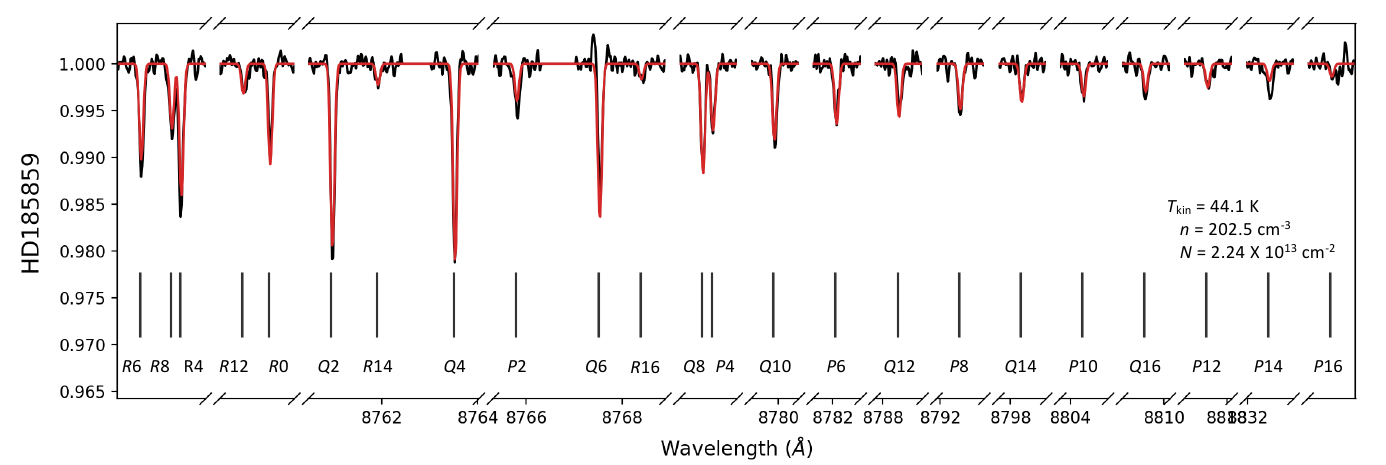}
    \includegraphics[width=0.96\textwidth]{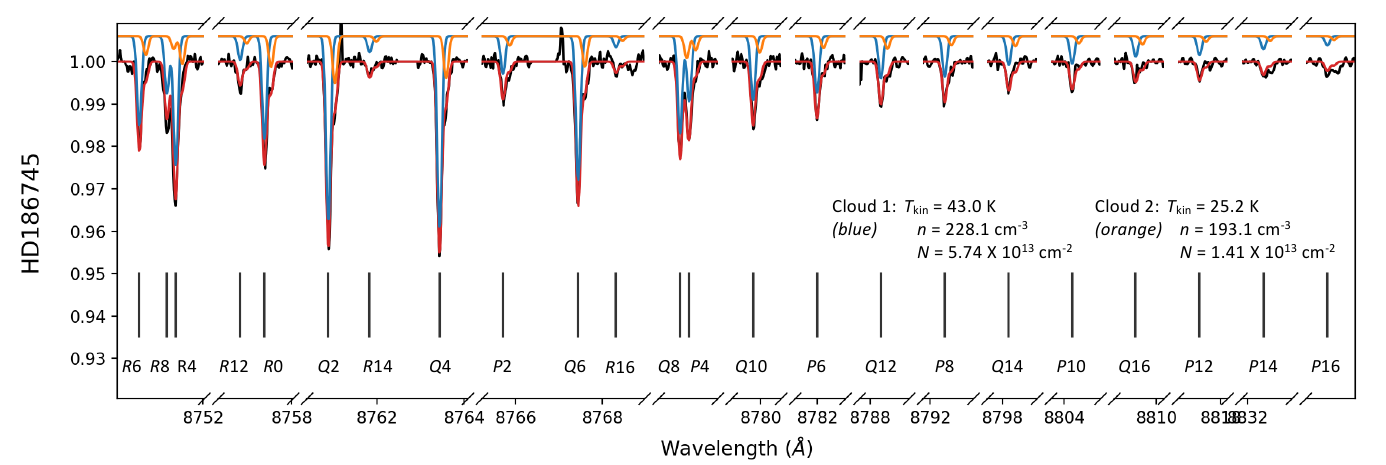}
    \includegraphics[width=0.96\textwidth]{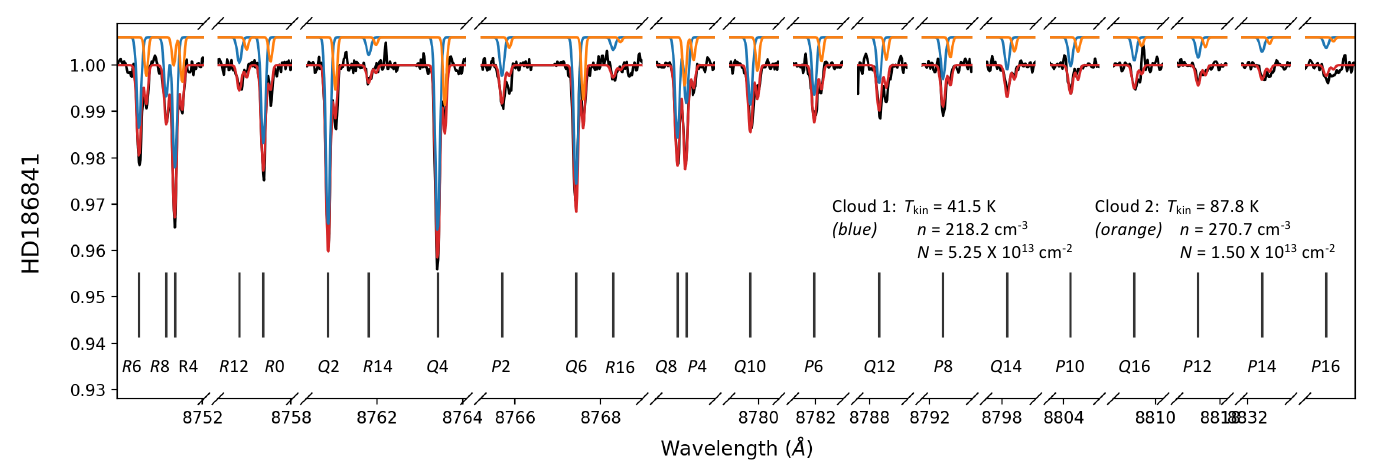}
    \includegraphics[width=0.96\textwidth]{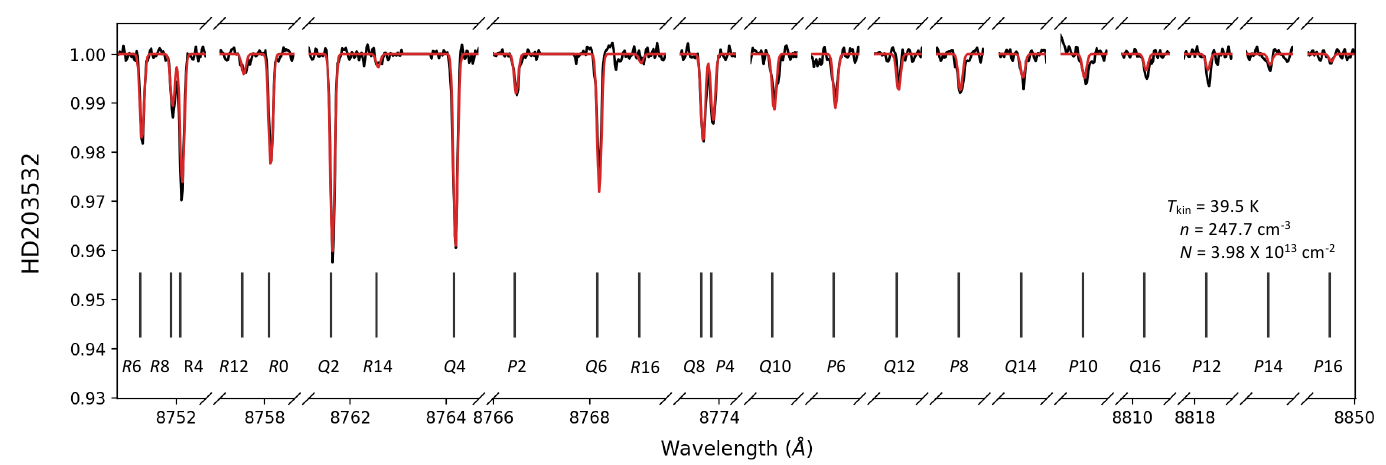}
\caption{Same as Fig.~\ref{Fig:fits_c2_app} but for HD~185859, HD~186745, HD~186841, and HD~203532. For HD~186745 and HD~HD186841, synthetic spectra for individual velocity components are shown in orange and blue.}    
\end{figure*}

\clearpage

\begin{figure*} [ht]
    \includegraphics[width=0.99\textwidth]{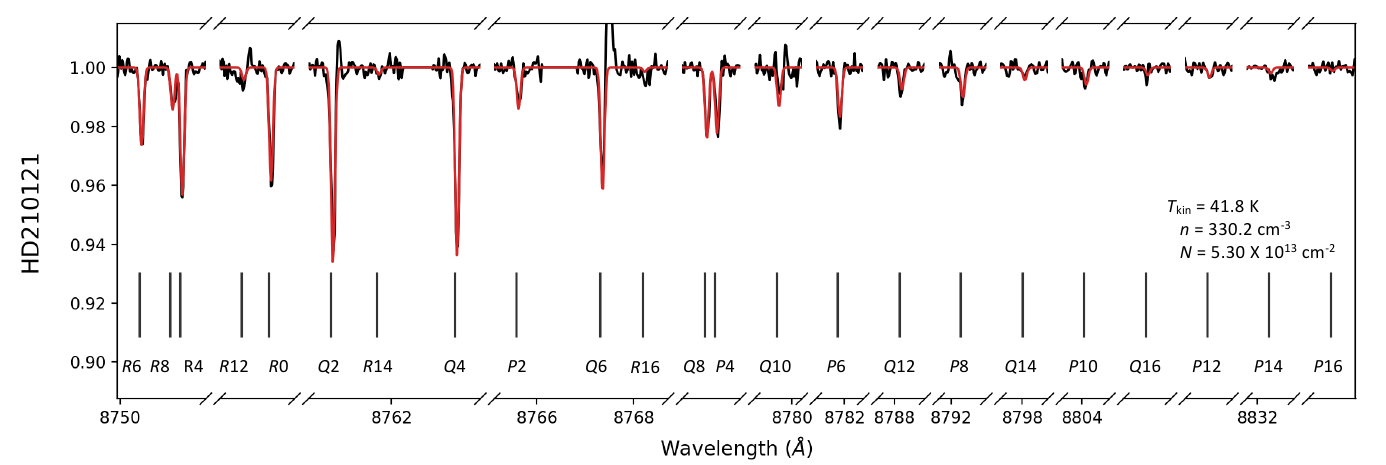}
\caption{Same as Fig.~\ref{Fig:fits_c2_app} but for HD~210121.}    
\end{figure*}

\section{Fitted C$_3$ models}\label{app_C3}
In this appendix we present plots of the best-fit model of the C$_3 \: \tilde{A} - \tilde{X}$ 000-000 electronic origin band compared to the normalised flux of the 27 sightlines where C$_3$ is detected or tentatively detected. Synthetic spectra for individual velocity components are plotted in different colours for the sightlines of HD~61827, HD~149404, HD~161056, and HD~186841. The excitation temperature of the lower $J \leq 14$ levels, $T_\textrm{low}$, and the column density of C$_3$ $N$, are listed in the legend of each panel. We refer the reader to Sect.~\ref{sec:C3} for more details on the modelling process.

\begin{figure} [ht]
    \centering
    \includegraphics[width=\columnwidth]{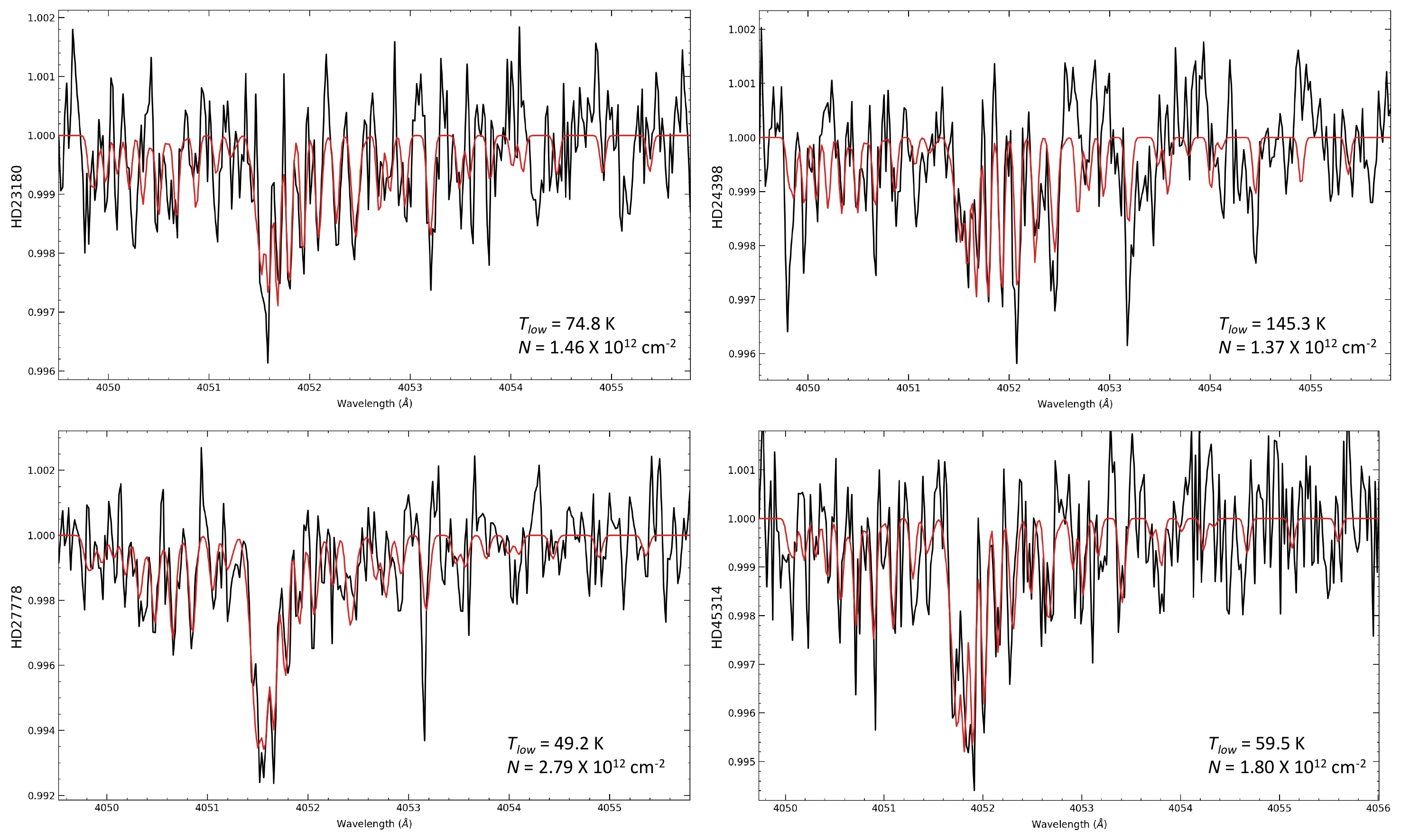}
    \caption{\label{Fig:fits_c3_app}Normalised flux (black) and fitted model (red) of the C$_3 \: \tilde{A} - \tilde{X}$ 000-000 electronic origin band for HD~23180, HD~24398, HD~27778, and HD~45314. The model parameters $T_\textrm{low}$ and $N$ are listed in the legend of each panel.  }
\end{figure}


\begin{figure*} [ht]
    \centering
    \resizebox{\columnwidth}{!}{\includegraphics[width=\columnwidth]{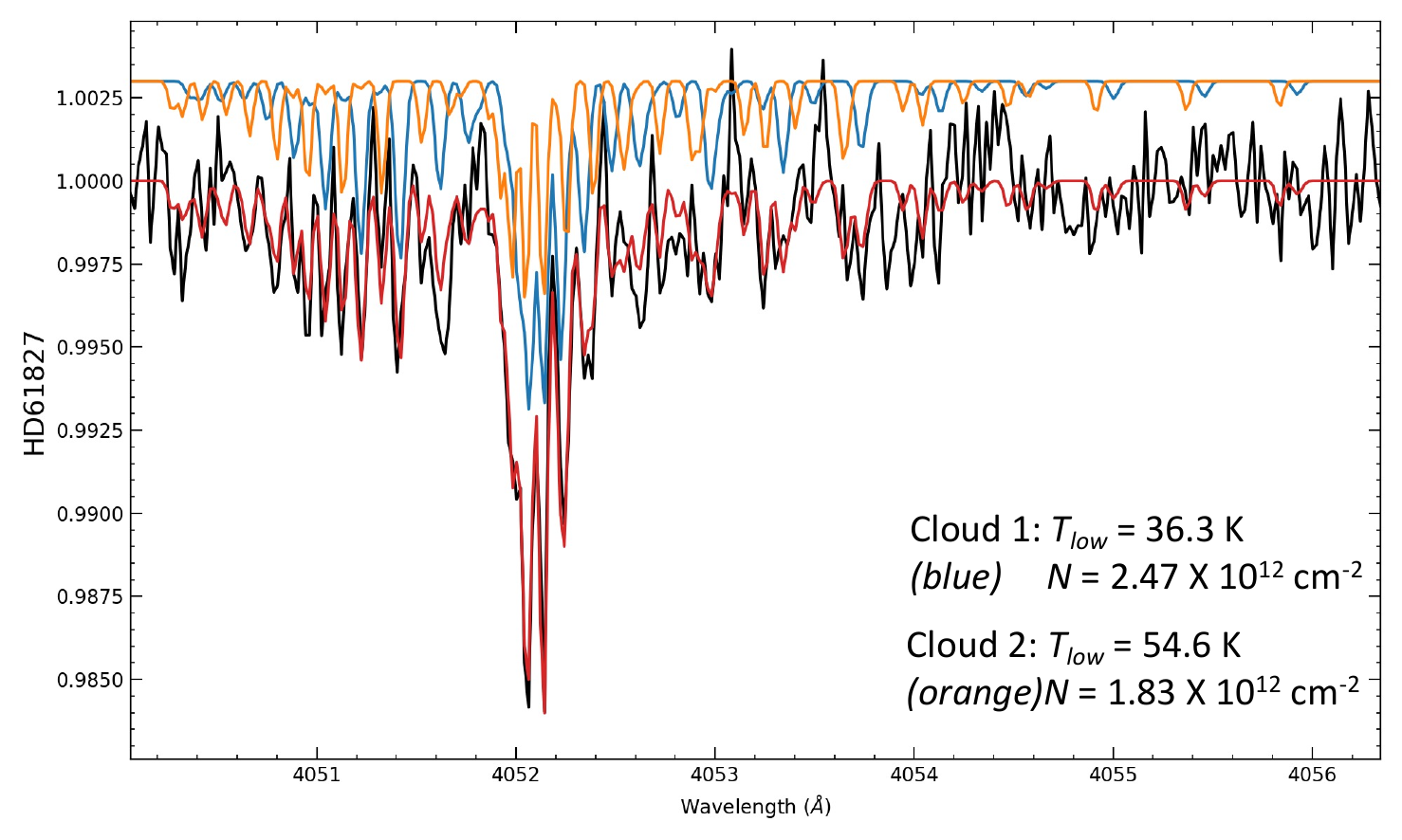}
    \includegraphics[width=\columnwidth]{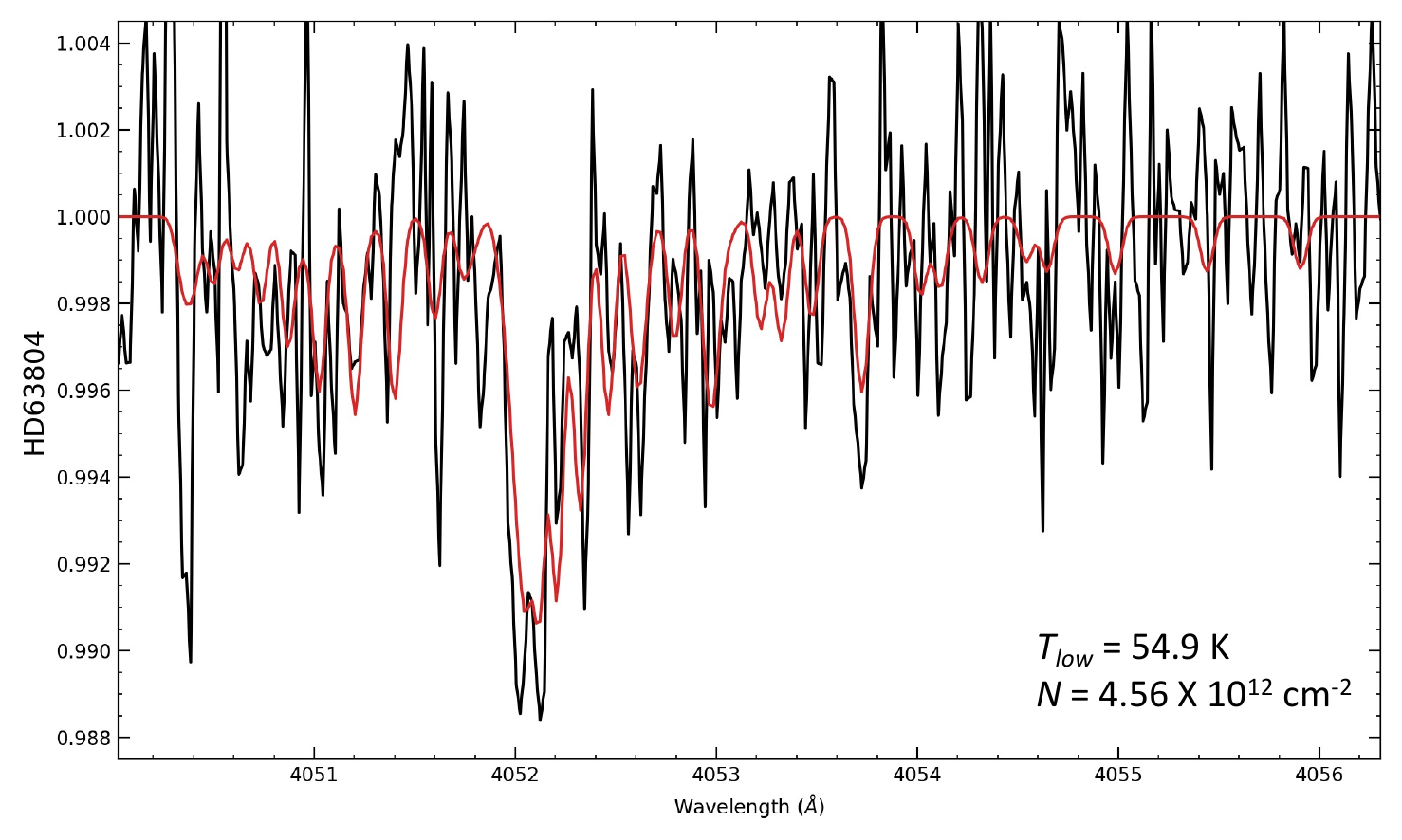}}
    \resizebox{\columnwidth}{!}{\includegraphics[width=\columnwidth]{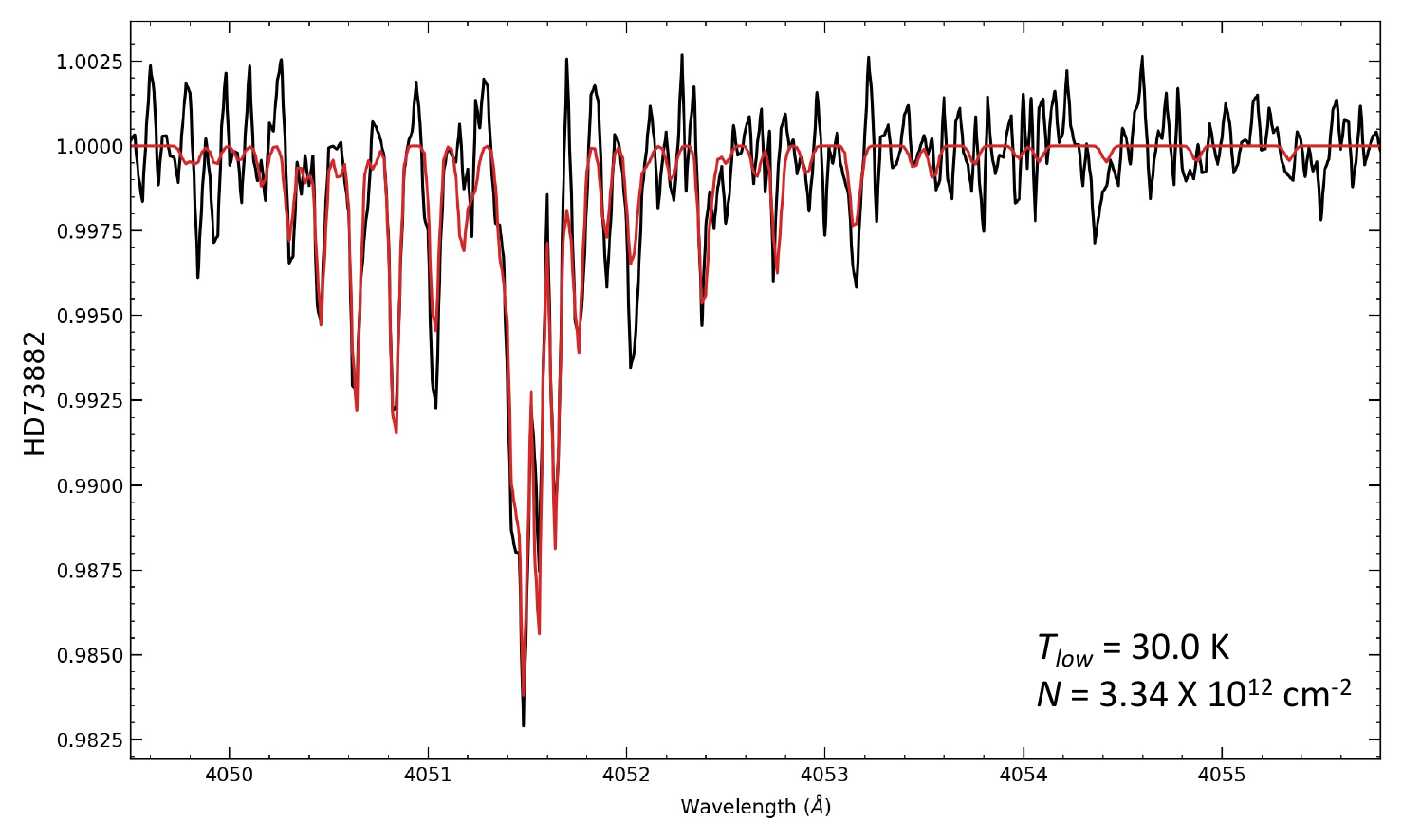}
    \includegraphics[width=\columnwidth]{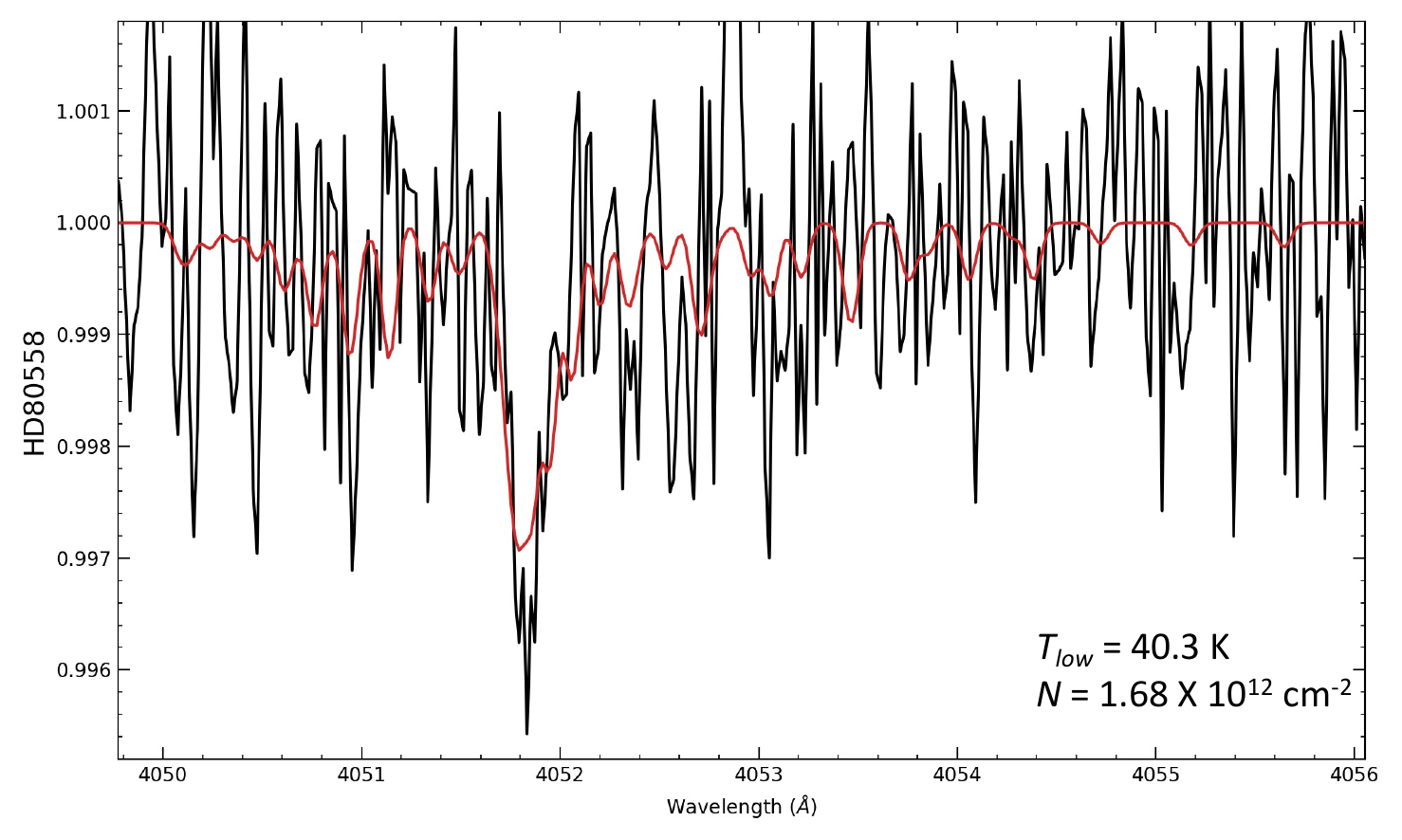}}
    \resizebox{\columnwidth}{!}{\includegraphics[width=\columnwidth]{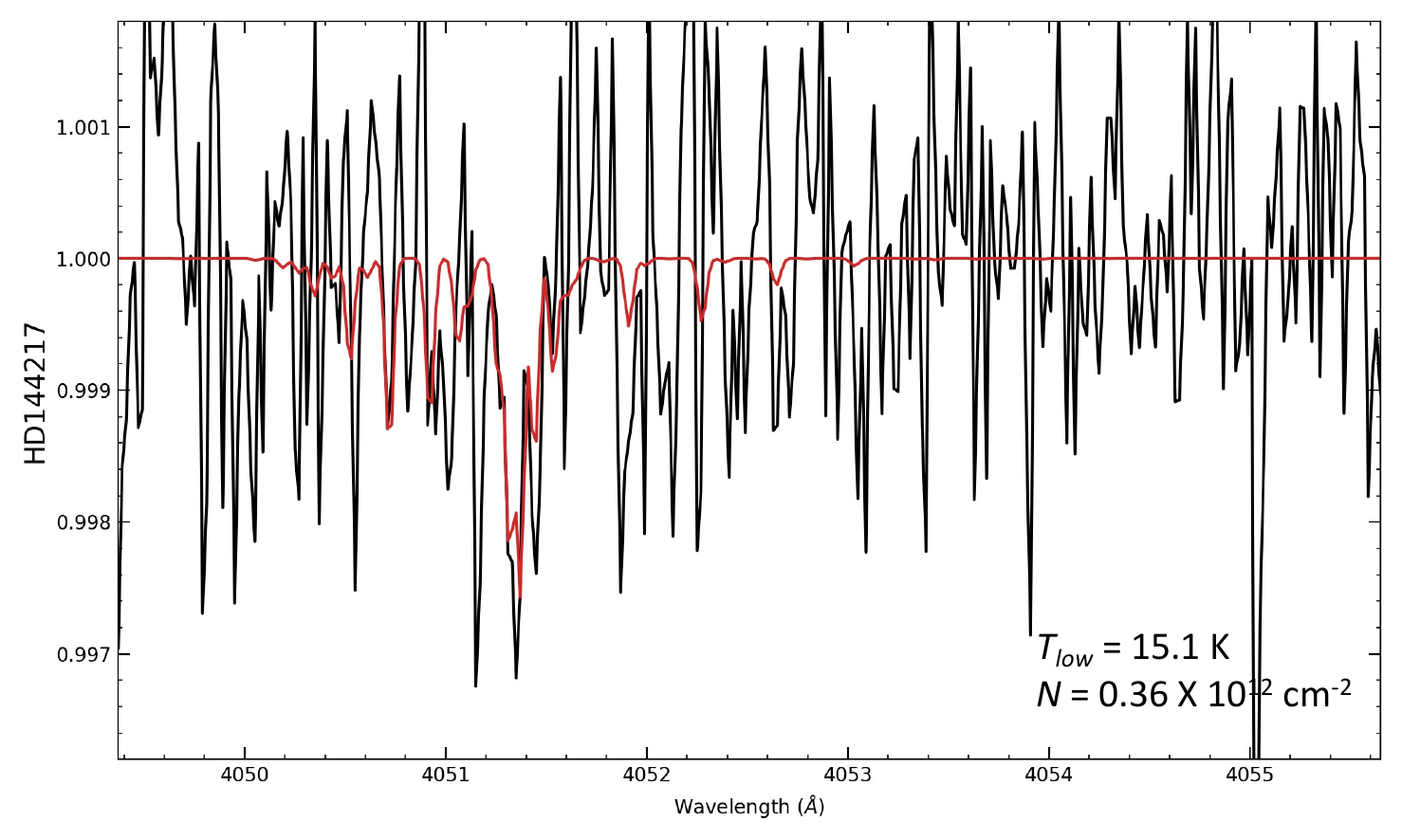}
    \includegraphics[width=\columnwidth]{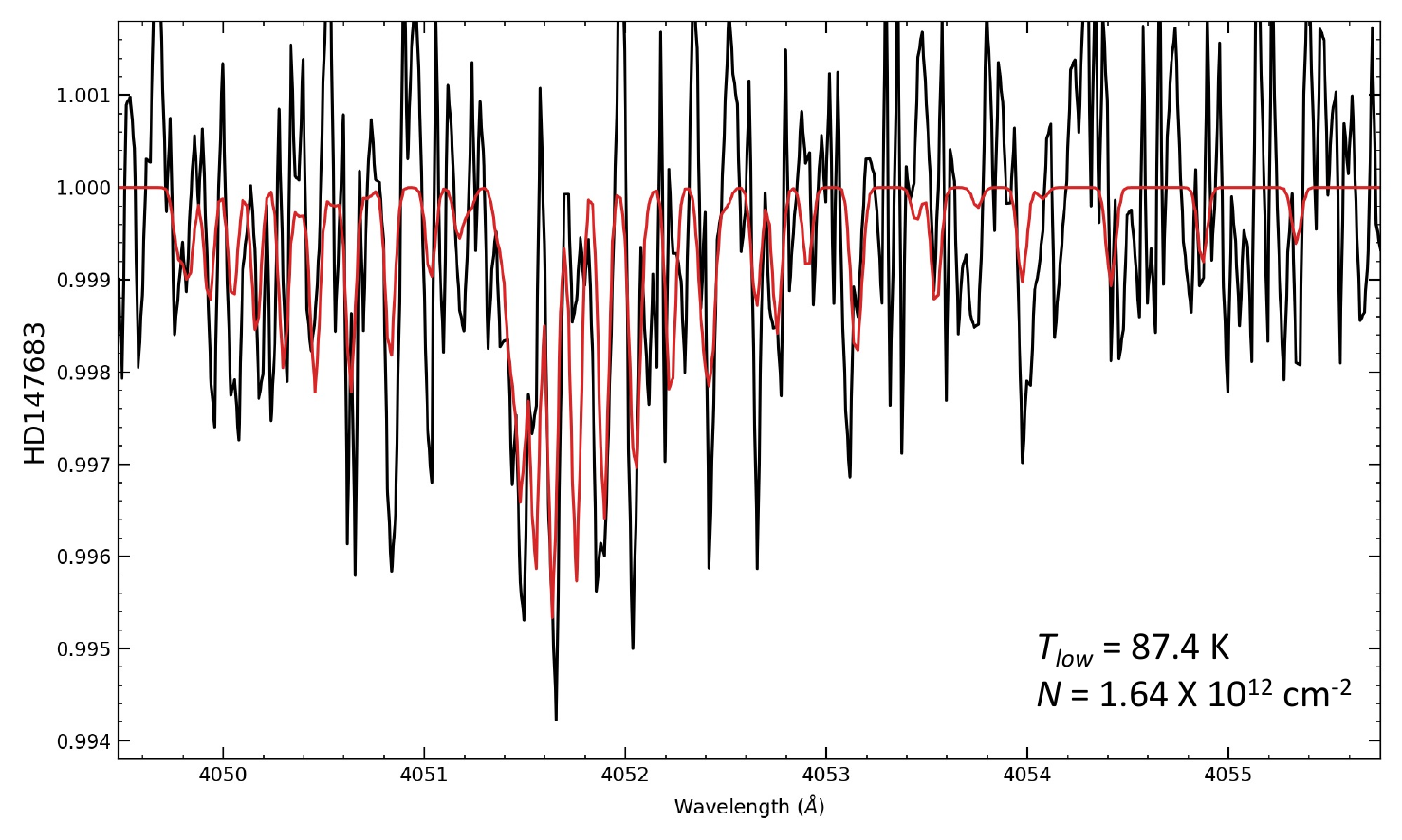}}
    \resizebox{\columnwidth}{!}{\includegraphics[width=\columnwidth]{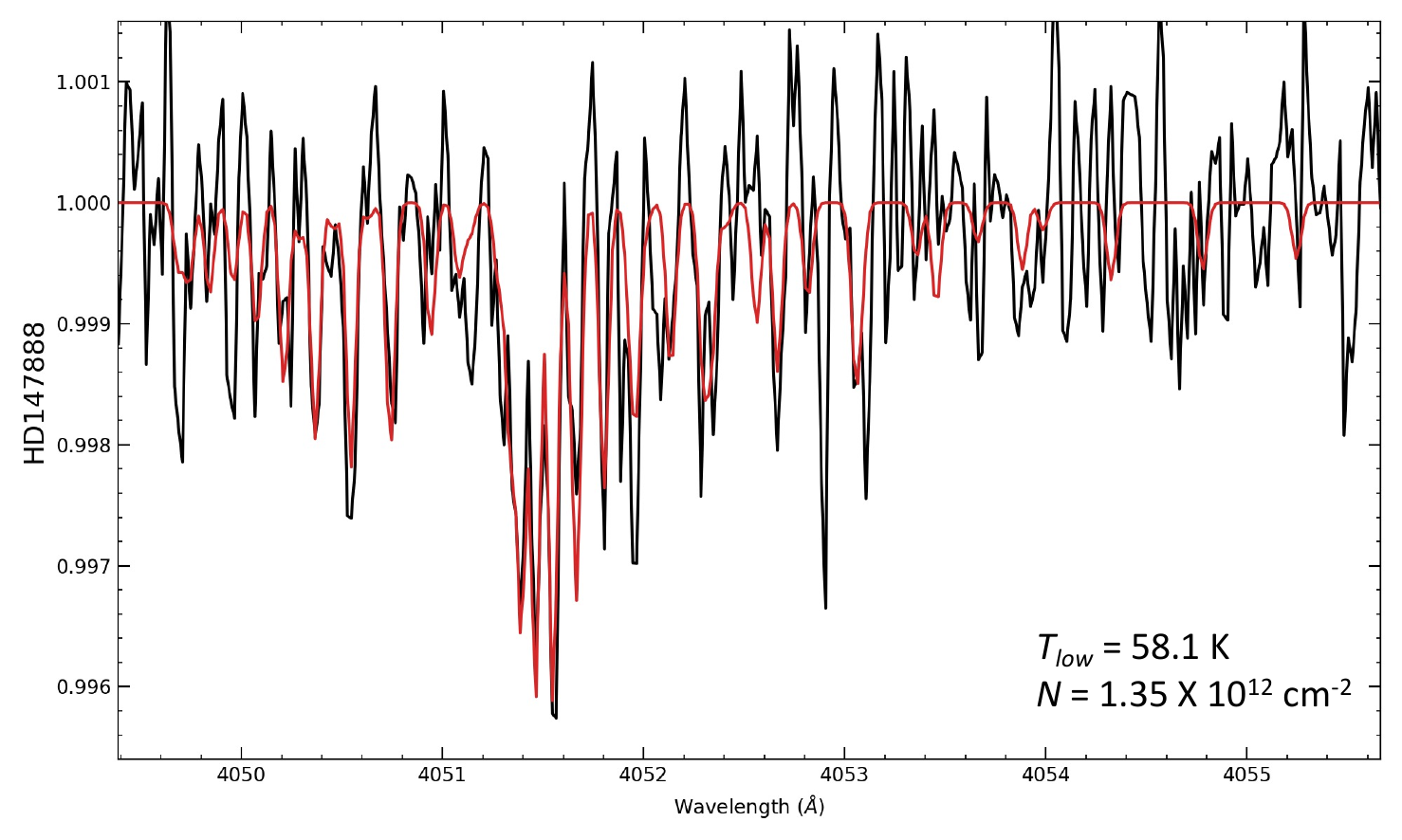}
    \includegraphics[width=\columnwidth]{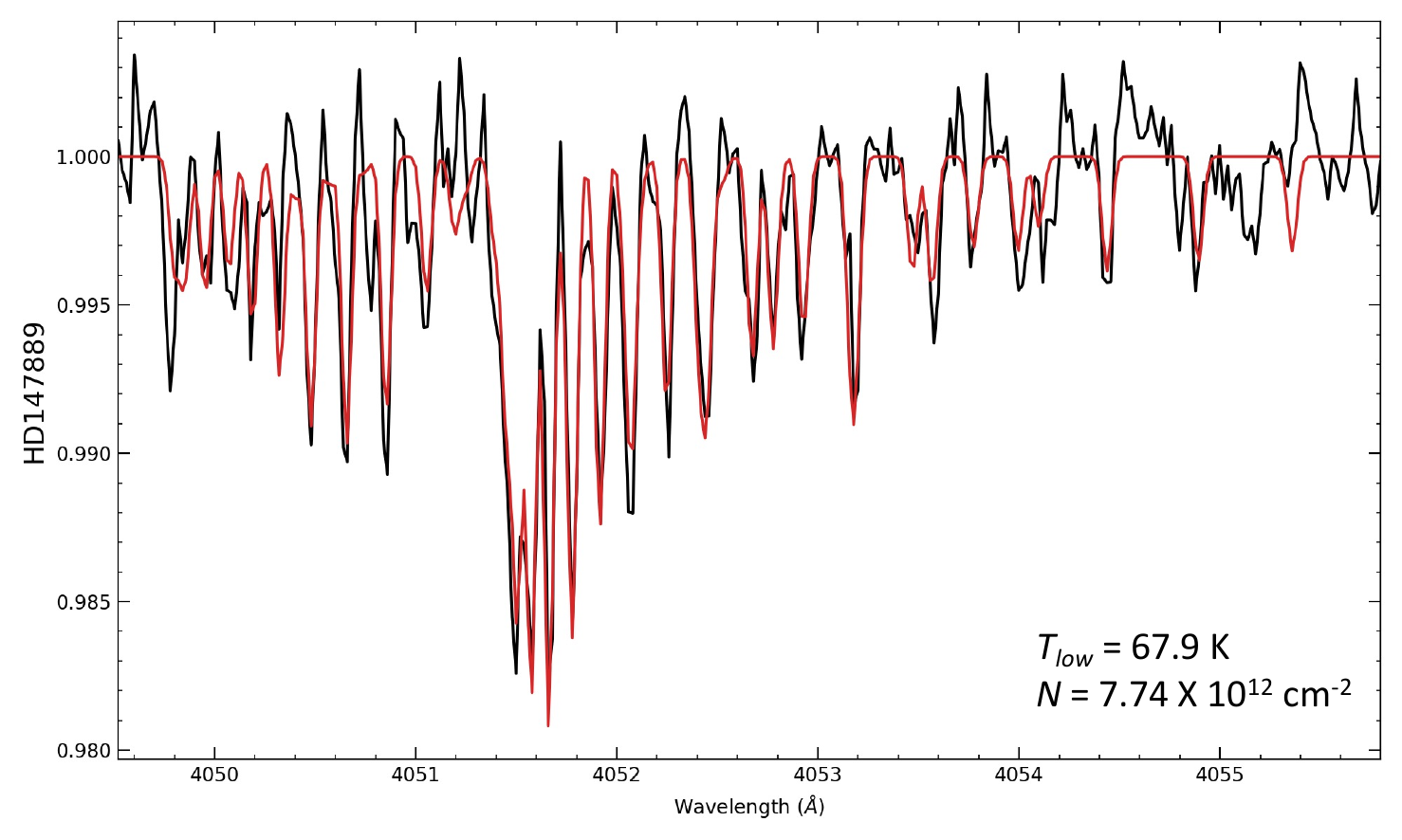}}
\caption{Same as Fig.~\ref{Fig:fits_c3_app} but for HD~61827, HD~63804, HD~73882, HD~80558, HD~144217, HD~147683, HD~147888, and HD~147889. For HD~61827, synthetic spectra for individual velocity components are shown in orange and blue.}
\end{figure*}

\begin{figure*} [ht]
    \centering
    \resizebox{\columnwidth}{!}{\includegraphics[width=\columnwidth]{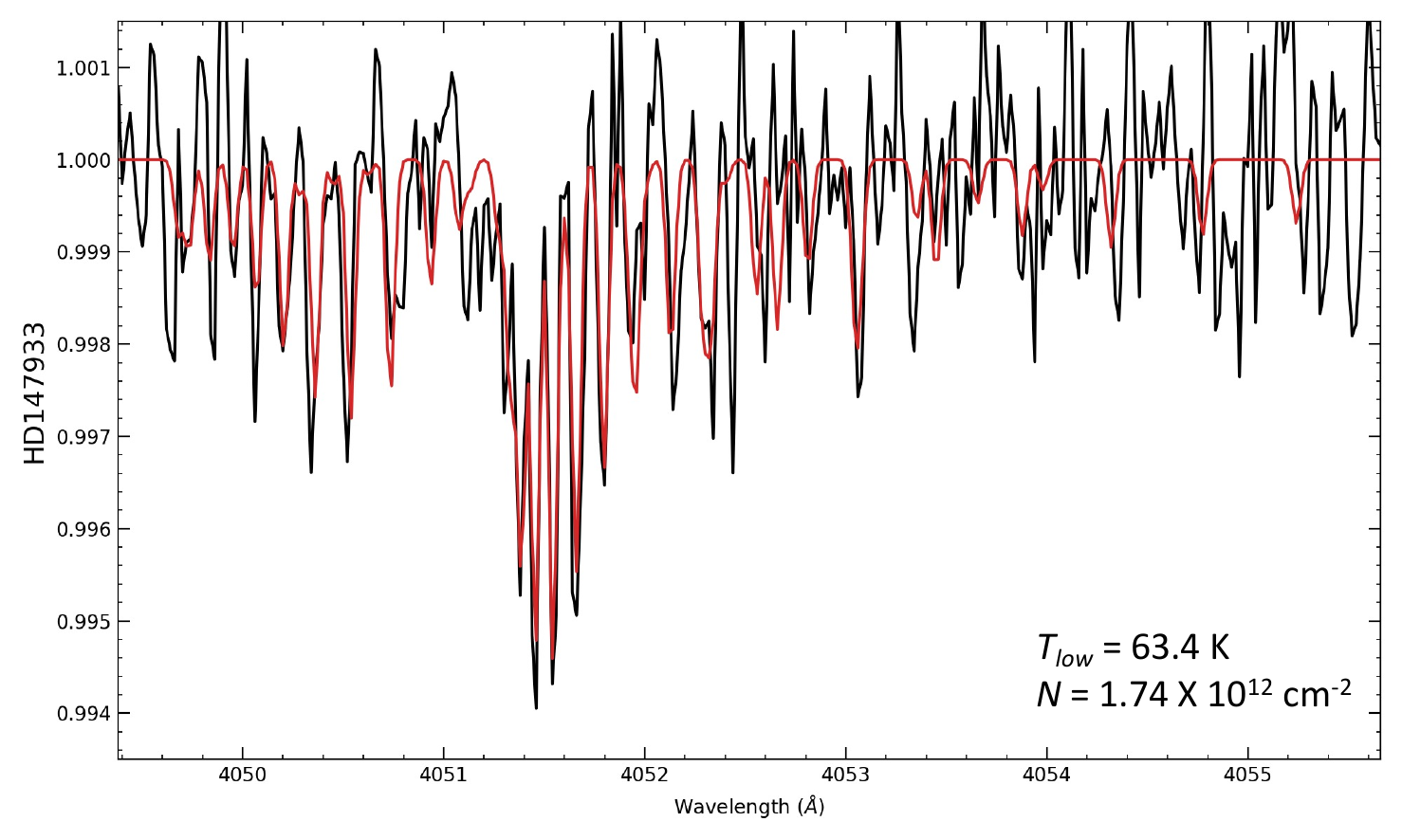}
    \includegraphics[width=\columnwidth]{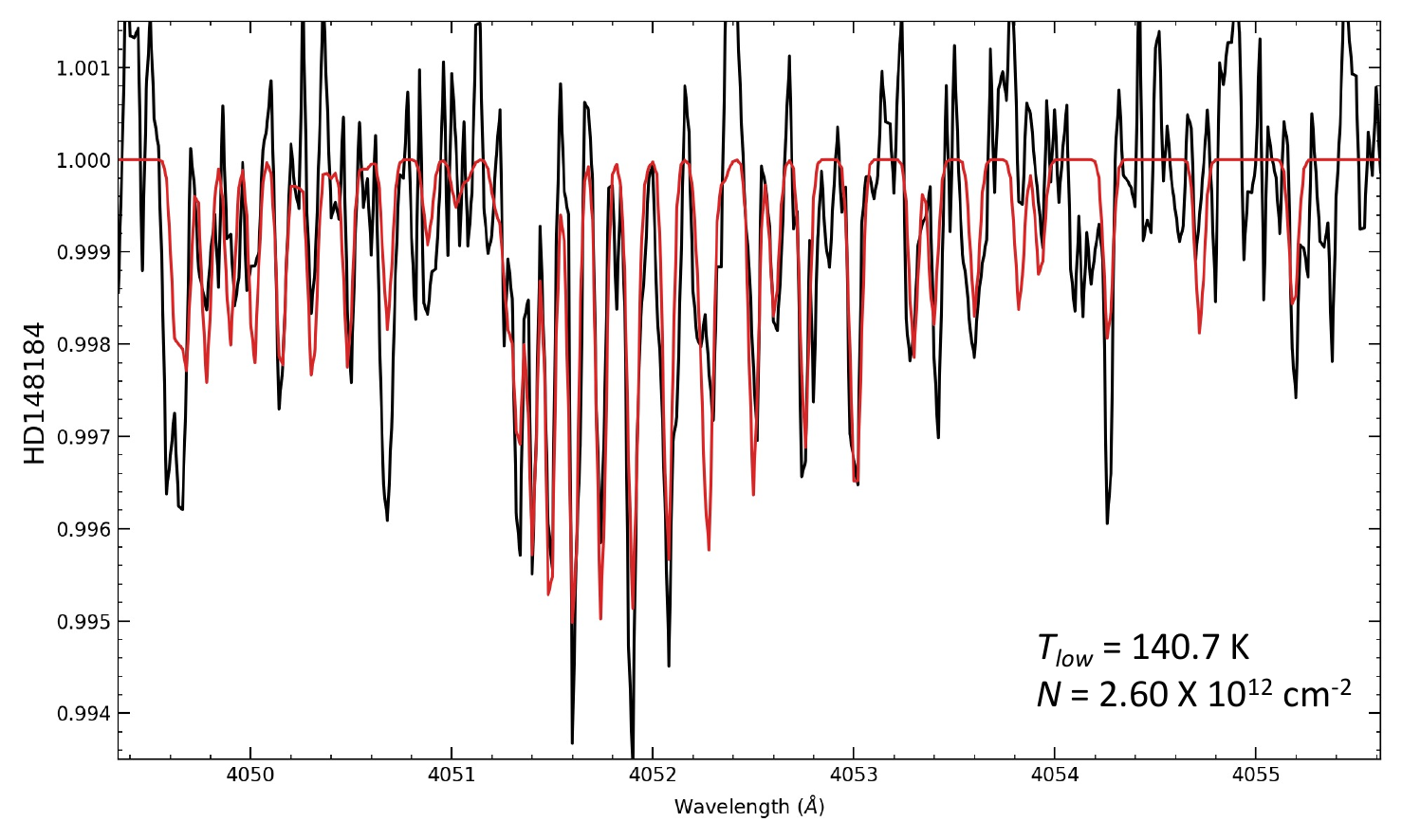}}
    \resizebox{\columnwidth}{!}{\includegraphics[width=\columnwidth]{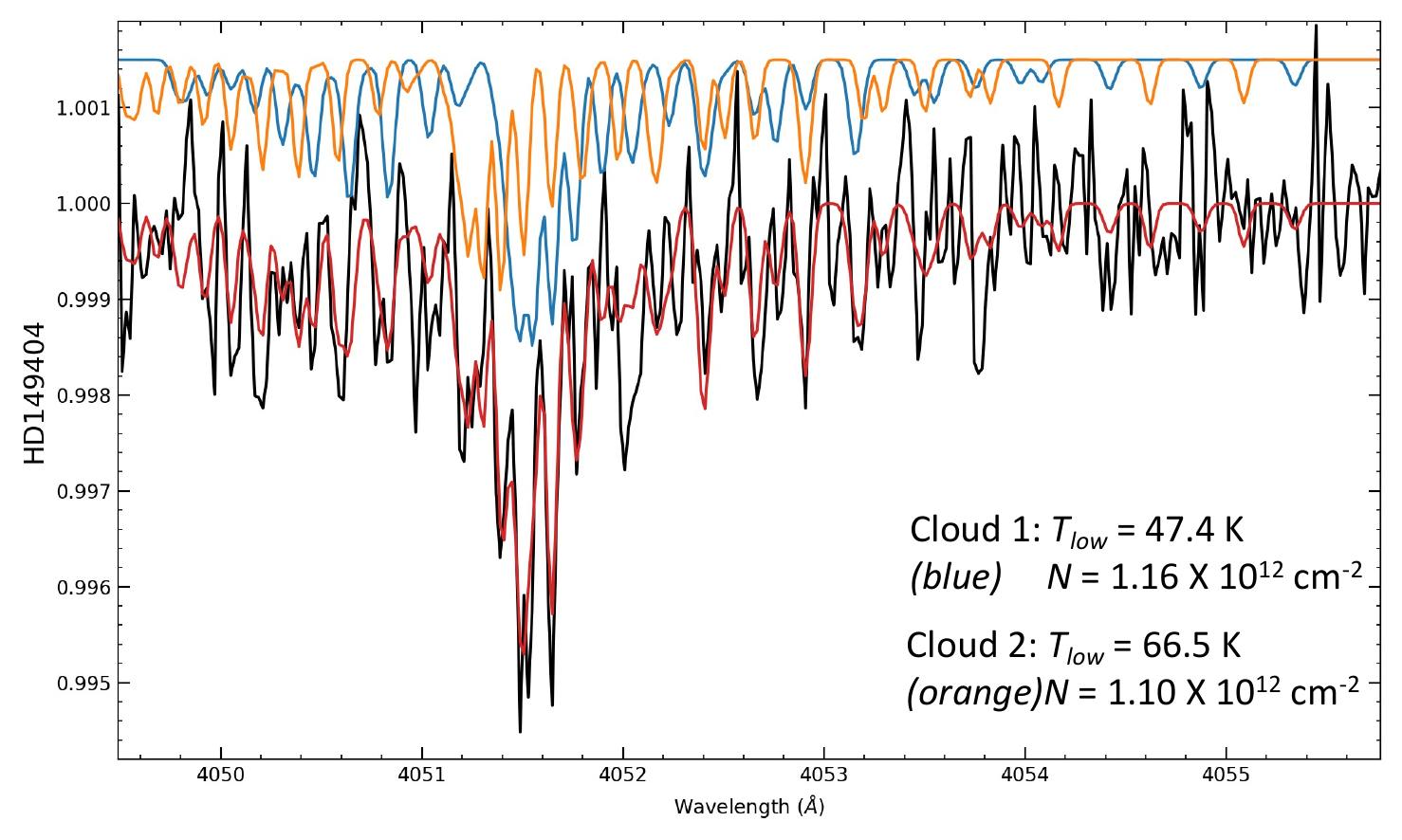}
    \includegraphics[width=\columnwidth]{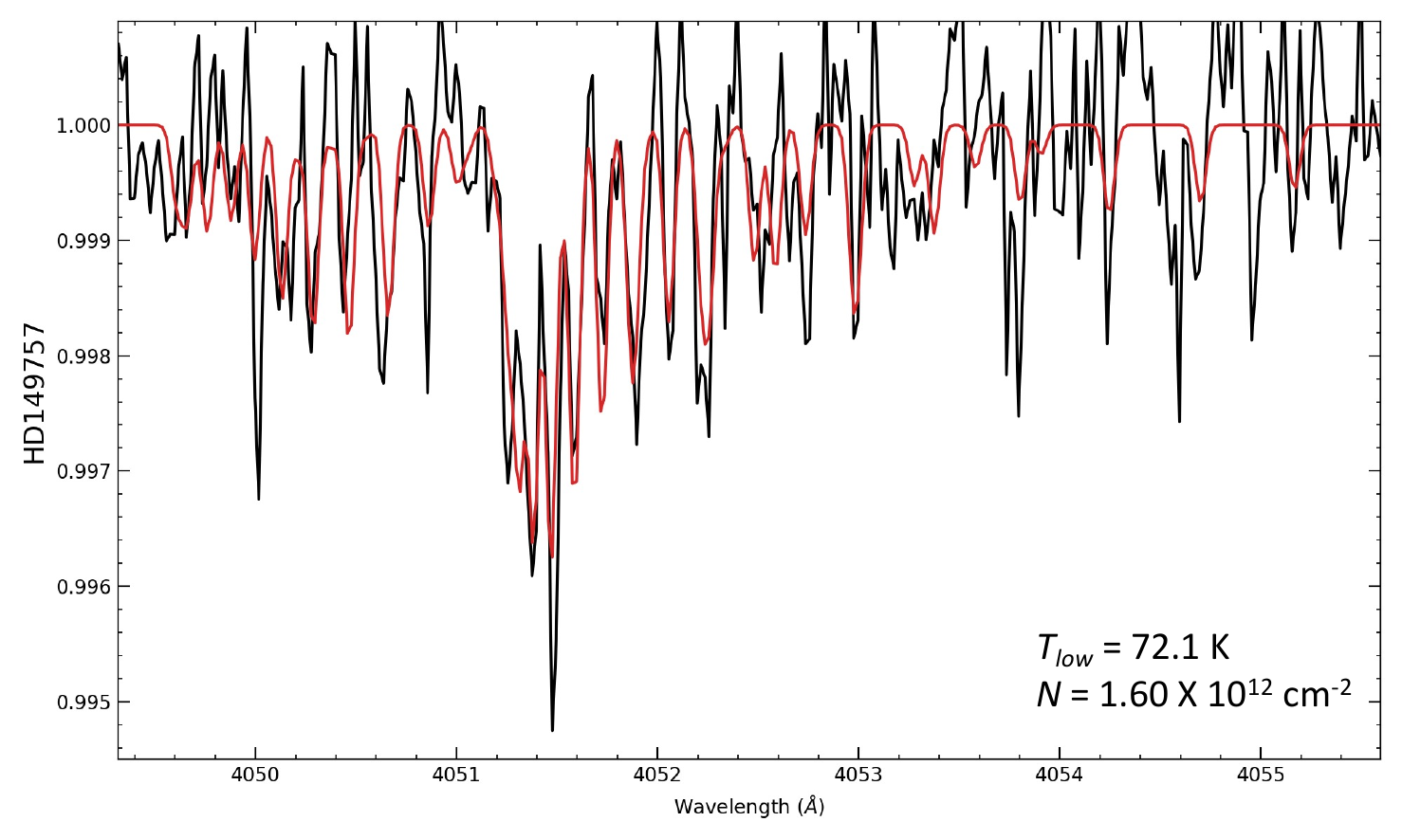}}
    \resizebox{\columnwidth}{!}{\includegraphics[width=\columnwidth]{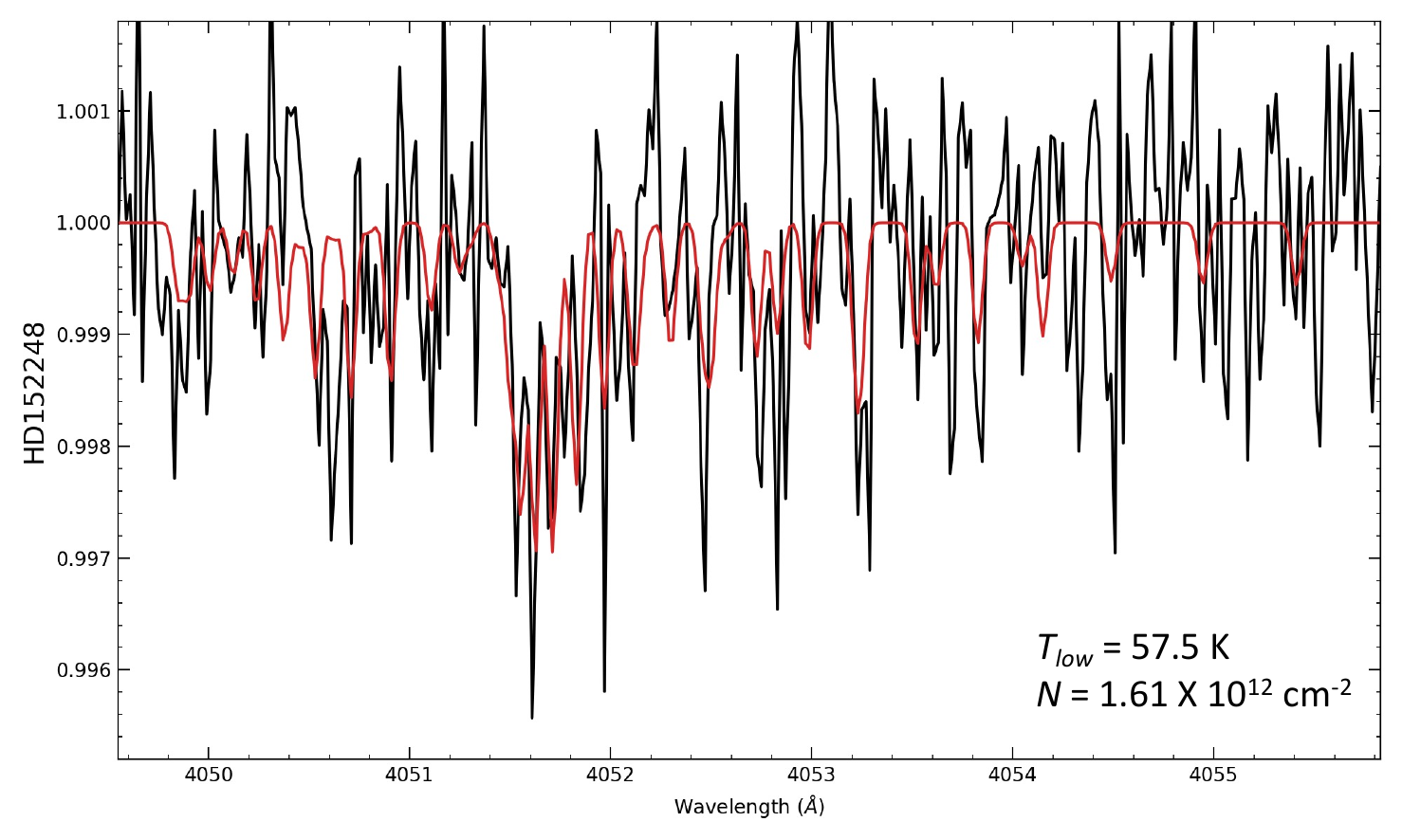}
    \includegraphics[width=\columnwidth]{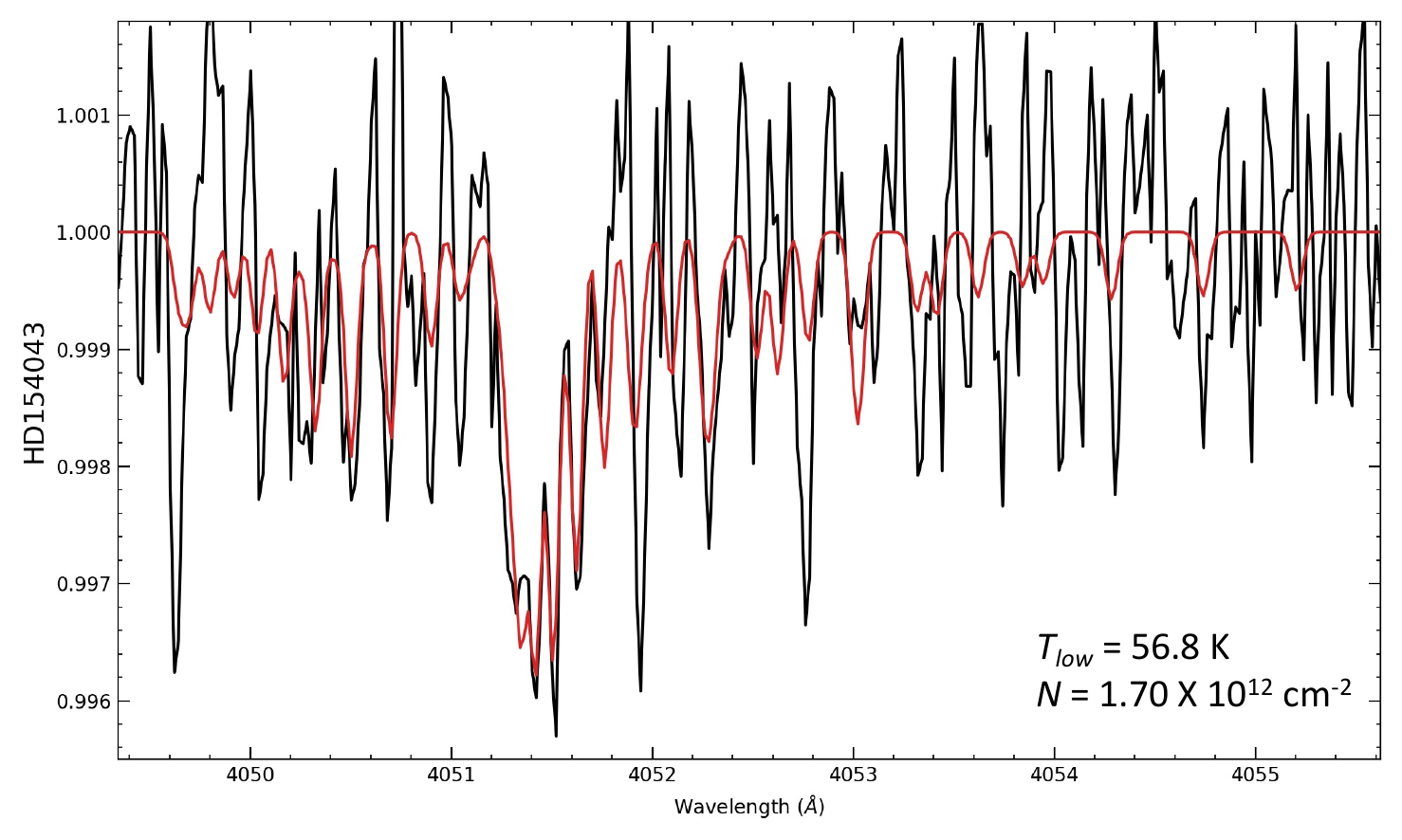}}
    \resizebox{\columnwidth}{!}{\includegraphics[width=\columnwidth]{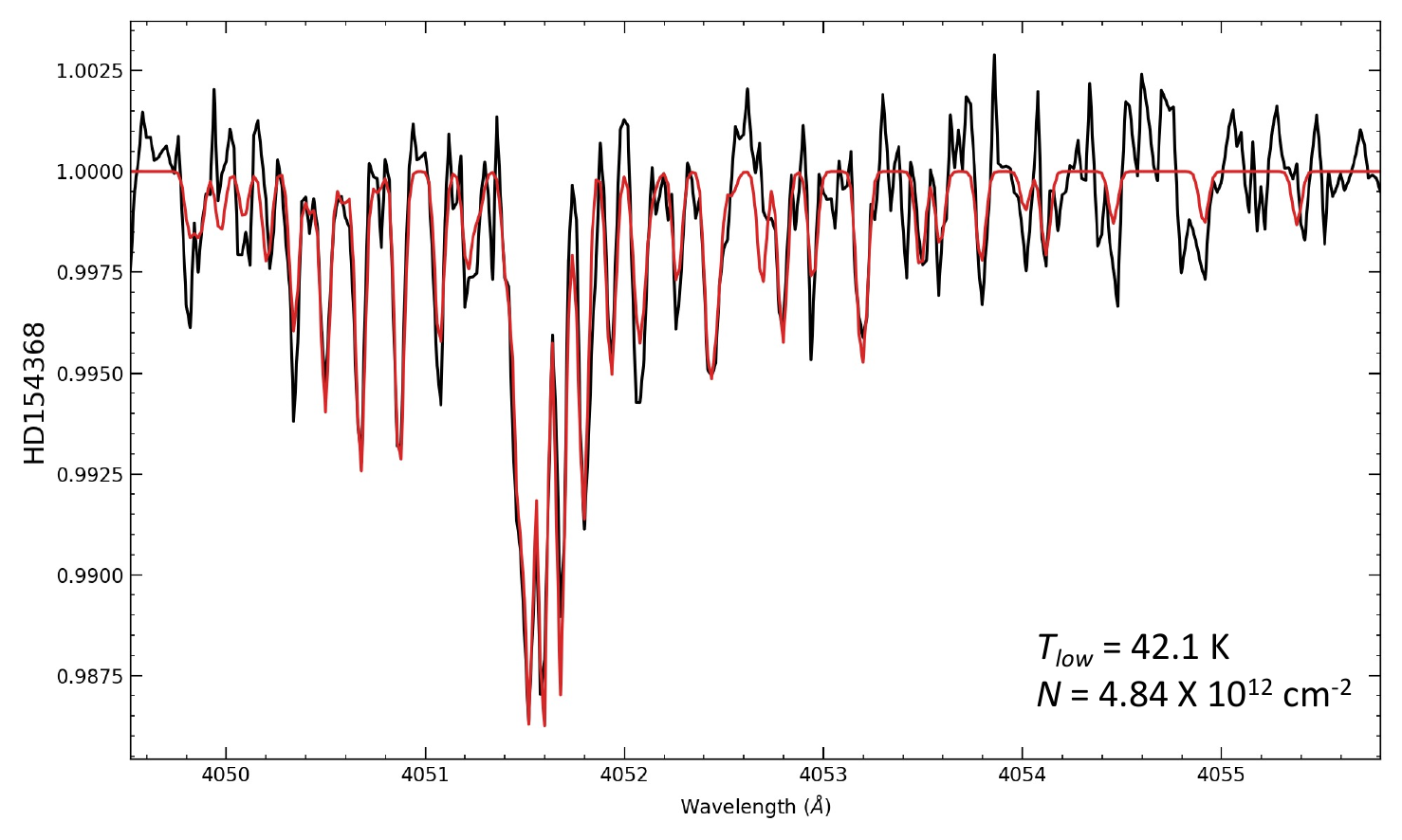}
    \includegraphics[width=\columnwidth]{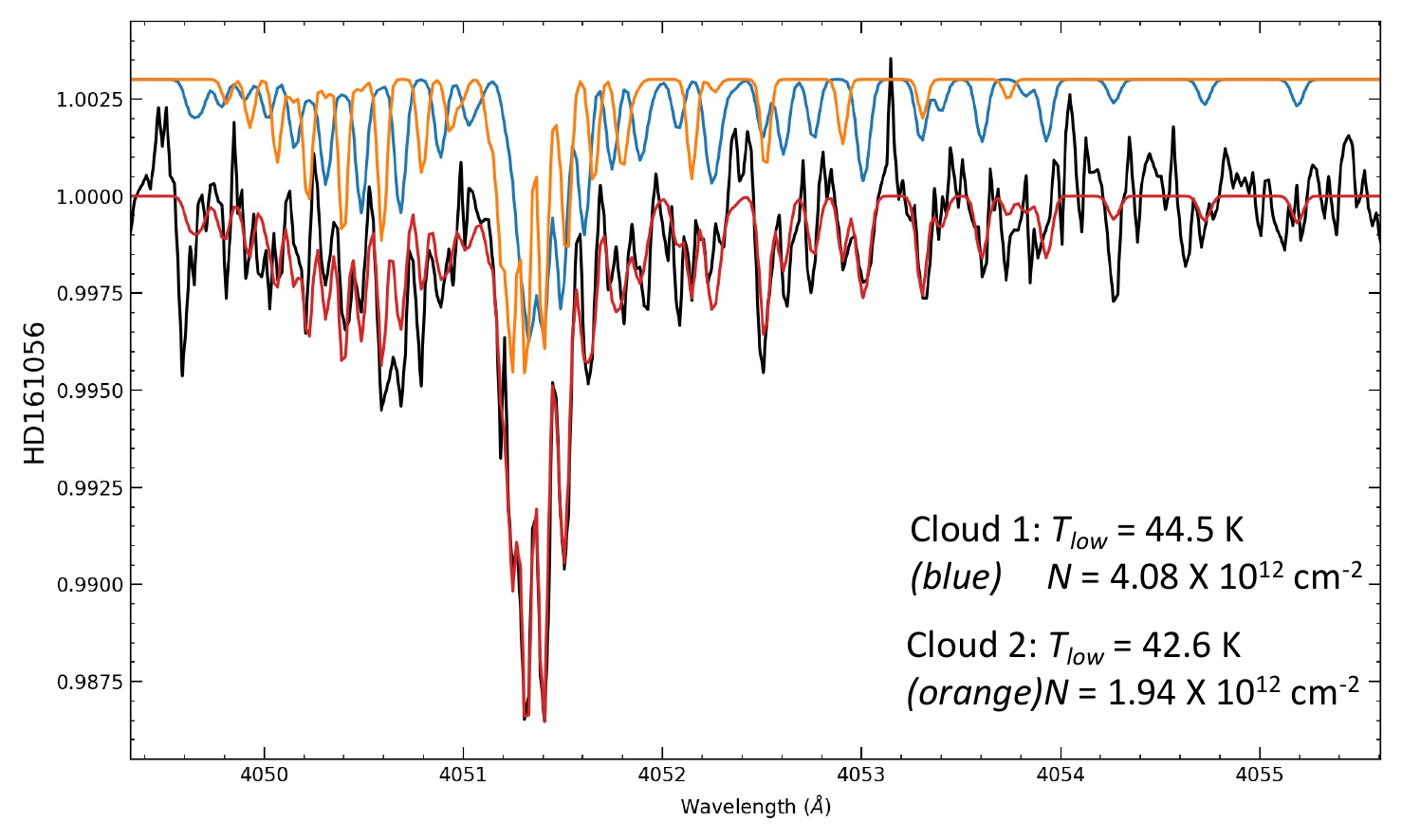}}
\caption{Same as Fig.~\ref{Fig:fits_c3_app} but for HD~147933, HD~148184, HD~149404, HD~149757, HD~152248, HD~154043, HD~154368, and HD~161056. For HD~149404 and HD~161056, synthetic spectra for individual velocity components are shown in orange and blue.}
\end{figure*}

\begin{figure*}
    \centering
    \resizebox{\columnwidth}{!}{\includegraphics[width=\columnwidth]{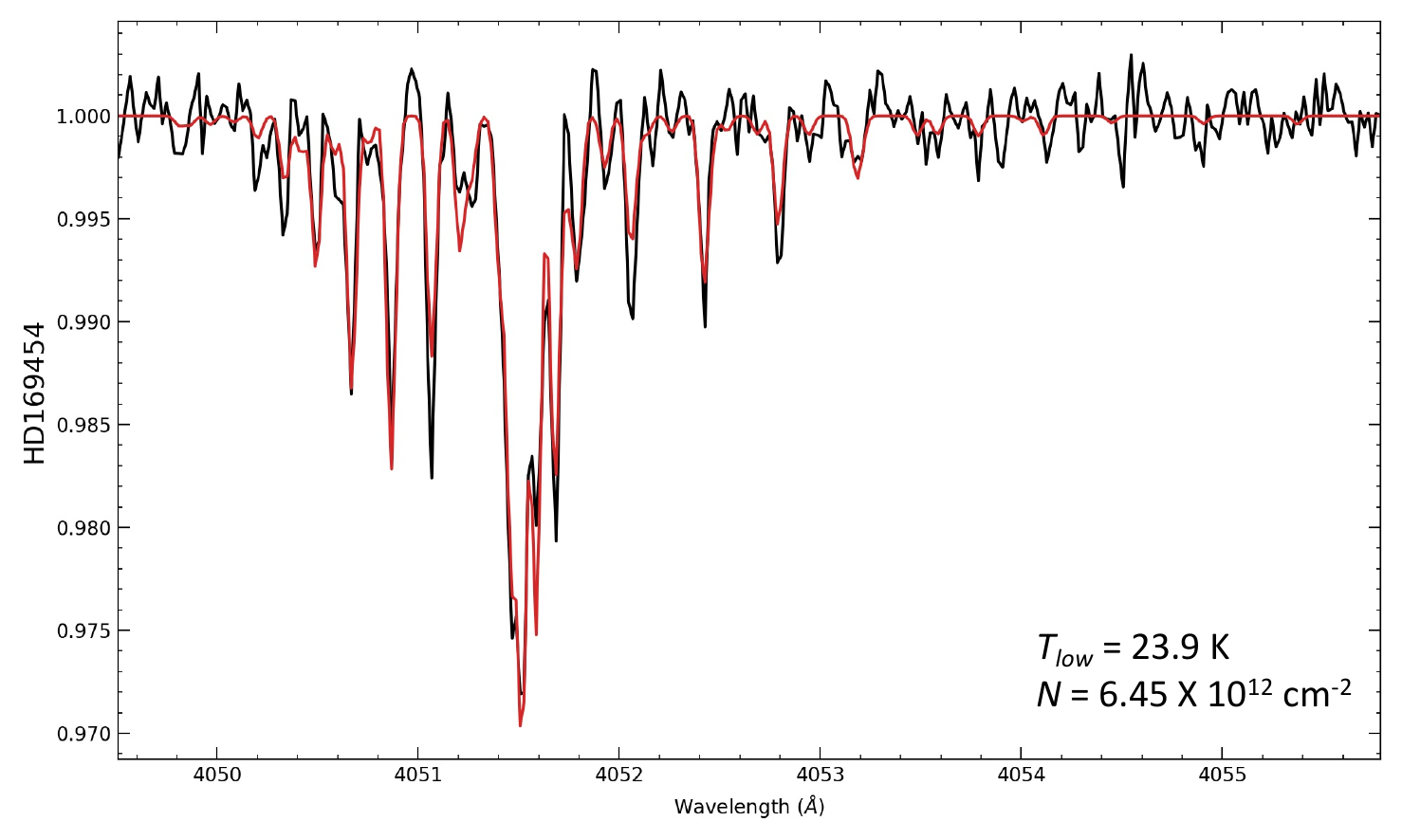}
    \includegraphics[width=\columnwidth]{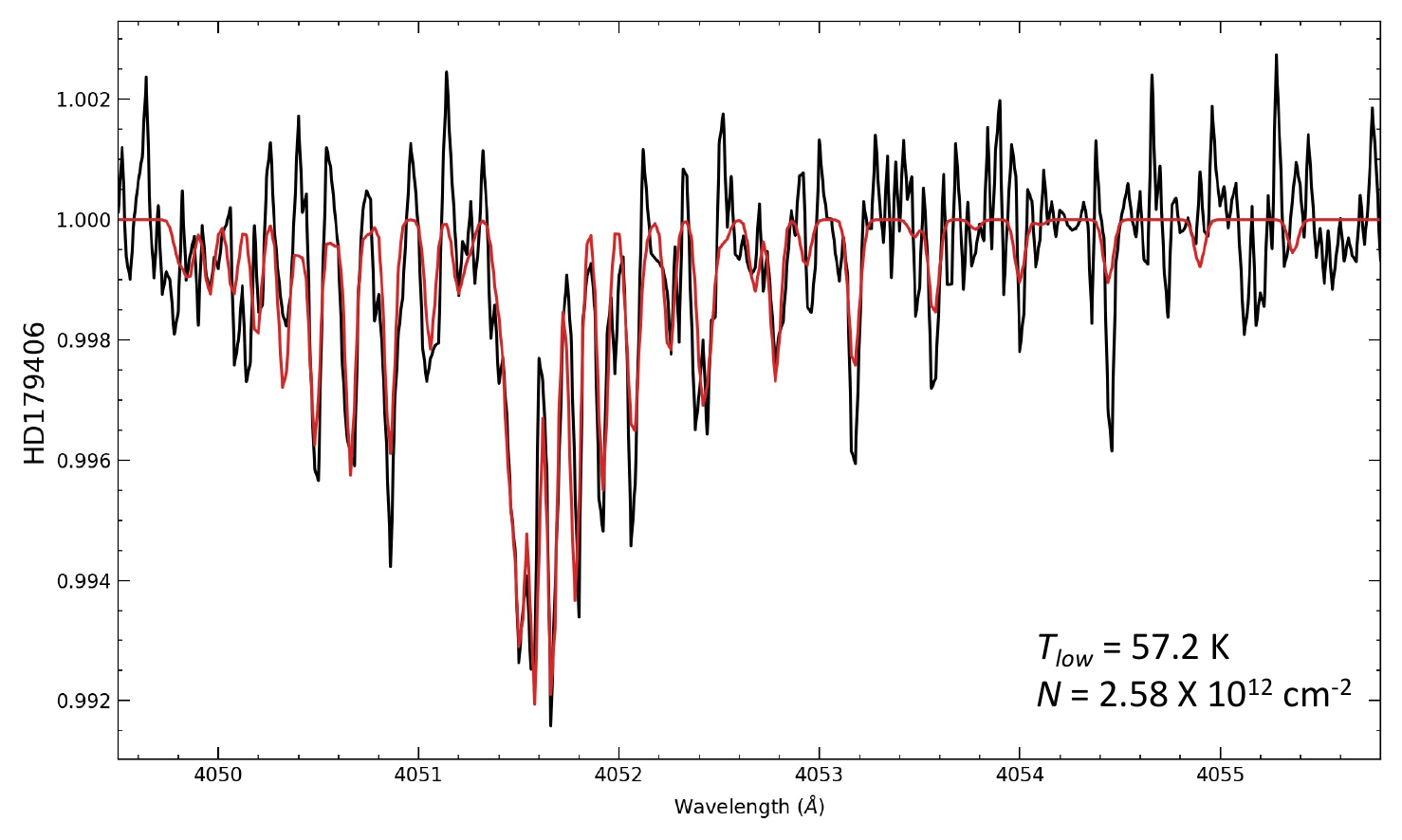}}
    \resizebox{\columnwidth}{!}{\includegraphics[width=\columnwidth]{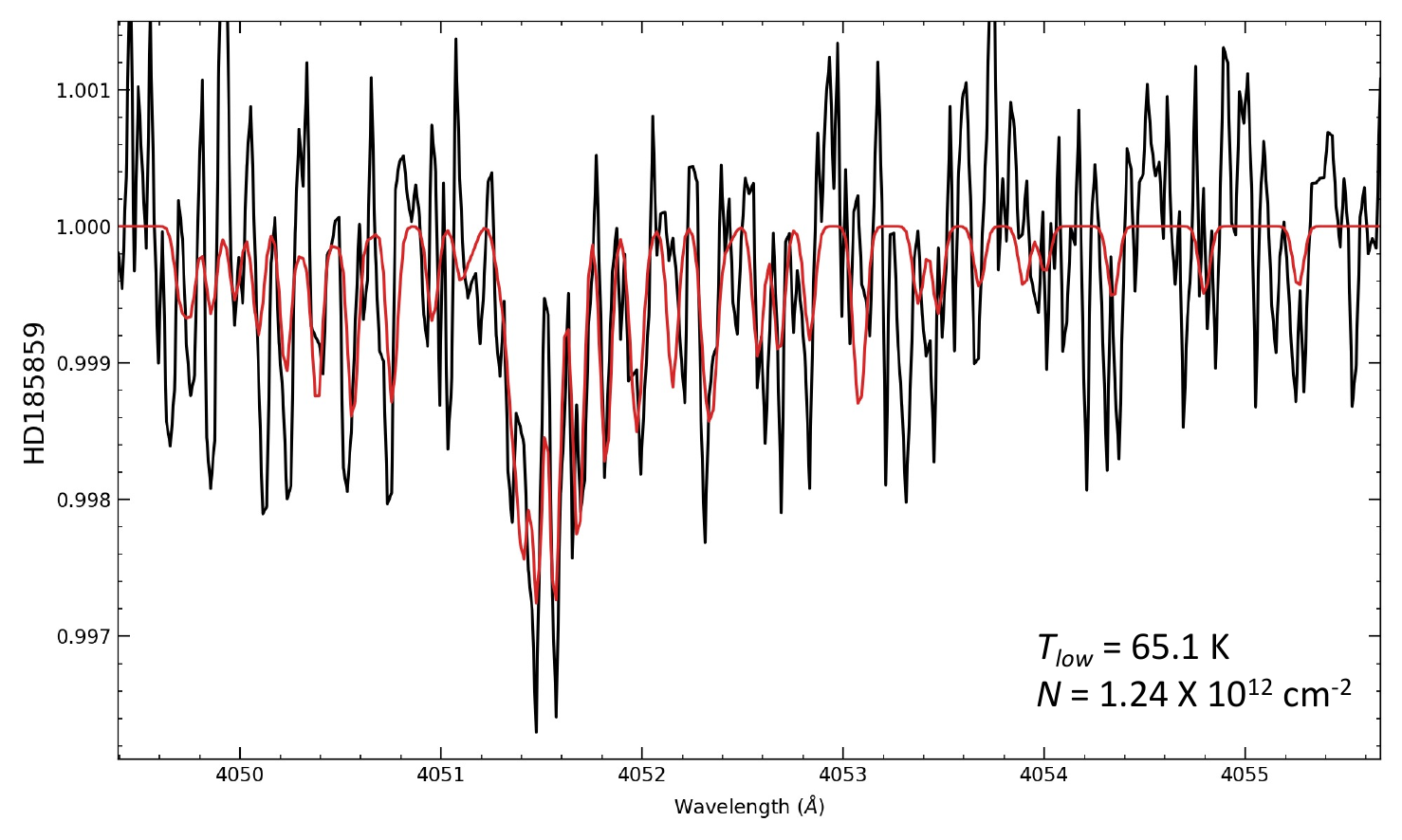}
    \includegraphics[width=\columnwidth]{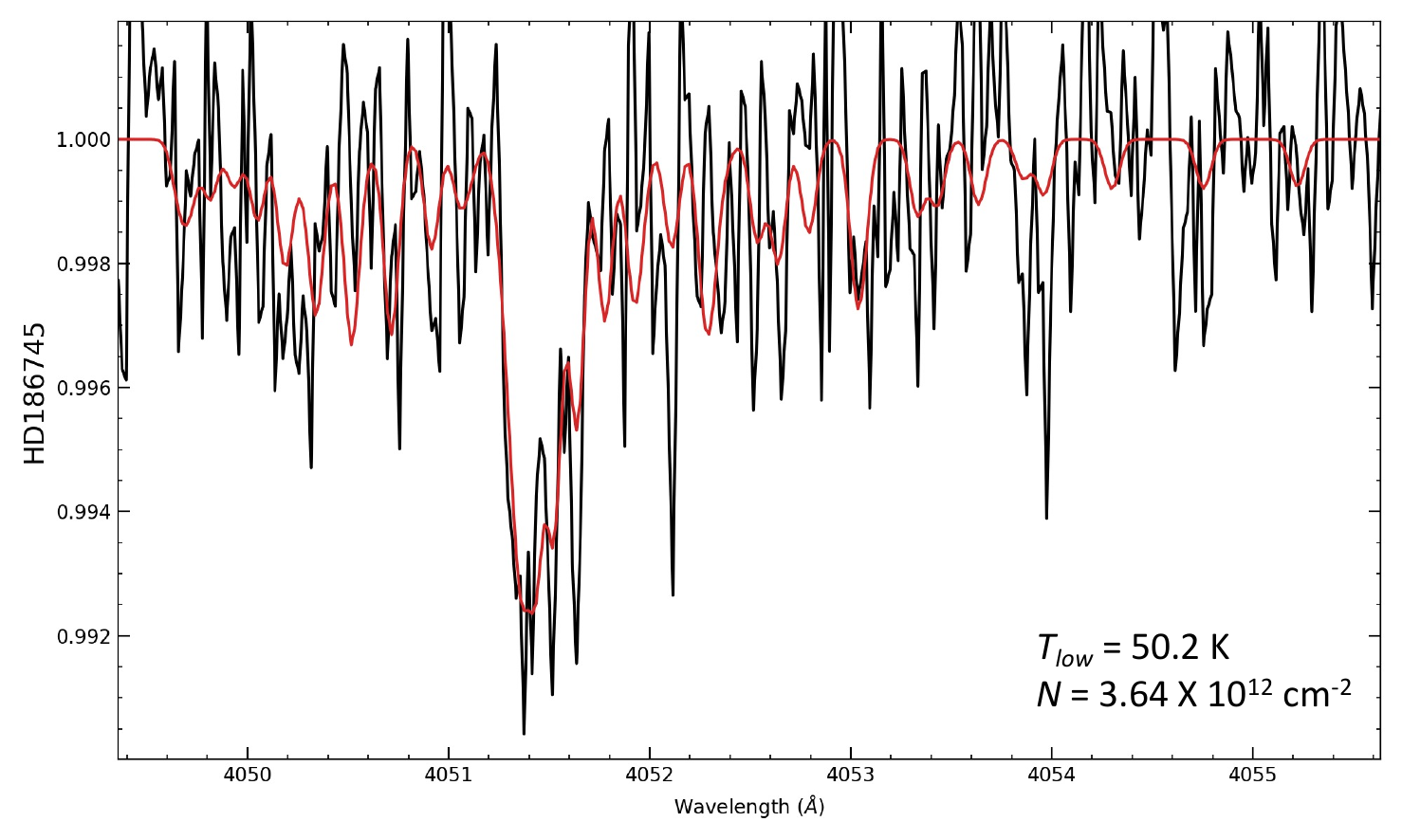}}
    \resizebox{\columnwidth}{!}{\includegraphics[width=\columnwidth]{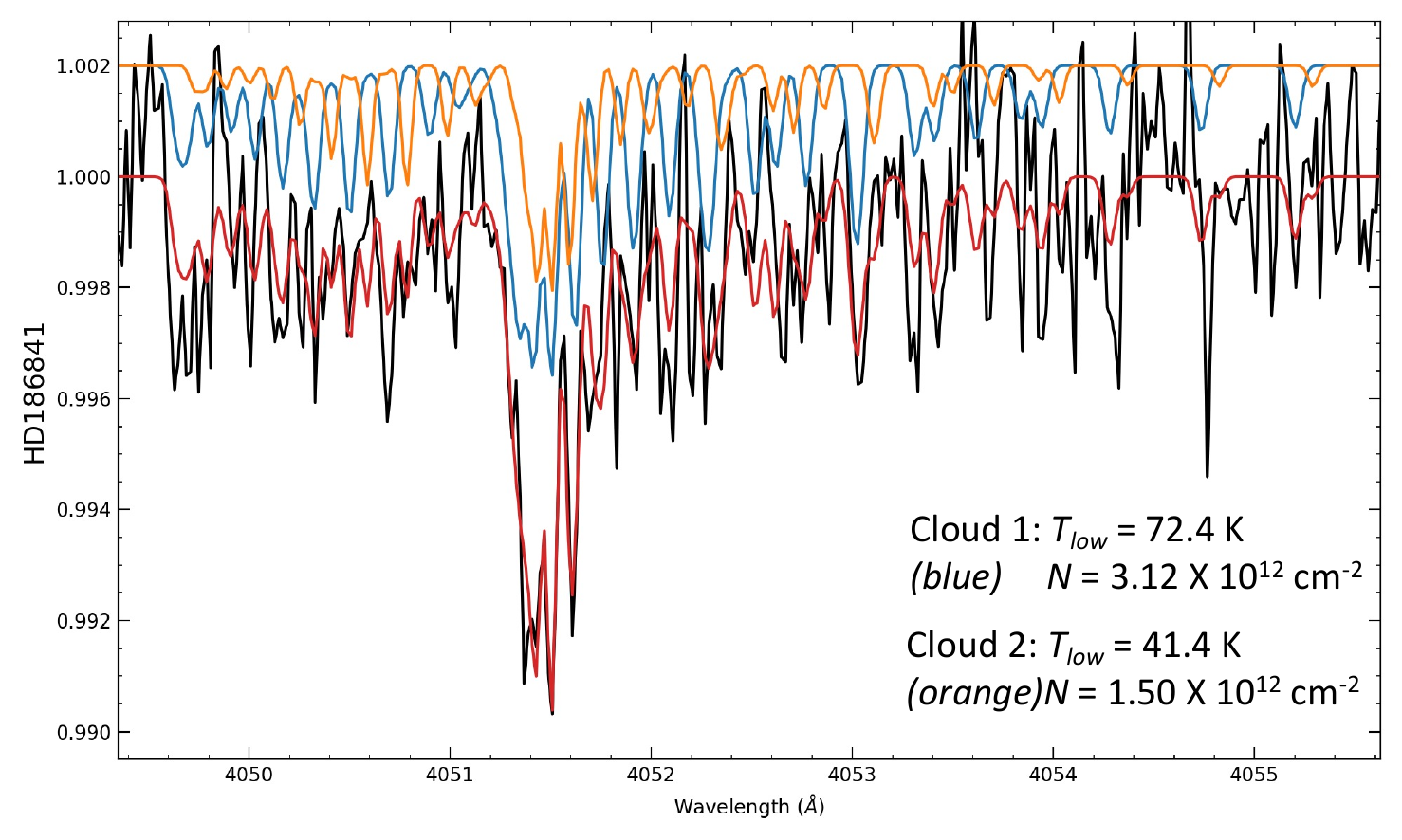}
    \includegraphics[width=\columnwidth]{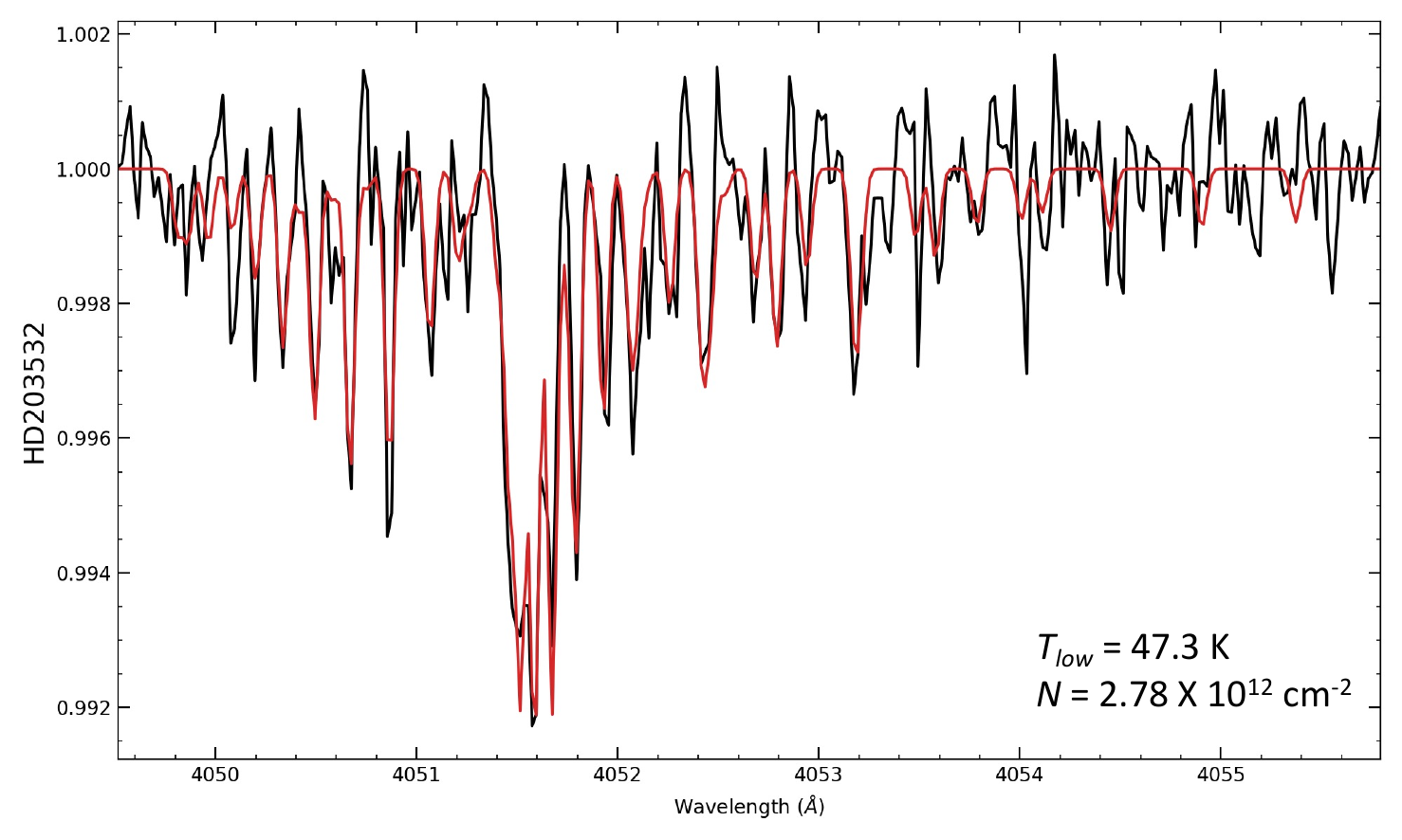}}
    \includegraphics[width=.5\columnwidth]{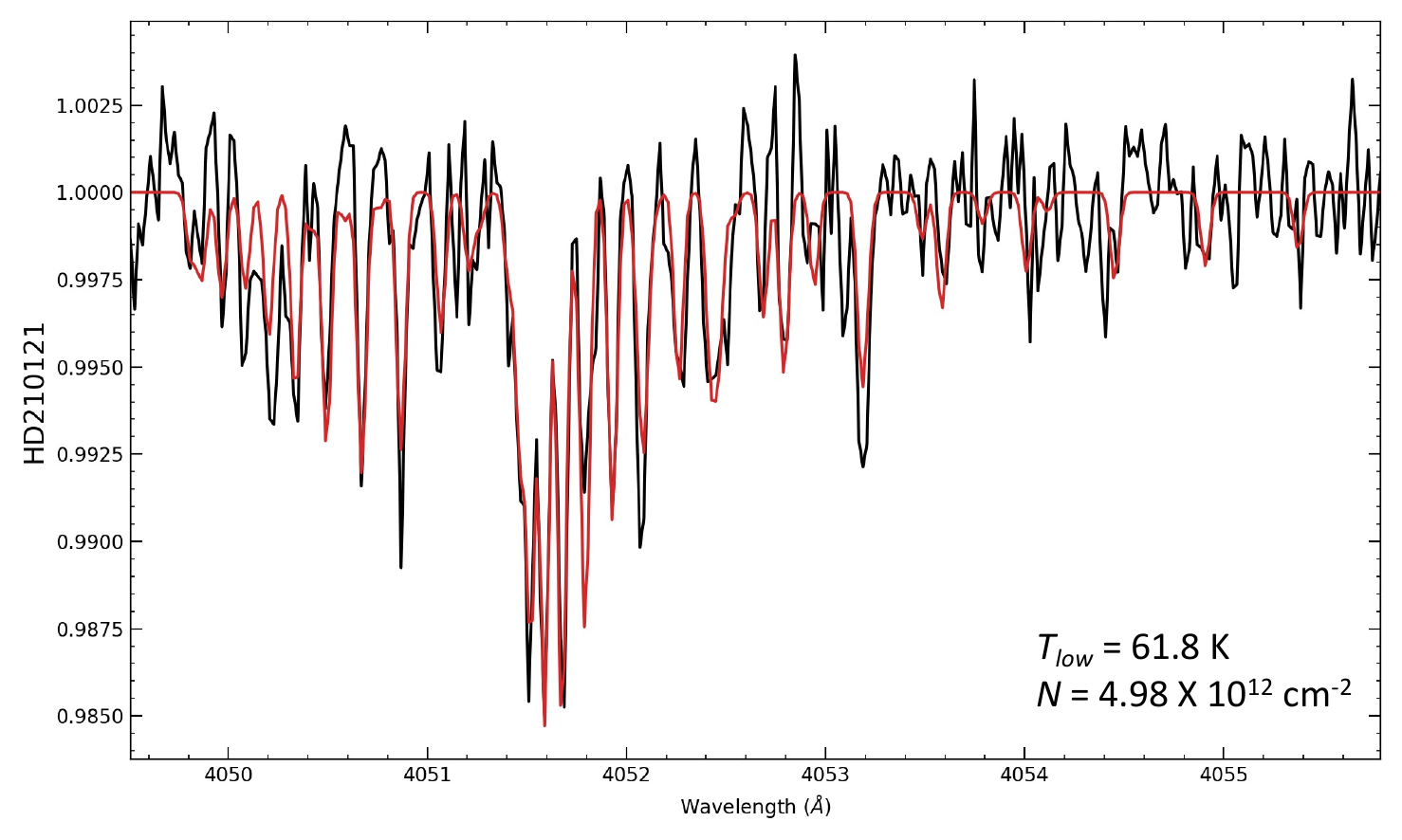}
\caption{Same as Fig.~\ref{Fig:fits_c3_app} but for HD~169454, HD~179406, HD~185859, HD~186745, HD~186841, HD~203532, and HD~210121. For HD~186841, synthetic spectra for individual velocity components are shown in orange and blue.}
\end{figure*}

\clearpage

\section{Physical conditions of C$_2$ sightlines}\label{app_C2_dense}

\begin{figure} [t]
    \includegraphics[width=\columnwidth]{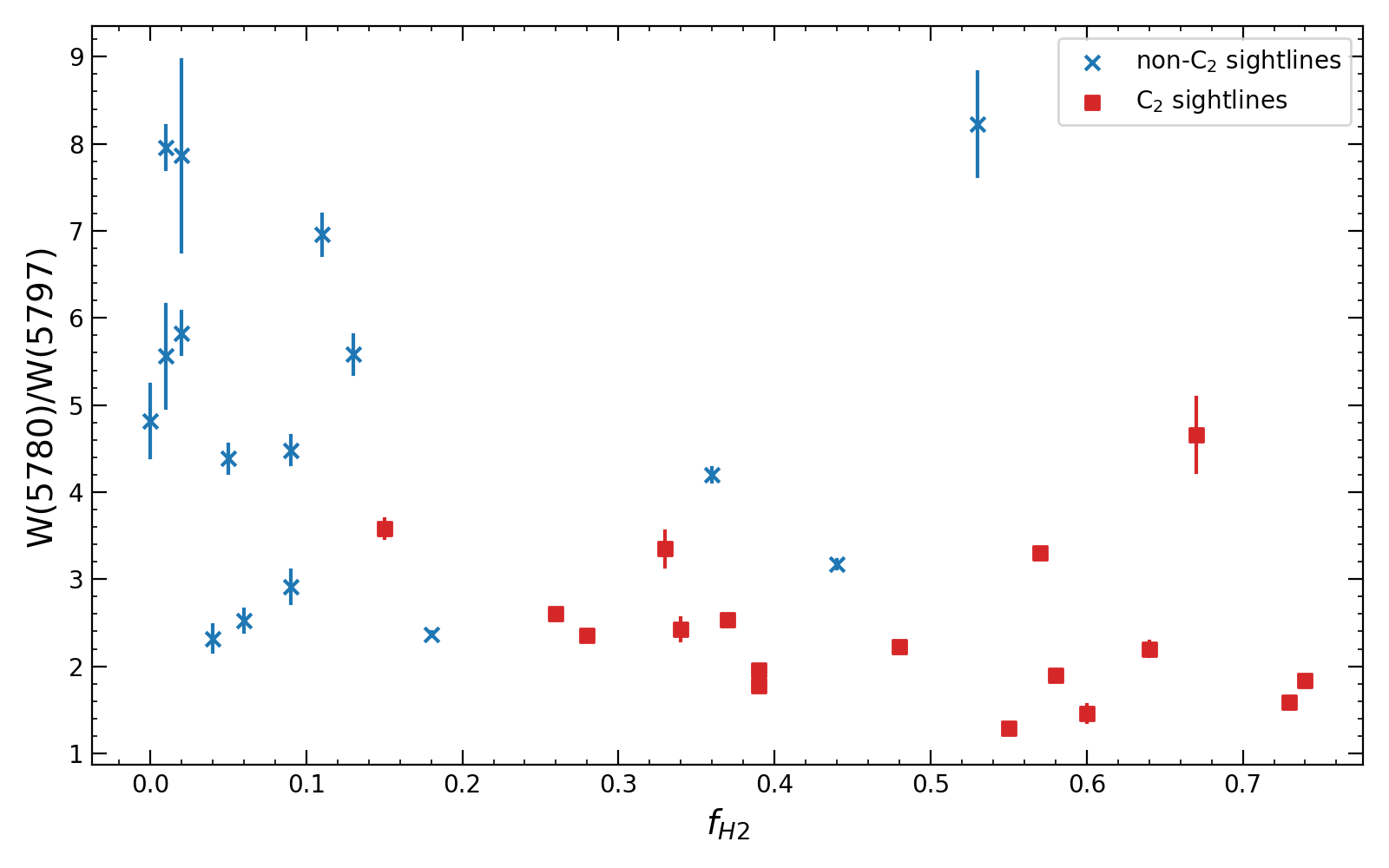}
    \caption{Behaviour of the $W$(5780)/$W$(5797) ratio at different $f_{H2}$ values. The DIB measurements are gathered from \cite{2017ApJ...850..194F}. The sightlines with and without C$_2$/C$_3$ detections are highlighted in different colours. We find the C$_2$ sightlines are mostly $\zeta$ types that trace the denser interiors of the ISM clouds.}
    \label{fig:C2_Dense}
\end{figure}

Figure \ref{fig:C2_Dense} compares the equivalent widths ratio between DIBs $\lambda\lambda$ 5780 and 5797 ($W$(5780)/$W$(5797)) among the C$_2$ and non-C$_2$ sightlines. The C$_2$ sightlines tend to have smaller $W$(5780)/$W$(5797) ratios and large mass fraction of molecular hydrogen ($f_\textrm{H2}$), both of which are characteristic for $\zeta$-type sightlines \citep[e.g.][]{2011A&A...533A.129V, 2013ApJ...774...72K}.

\begin{longtable}{l|ccccc|ccc} 
        \caption{\label{table: C2 Detection}Best-fit parameters of the C$_2$ model in our 40 sightlines, and literature values for comparison.  } \\
        \hline
                 & \multicolumn{5}{c|}{This Work} & \multicolumn{3}{c}{Literature} \\
        Sightline\tablefootmark{a} & $T_\textrm{kin}$ & $n$ & $N$ & $b$ & $v_\textrm{off}$\tablefootmark{b} & $T_\textrm{kin}$ & $n$ & Ref \\
     
                 & (K) & ($\textrm{cm}^{-3}$) & ($10^{13}\ \textrm{cm}^{-2})$ & (km $\cdot$ s$^{-1}$) & (km $\cdot$ s$^{-1}$) & (K) & ($\textrm{cm}^{-3}$)  & \\ 
        \hline
        HD~22951\tablefootmark{c} & 50.0$\substack{+28.2 \\ -5.4}$ & 123.6$\substack{+11.5 \\ -10.3}$ & 0.84$\substack{+0.06 \\ -0.06}$ & 2.67$\substack{+0.30 \\ -0.27}$ & 12.9$\substack{+1.2 \\ -1.2}$ \\

        HD~23016\tablefootmark{c} & 17.0$\substack{+11.4 \\ -2.0}$ & 177.9$\substack{+18.0 \\ -11.6}$ & 0.59$\substack{+0.07 \\ -0.04}$ & 0.32$\substack{+0.03 \\ -0.04}$ & 18.0$\substack{+1.1 \\ -1.1}$ \\ 

        HD~23180 & 37.7$\substack{+2.3 \\ -2.5}$ & 193.1$\substack{+4.4 \\ -5.3}$ & 2.49$\substack{+0.04 \\ -0.06}$ & 2.37$\substack{+0.12 \\ -0.06}$ & 13.3$\substack{+1.1 \\ -1.1}$ & 50 & 200$\substack{+250\\-50}$ & [1]\\

        HD~24398 & 51.1$\substack{+5.8 \\ -3.5}$ & 168.4$\substack{+5.8 \\ -5.6}$ & 2.08$\substack{+0.05 \\ -0.05}$ & 1.53$\substack{+0.08 \\ -0.09}$ & 13.9$\substack{+1.1 \\ -1.1}$ & 50 & 150$\substack{+200\\-50}$ & [1]\\

        HD~27778 & 40.1$\substack{+1.9 \\ -1.9}$ & 202.8$\substack{+5.2 \\ -4.9}$ & 3.39$\substack{+0.05 \\ -0.06}$ & 2.40$\substack{+0.07 \\ -0.07}$ & 15.3$\substack{+1.1 \\ -1.1}$ & 50$\pm$10 & 200$\substack{+25 \\ -25}$ & [1] \\
                
        HD~41117 & 50.9$\substack{+1.0 \\ -1.1}$ & 290.0$\substack{+7.6 \\ -7.3}$ & 1.12$\substack{+0.03 \\ -0.04}$ & 1.45$\substack{+0.11 \\ -0.09}$ & 15.1$\substack{+1.1 \\ -1.1}$ \\ 
                       & 49.0$\substack{+19.4 \\ -1.6}$ & 190.7$\substack{+33.9 \\ -7.7}$ & 0.69$\substack{+0.04 \\ -0.04}$ & 2.56$\substack{+0.22 \\ -0.23}$ & 9.7$\substack{+1.2 \\ -1.2}$ \\
                
                
        HD~45314 & 39.5$\substack{+2.4 \\ -2.4}$ & 257.6$\substack{+9.1 \\ -9.5}$ & 2.18$\substack{+0.06 \\ -0.05}$ & 2.52$\substack{+0.09 \\ -0.10}$ & 18.3$\substack{+1.1 \\ -1.1}$ \\ 
                
        HD~54239\tablefootmark{c} & 69.9$\substack{+12.6 \\ -8.1}$ & 899.3$\substack{+100.7 \\ -448.8}$ & 0.41$\substack{+0.05 \\ -0.04}$ & 1.07$\substack{+0.37 \\ -0.49}$ & 17.6$\substack{+1.2 \\ -1.2}$ \\
                
        HD~61827 & 43.2$\substack{+1.0 \\ -1.0}$ & 341.5$\substack{+9.1 \\ -7.7}$ & 4.51$\substack{+0.05 \\ -0.05}$ & 1.73$\substack{+0.04 \\ -0.04}$ & 41.8$\substack{+1.0 \\ -1.0}$ \\
                    & 50.8$\substack{+0.6 \\ -0.7}$ & 290.6$\substack{+4.6 \\ -3.8}$ & 3.80$\substack{+0.05 \\ -0.05}$ & 1.53$\substack{+0.05 \\ -0.05}$ & 35.8$\substack{+1.0 \\ -1.0}$ \\
                
        HD~63804 & 46.7$\substack{+0.7 \\ -0.7}$ & 277.9$\substack{+4.0 \\ -4.3}$ & 7.21$\substack{+0.07 \\ -0.06}$ & 3.47$\substack{+0.04 \\ -0.05}$ & 38.5$\substack{+1.0 \\ -1.0}$ \\ 
                
        HD~73882 & 16.6$\substack{+0.3 \\ -0.3}$ & 452.7$\substack{+8.4 \\ -6.2}$ & 4.27$\substack{+0.03 \\ -0.03}$ & 0.75$\substack{+0.04 \\ -0.04}$ & 22.1$\substack{+1.0 \\ -1.0}$ & 20$\pm$5 & 350$\substack{+500 \\ -150}$ & [1]\\ 
                
        HD~80558 & 65.5$\substack{+3.0 \\ -2.4}$ & 254.6$\substack{+9.5 \\ -9.6}$ & 2.27$\substack{+0.04 \\ -0.05}$ & 2.05$\substack{+0.08 \\ -0.07}$ & 19.5$\substack{+1.1 \\ -1.1}$ \\
                
        HD~112272 & 48.5$\substack{+5.7 \\ -3.2}$ & 169.3$\substack{+6.4 \\ -6.4}$ & 1.46$\substack{+0.04 \\ -0.04}$ & 1.39$\substack{+0.12 \\ -0.11}$ & 5.1$\substack{+1.1 \\ -1.1}$ \\ 
                       & 47.6$\substack{+8.2 \\ -9.1}$ & 136.2$\substack{+7.4 \\ -8.8}$ & 0.92$\substack{+0.04 \\ -0.05}$ & 1.41$\substack{+0.19 \\ -0.17}$ & -9.8$\substack{+1.1 \\ -1.1}$ \\

        HD~147084 & 42.2$\substack{+0.2 \\ -12.3}$ & 132.4$\substack{+0.2 \\ -72.9}$ & 2.01$\substack{+0.01 \\ -0.85}$ & 1.13$\substack{+2.05 \\ -0.01}$ & -7.1$\substack{+1.3 \\ -1.0}$ & 35$\pm$10 & 150$\substack{+50 \\ -25}$ & [2]\\ 
         & 37.4$\substack{+0.1 \\ -9.0}$ & 601.6$\substack{+198.4 \\ -82.2}$ & 3.13$\substack{+0.01 \\ -0.43}$ & 0.22$\substack{+0.01 \\ -0.03}$ & -6.0$\substack{+1.0 \\ -1.0}$ &  &  &  \\

        HD~147683 & 26.4$\substack{+3.6 \\ -2.1}$ & 230.9$\substack{+9.4 \\ -8.6}$ & 2.21$\substack{+0.07 \\ -0.07}$ & 1.12$\substack{+0.14 \\ -0.15}$ & -1.1$\substack{+1.1 \\ -1.1}$ & 55$\pm$5 & 95 & [3] \\ 
                
        HD~147888 & 41.7$\substack{+1.6 \\ -1.5}$ & 216.1$\substack{+4.4 \\ -4.1}$ & 1.89$\substack{+0.03 \\ -0.03}$ & 1.30$\substack{+0.05 \\ -0.06}$ & -8.4$\substack{+1.0 \\ -1.0}$ & 55$\pm$5 & 200$\substack{+50 \\ -25}$ & [1]\\ 
                
        HD~147889 & 40.9$\substack{+0.3 \\ -0.4}$ & 241.3$\substack{+0.6 \\ -1.6}$ & 12.52$\substack{+0.02 \\ -0.08}$ & 0.95$\substack{+0.05 \\ -0.01}$ & -7.8$\substack{+1.0 \\ -1.0}$ & 39$\pm$2 & 199$\pm$7& [4]\\ 
                
        HD~147933 & 34.9$\substack{+1.5 \\ -1.3}$ & 209.1$\substack{+3.9 \\ -4.1}$ & 2.51$\substack{+0.04 \\ -0.04}$ & 1.12$\substack{+0.07 \\ -0.06}$ & -8.6$\substack{+1.0 \\ -1.0}$ & & & [5]\\ 
                
        HD~148184 & 44.8$\substack{+2.0 \\ -2.3}$ & 205.6$\substack{+4.8 \\ -4.4}$ & 3.13$\substack{+0.05 \\ -0.05}$ & 0.42$\substack{+0.04 \\ -0.03}$ & -11.7$\substack{+1.0 \\ -1.0}$ & 45$\pm$12 & 209$\pm$60 & [4]\\ 
      & & & & & & 50$\pm$15 & 150$\substack{+100 \\ -25}$ & [1]\\

        HD~148937 & 44.7$\substack{+5.2 \\ -10.1}$ & 153.6$\substack{+9.4 \\ -9.0}$ & 1.21$\substack{+0.05 \\ -0.06}$ & 2.52$\substack{+0.17 \\ -0.15}$ & -14.6$\substack{+1.1 \\ -1.1}$ \\
                
        HD~149404 & 42.8$\substack{+4.0 \\ -3.4}$ & 186.4$\substack{+6.5 \\ -7.8}$ & 0.84$\substack{+0.03 \\ -0.04}$ & 1.00$\substack{+0.15 \\ -0.13}$ & -0.1$\substack{+1.1 \\ -1.1}$ \\ 
                       & 56.3$\substack{+4.8 \\ -2.5}$ & 260.4$\substack{+21.2 \\ -9.7}$ & 0.80$\substack{+0.03 \\ -0.03}$ & 1.31$\substack{+0.12 \\ -0.14}$ & -18.9$\substack{+1.1 \\ -1.1}$ \\
                
        HD~149757 & 42.6$\substack{+1.1 \\ -1.9}$ & 184.3$\substack{+4.3 \\ -4.6}$ & 1.88$\substack{+0.04 \\ -0.04}$ & 0.99$\substack{+0.09 \\ -0.09}$ & -14.7$\substack{+1.0 \\ -1.0}$ & 43$\pm$19 & 177$\pm$55 & [6] \\ 
        & & & & & & 20$\pm$10 & 250$\substack{+750 \\ -50}$ & [7]\\
        & & & & & & 30$\pm$10 & 250$\substack{+250 \\ -75}$ & [7]\\
        & & & & & & 50$\pm$ 5 & 150$\substack{+25\\-25}$ & [1]\\
  
        HD~152248 & 34.5 $\substack{+3.6 \\ -3.4}$ & 190.5$\substack{+7.8 \\ -6.6}$ & 1.04$\substack{+0.04 \\ -0.03}$ & 1.08$\substack{+0.12 \\ -0.14}$ & 2.0$\substack{+1.1 \\ -1.1}$ \\ 
                
        HD~154043 & 43.9$\substack{+1.4 \\ -1.8}$ & 204.2$\substack{+4.5 \\ -4.0}$ & 2.22$\substack{+0.04 \\ -0.03}$ & 1.75$\substack{+0.05 \\ -0.06}$ & -11.3$\substack{+1.0 \\ -1.0}$ \\ 

        HD~154368 & 24.1$\substack{+0.6 \\ -0.7}$ & 226.4$\substack{+2.0 \\ -1.8}$ & 5.11$\substack{+0.04 \\ -0.04}$ & 0.97$\substack{+0.04 \\ -0.04}$ & -2.8$\substack{+1.0 \\ -1.0}$ & 30$\pm$2 & 204$\pm$8 & [6]\\
        & & & & & & 20$\pm$5 & 150$\substack{+50 \\ -25}$ & [1]\\
                
        HD~161056 & 45.5$\substack{+0.8 \\ -1.2}$ & 201.2$\substack{+2.1 \\ -1.9}$ & 5.30$\substack{+0.04 \\ -0.03}$ & 2.24$\substack{+0.03 \\ -0.03}$ & -13.1$\substack{+1.0 \\ -1.0}$ \\ 
                 & 50.7$\substack{+0.6 \\ -0.8}$ & 291.6$\substack{+20.4 \\ -3.8}$ & 1.92$\substack{+0.03 \\ -0.03}$ & 1.60$\substack{+0.05 \\ -0.06}$ & -19.0$\substack{+1.0 \\ -1.0}$ \\
                
        HD~165319 & 79.1$\substack{+12.6 \\ -5.5}$ & 197.7$\substack{+8.6 \\ -26.8}$ & 1.15$\substack{+0.03 \\ -0.10}$ & 1.42$\substack{+0.49 \\ -0.09}$ & -1.5$\substack{+1.1 \\ -1.1}$ \\
                
        HD~167971 & 39.9$\substack{+6.9 \\ -1.0}$ & 171.4$\substack{+5.4 \\ -4.5}$ & 1.85$\substack{+0.06 \\ -0.07}$ & 3.83$\substack{+0.21 \\ -0.17}$ & -12.9$\substack{+1.2 \\ -1.1}$ \\ 
                & 37.0$\substack{+20.3 \\ -1.7}$ & 164.0$\substack{+11.4 \\ -8.1}$ & 0.80$\substack{+0.06 \\ -0.05}$ & 2.84$\substack{+0.33 \\ -0.33}$ & -1.9$\substack{+1.2 \\ -1.2}$ &  & & [5]\\ 
                & 54.8$\substack{+1.0 \\ -1.2}$ & 320.1$\substack{+21.2 \\ -7.7}$ & 1.80$\substack{+0.05 \\ -0.05}$ & 2.35$\substack{+0.12 \\ -0.10}$ & 11.1$\substack{+1.1 \\ -1.1}$ \\

        HD~169454 & 18.0$\substack{+0.2 \\ -0.2}$ & 369.0$\substack{+2.1 \\ -2.1}$ & 6.74$\substack{+0.03 \\ -0.03}$ & 0.66$\substack{+0.03 \\ -0.03}$ & -8.7$\substack{+1.0 \\ -1.0}$ & 19$\pm$2 & 326$\pm$17 & [4]\\ 
        & & & & & & 20$\pm$5 & 200$\substack{+50 \\ -25}$ & [1]\\
                & & & & & & 25 & 400$\pm$100 & [8]\\
  
        HD~170740 & 33.2$\substack{+2.8 \\ -1.9}$ & 172.1$\substack{+4.3 \\ -3.7}$ & 1.59$\substack{+0.04 \\ -0.03}$ & 1.76$\substack{+0.06 \\ -0.09}$ & -8.8$\substack{+1.1 \\ -1.1}$ & 21$\pm$9 & 161$\pm$16 & [6]\\ 
                
        HD~170938 & 54.0$\substack{+9.5 \\ -4.3}$ & 141.9$\substack{+7.3 \\ -5.2}$ & 1.31$\substack{+0.05 \\ -0.05}$ & 2.49$\substack{+0.16 \\ -0.16}$ & -9.3$\substack{+1.1 \\ -1.1}$ \\ 
                
        HD~171957 & 44.0$\substack{+3.0 \\ -6.3}$ & 194.5$\substack{+8.7 \\ -8.7}$ & 0.98$\substack{+0.03 \\ -0.04}$ & 0.98$\substack{+0.16 \\ -0.15}$ & -4.7$\substack{+1.1 \\ -1.1}$ \\ 
                
        HD~179406 & 35.7$\substack{+0.7 \\ -0.8}$ & 251.1$\substack{+3.3 \\ -3.3}$ & 4.95$\substack{+0.04 \\ -0.05}$ & 1.39$\substack{+0.04 \\ -0.04}$ & -12.0$\substack{+1.0 \\ -1.0}$ & 38$\pm$9 & 251$\pm$83 & [4]\\

        HD~183143 & 18.0$\substack{+13.5 \\ -5.9}$ & 188.4$\substack{+20.3 \\ -13.3}$ & 0.70$\substack{+0.19 \\ -0.02}$ & 2.13$\substack{+0.12 \\ -0.39}$ & -10.4$\substack{+1.2 \\ -1.3}$ &  & & [5]\\ 
                 & 96.3$\substack{+3.8 \\ -5.9}$ & 240.2$\substack{+28.5 \\ -24.6}$ & 1.00$\substack{+0.06 \\ -0.05}$ & 1.98$\substack{+0.21 \\ -0.21}$ & 4.7$\substack{+1.1 \\ -1.2}$ \\
                
        HD~185418 & 31.6$\substack{+5.3 \\ -4.8}$ & 131.3$\substack{+6.9 \\ -6.0}$ & 1.48$\substack{+0.06 \\ -0.06}$ & 2.74$\substack{+0.18 \\ -0.18}$ & -10.6$\substack{+1.1 \\ -1.1}$ &  \\ 
                
        HD~185859 & 44.1$\substack{+1.0 \\ -0.9}$ & 202.5$\substack{+3.7 \\ -3.2}$ & 2.24$\substack{+0.03 \\ -0.03}$ & 1.18$\substack{+0.05 \\ -0.05}$ & -7.5$\substack{+1.0 \\ -1.0}$ \\
                
        HD~186745 & 43.0$\substack{+0.8 \\ -0.9}$ & 228.1$\substack{+2.7 \\ -2.8}$ & 5.74$\substack{+0.05 \\ -0.05}$ & 1.95$\substack{+0.04 \\ -0.03}$ & -10.6$\substack{+1.0 \\ -1.0}$ \\ 
                       & 25.2$\substack{+3.6 \\ -2.6}$ & 193.1$\substack{+7.2 \\ -7.2}$ & 1.41$\substack{+0.05 \\ -0.05}$ & 1.89$\substack{+0.12 \\ -0.13}$ & -5.4$\substack{+1.1 \\ -1.1}$ \\

        HD~186841 & 41.5$\substack{+1.0 \\ -0.9}$ & 218.2$\substack{+2.5 \\ -2.5}$ & 5.25$\substack{+0.05 \\ -0.04}$ & 1.84$\substack{+0.04 \\ -0.03}$ & -11.4$\substack{+1.0 \\ -1.0}$ \\ 
                       & 87.8$\substack{+1.2 \\ -13.1}$ & 270.7$\substack{+11.3 \\ -10.3}$ & 1.50$\substack{+0.05 \\ -0.03}$ & 0.92$\substack{+0.10 \\ -0.14}$ & -5.6$\substack{+1.1 \\ -1.1}$ \\

        HD~203532 & 39.5$\substack{+1.0 \\ -1.7}$ & 247.7$\substack{+3.6 \\ -4.5}$ & 3.98$\substack{+0.03 \\ -0.05}$ & 1.30$\substack{+0.05 \\ -0.03}$ & 15.2$\substack{+1.0 \\ -1.0}$ & 30$\pm$10 & 210 & [3]\\
                
        HD~210121 & 41.8$\substack{+0.8 \\ -0.9}$ & 330.2$\substack{+7.0 \\ -6.7}$ & 5.30$\substack{+0.06 \\ -0.06}$ & 0.87$\substack{+0.05 \\ -0.06}$ & -13.7$\substack{+1.0 \\ -1.0}$ & 45$\pm$5 & 200$\substack{+50 \\ -50}$ & [1] \\
        \hline
  HD 23478 & & & & & & 30$\pm$ 10       & 190 & [3]\\
  HD 24534 & & & & & & 45$\pm$10 & 200$\substack{+50\\-50}$     & [1]\\
  HD 28975 & & & & & & 30       & 125 & [9]\\
  HD 29647 & & & & & & 10       & $\ge$ 175      & [9]\\
           & & & & & & 14$\pm$8 & 350$\substack{+400\\-150}$    & [10]\\
  HD 62542 & & & & & & 40$\pm$5 & 250$\substack{+50\\-25}$       & [11]\\
           & & & & & & 30$\pm$ 10 & 350$\substack{+400\\-100}$ & [1]\\
  HD 76341 & & & & & & 34$\pm$20 & 300$\pm$200 & [4]\\
  HD 80077 & & & & & & 35        & 600$\pm$300 & [8]\\
           & & & & & & 20$\pm$ 5 & 200$\substack{+100\\-25}$    & [1]\\
  HD 114213 & & & & & & 25 & 400$\pm$100 & [8]\\
  HD 115842 & & & & & & 53$\pm$30 & 242$\pm$217 & [6]\\
  HD 136239 & & & & & & 29$\pm$5  & 226$\pm$30  & [6]\\
  HD 148379 & & & & & & 38$\pm$27 & 152$\pm$77 & [6]\\
  HD 151932 & & & & & & 33$\pm$8 & 163$\pm$ 20 & [6]\\
  HD 152236 & & & & & & 27$\pm$7 & 176$\pm$24   & [6]\\
  HD 154445 & & & & & & 18$\pm$18 & 101$\pm$17 & [6]\\
  HD 163800 & & & & & & 38$\pm$15 & 169$\pm$43 & [4]\\
  HD 172028 & & & & & & 30 & 300$\pm$100 & [8]\\
            & & & & & & 40$\pm$10       & 250$\substack{+100\\-50}$& [1]\\
  HD 177989 & & & & & & 20        & 210  & [3]\\
  HD 192035     & & & & & & 25$\pm$ & 350        & [3]\\
  HD 198781   & & & & & & 55$\pm$45 & $\le$45 & [3]\\
  HD 200775     & & & & & & 15$\pm$5  & 200$\substack{+50\\-25}$        & [1]\\
  HD 204827     & & & & & & 40$\pm$5 & 300$\substack{+50\\-50}$ & [1]\\
  HD 206267     & & & & & & 35$\pm$5 & 250$\substack{+100\\-25}$        & [1]\\
              & & & & & & 30$\pm$10     & 280 & [3]\\
  HD 207198     & & & & & & 60$\pm$10 & 200$\substack{+25\\-50}$        & [1]\\
  HD 207308     & & & & & & 35$\pm$15   & 190 & [3]\\
  HD 207538     & & & & & & 35$\pm$15 & 150$\substack{+25\\-25}$        & [1]\\
  HD 208266     & & & & & & 10 & 210 & [3]\\
  HD 210839     & & & & & & 30 & 100$\substack{+50\\-25}$ & [1]\\
  HD 220057     & & & & & & 90 & 165 & [3]\\
  BD+66$^{\circ}$ 1661 & & & & & & 35$\pm$10 & 700$\pm$200 & [12] \\
  BD+66$^{\circ}$ 1674 & & & & & & 60$\pm$10 & 800$\pm$400 & [12]\\
  BD+66$^{\circ}$ 1675 & & & & & & 35$\pm$10 & 700$\pm$200 & [12]\\ 
  Cyg OB2 5     & & & & & & 50 & 300$\pm$100 & [13]\\
              & & & & & & 35$\pm$15     & 150$\substack{+50\\-25}$ & [1]\\
  Cyg OB2 12  & & & & & & 35 & 300$\pm$ 50 & [13]\\
              & & & & & & 40$\pm$15     & 200$\substack{+75\\-50}$ & [1]\\
  Cyg OB2 12, Comp. 2 & & & & & & 30$\pm$5 & 100$\pm$7 & [14]\\
  Cyg OB2 12, Comp. 3 & & & & & & 25$\pm$5 & 125$\pm$7 & [14]\\
  SN2008fp      & & & & & & 30 & 250 & [15]\\
  Cernis 52     & & & & & & 40$\pm$10 & 250$\pm$50       & [16]\\
  Sk~143                & & & & & & 25 & 600 & [17]\\
\end{longtable}
\tablebib{%
[1] \citet{2007ApJS..168...58S}; 
[2] \citet{1989ApJ...340..273V}; 
[3] \citet{2012ApJ...761...38H}; 
[4] \citet{2010MNRAS.402.2548K}; 
[5] \citet{2003ApJ...584..339T};  
[6] \citet{2010MNRAS.408.1590K}; 
[7] \citet{1997MNRAS.290...41C} 
[8]  \citet{1999A&A...351..657G}; 
[9] \citet{2021ApJ...914...59F}; 
[10] \citet{1983ApJ...271L..95H}; 
[11] \citet{2020ApJ...897...36W}; 
[12] \citet{2004A&A...425..151G}; 
[13] \citet{2001A&A...375..553G}; 
[14] \citet{2019ApJ...881..143H}; 
[15] \citet{2014A&A...565A..61C}; 
[16] \citet{2011MNRAS.411.1857I}; 
[17] \citet{2013MNRAS.428.1107W}}
  \tablefoot{The last part of the table (below the line) contains literature values only, and only for sightlines for which both $T_{\rm kin}$ and $n$ are determined.\\
\tablefoottext{a}{For sightlines that harbour multiple velocity components, the best-fit results are listed for each component in separate entries.}\\
  \tablefoottext{b}{Relative to the barycentric frame.}\\
\tablefoottext{c}{Tentative detection where no individual transition reached the 3$\sigma$ detection limit.}\\
}

\end{appendix}
\end{document}